\documentclass[11pt, a4paper]{article}

\usepackage{amsmath}
\usepackage{amsfonts}
\usepackage{amssymb}
\usepackage{mathrsfs}
\usepackage{bm}
\usepackage{graphicx, rotating}
\usepackage{epstopdf}
\usepackage{epsfig}
\usepackage{latexsym}
\usepackage{color}
\usepackage[dvipsnames]{xcolor}
\usepackage{cite}
\usepackage{slashed}
\usepackage{hyperref}
\usepackage{comment}
\usepackage[utf8]{inputenc}
\usepackage{soul}
\usepackage{multirow}
\usepackage{graphics,subfigure}

\hypersetup{colorlinks, citecolor=bluscuro, linkcolor=black, urlcolor=bluscuro}
\definecolor{bluscuro}{rgb}{0.15, 0.2, .85}

\newcommand{\be}{\begin{equation}}
\newcommand{\ee}{\end{equation}}
\newcommand{\bea}{\begin{eqnarray}}
\newcommand{\eea}{\end{eqnarray}}

\def\simlt{\stackrel{<}{{}_\sim}}
\def\simgt{\stackrel{>}{{}_\sim}}

\newcommand{\GeV}{\,\mathrm{GeV}}

\newcommand\blfootnote[1]{%
  \begingroup
  \renewcommand\thefootnote{}\footnote{#1}%
  \addtocounter{footnote}{-1}%
  \endgroup
}

 \def\bea{\begin{eqnarray}}
  \def\eea{\end{eqnarray}}
 \def\al{\alpha}

 \def\De{\Delta}
 
 \def\ep{\varepsilon}
 
 \def\th{\theta}

 \def\Om{\Omega}

	\def \beq {\begin{equation}}
	\def \eeq {\end{equation}}
	\def \ba {\begin{array}}
	\def \ea {\end{array}}
	\def\simlt{\stackrel{<}{{}_\sim}}
	\def\simgt{\stackrel{>}{{}_\sim}}
	\def \ecart {\noalign{\medskip}}
	\def \dis {\displaystyle}

	\def \Om {\Omega}
	\def \lam {\lambda}
	
	\def \ZZ {\mathbb Z}
	\def \al {\alpha}
	\def \be {\beta}

	\def \De {\Delta}
	\def \ep {\epsilon}

	\def \M {\mathcal{M}}

\newcommand{\orcid}{\includegraphics{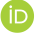}}
\newcommand{\orcidlink}[1]{\href{https://orcid.org/#1}{{\orcid}}}

\begin{document}

\begin{titlepage}
\begin{flushright}
IFT-UAM/CSIC-22-12
\end{flushright}
\begin{center} ~~\\
\vspace{0.5cm} 
\Large { \bf\Large Fine-Tuning in the 2HDM} 
\vspace*{1.5cm}

\normalsize{
{\bf 
A. Bernal\orcidlink{0000-0003-3371-5320}\blfootnote{alexander.bernal@csic.es},
J. A. Casas\orcidlink{0000-0001-5538-1398}\blfootnote{
j.alberto.casas@gmail.com},
and J.M. Moreno\orcidlink{0000-0002-2941-0690}
\blfootnote{
jesus.moreno@csic.es}
 } \\
\smallskip  \medskip
\emph{Instituto de F\'\i sica Te\'orica, IFT-UAM/CSIC,}\\
\emph{Universidad Aut\'onoma de Madrid, Cantoblanco, 28049 Madrid, Spain}}

\medskip

\vskip0.6in 

\end{center}

\centerline{ \large\bf Abstract }
\vspace{.5cm}
\noindent

The Two-Higgs Doublet Model (2HDM) is one of the most popular and natural extensions of the Higgs sector; but it has two potential fine-tuning problems, related to the electroweak (EW) breaking and the requirement of  alignment with the SM Higgs boson.
We have quantified the fine-tunings obtaining analytical expressions, both in terms of the initial 2HDM parameters and the physical ones (masses, mixing angles, etc.).
We also provide simple approximate expressions. 
We have taken into account that the fine-tunings are not independent and 
removed  the ``double counting" by projecting the variations of the alignment onto the constant-$v^2$ hypersurface.
The EW and the alignment fine-tunings become severe in different, even opposite, regions of the parameter space, namely in the regimes of large and small extra-Higgs masses, respectively; emerging an intermediate region,  $500\ {\rm GeV} \simlt \{m_H, m_A, m_{H^\pm}\} \simlt 700\ {\rm GeV}$, where both are acceptably small. 
We also discuss a remarkable trend that is not obvious at first glance. Namely, for large $\tan\beta$ both the EW and the alignment fine-tunings become mitigated.  In consequence, the 2HDM becomes quite natural for $\tan\beta\geq {\cal O} (10)$, even if $m_H, m_A, m_{H^\pm}$ are as large as 1500 GeV. We explain why this is not the case for the 2HDM stemming from supersymmetry.
We have illustrated all these trends by numerically analyzing several representative scenarios.

\vspace*{2mm}
\end{titlepage}


\section{Introduction}
\label{sec:Intro}

\quad The Two-Higgs Doublet Model (2HDM) \cite{Lee:1973iz,Gunion:1989we,Branco:2011iw} is one of the most popular extensions of the Higgs sector. It naturally stems in prominent beyond-the-Standard-Model (BSM) scenarios, such as grand unification \cite{Georgi:1974sy,Ross:84}, supersymmetry \cite{Martin:1997ns} and axion models \cite{Peccei:1977hh}. Besides, it has the potential to improve the performance of the conventional Higgs sector for important issues, e.g. the appealing possibility of Higgs-portal dark matter (see \cite{Barbieri:2006dq,LopezHonorez:2006gr,Arcadi:2019lka, Bell:2017rgi, Berlin:2015wwa, Cabrera:2020lmg}). On top of all this, the 2HDM represents the simplest non-trivial extension of the SM Higgs sector, and thus a test bed of BSM physics, which could be probed in present and future experiments.

The scalar sector of the 2HDM consists of two scalar doublets, which leads to five physical scalar fields ($h, H^0, H^\pm, A$), unlike the sole SM Higgs boson. On the other hand, the current experimental data require that the properties of one of these fields (typically the lightest CP-even Higgs mass eigenstate, $h$), such as quantum numbers, mass and couplings, are equal or very similar to those of the SM Higgs boson. This is called the alignment limit\cite{Gunion:2002zf, Craig:2012vn,Craig:2013hca,Asner:2013psa,Carena:2013ooa,Haber:2013mia}, which is essentially imposed by the observation.

Apart from its theoretical appeal, the 2HDM has two potential fine-tuning problems. The first one is related to the 
 electroweak (EW) breaking, when the magnitude of the VEV, $v^2=(246\ {\rm GeV})^2$, is much smaller than  the  squared-mass terms entering the theory.  The second one is related to the alignment condition. 
 The main goal of this paper is to explore
these fine-tunings, looking for regions of the parameter space where both are mild or irrelevant. 

In this analysis we will adopt an ``agnostic" point of view, in the sense that we will not consider additional symmetries which constrain the model (beyond the usual $\ZZ_2$ parity to avoid FCNC). There have been several analyses exploring this alternative direction, in particular to get an exact or approximate alignment \cite{BhupalDev:2014bir,BhupalDev:2017txh,
Draper:2020tyq,Haber:2021zva,Lane:2018ycs, Eichten:2021qbm}.
 As pointed out in ref. \cite{Draper:2020tyq}, an approximate alignment can arise from a softly broken global symmetry of the scalar potential, but this requires to extend the Yukawa sector, e.g. with vector-like top quark partners.

\vspace{0.2 cm}
In section \ref{2HDM} we present the generic 2HDM, fixing the notation and providing analytical expressions for the Higgs VEVs and the alignment parameter. In section \ref{FT2HDM} we explain the method to evaluate the fine-tunings and discuss the expected regions where they become severe. In section \ref{EvFT} we provide analytical expression for all the potential fine-tunings, both in terms of the initial 2HDM parameters and the physical ones (masses,  mixing angles, etc.) We also explain how to remove the “double counting” of tunings by projecting  the  variations  of  the alignment  onto  the  constant-$v^2$ hypersurface. In addition, simple approximate expressions are also provided. 
The analysis shows that
the electroweak and the alignment fine-tunings become severe in different regions of the parameter space.
In section \ref{NumAn} we illustrate all these trends by numerically analyzing several representative scenarios. This allows to identify the regions where both fine-tunings are acceptably mild or even irrelevant.
Finally, in section \ref{Conclusions} we present our conclusions. The complete analytical expressions for all the fine-tunings are given in the Appendix.

\section{The Two-Higgs Doublet Model} \label{2HDM}

Let us briefly review the general formulation of the 2HDM following the notation and conventions of ref.~\cite{Bernon:2015qea}. Denoting by
\beq\label{Z2Basis} 
\Phi_1=\left(\ba{c} \Phi_1^+\\ \ecart\dis \Phi_1^0 \ea\right),\quad \Phi_2=\left(\ba{c} \Phi_2^+\\ \ecart\dis \Phi_2^0 \ea\right) 
\eeq
the two complex $Y=1/2,\, SU(2)_L$ doublet scalar fields, the most general gauge-invariant renormalizable scalar potential is given by:
\begin{eqnarray} 
\label{pot}
{V}&=& m_{11}^2\Phi_1^\dagger\Phi_1+m_{22}^2\Phi_2^\dagger\Phi_2
-[m_{12}^2\Phi_1^\dagger\Phi_2+{\rm h.c.}]+\nonumber\\[8pt]
&&\frac{1}{2}\ ,\lambda_1(\Phi_1^\dagger\Phi_1)^2
+\frac{1}{2}\ ,\lambda_2(\Phi_2^\dagger\Phi_2)^2
+\lambda_3(\Phi_1^\dagger\Phi_1)(\Phi_2^\dagger\Phi_2)
+\lambda_4(\Phi_1^\dagger\Phi_2)(\Phi_2^\dagger\Phi_1)+\nonumber\\[8pt]
&&\left\{\frac{1}{2}\ ,\lambda_5(\Phi_1^\dagger\Phi_2)^2
+\big[\lambda_6(\Phi_1^\dagger\Phi_1)
+\lambda_7(\Phi_2^\dagger\Phi_2)\big]
\Phi_1^\dagger\Phi_2+{\rm h.c.}\right\}\,,
\end{eqnarray} 
where the $m_{11}^2,\, m_{22}^2$ mass terms and the $\lam_{1,2,3,4}$ couplings are real, while  $m_{12}^2$ and $\lam_{5,6,7}$ could be complex.  The absence of flavour changing neutral currents (FCNCs) essentially requires each type of quark to couple to just one scalar doublet \cite{Glashow:1976nt, Paschos:1976ay}. This is accomplished by means of a softly-broken $\ZZ_2$ symmetry, which implies $\lambda_6 = \lambda_7=0$, while a non-zero $m_{12}^2$ value is still possible, see table \ref{table:2HDM_Types} below. 

Furthermore, to avoid dangerous CP-violating phenomena, we will assume throughout the paper that $m_{12}^2$, $\lam_5$ are real.
In addition, stability bounds on the quartic coefficients imply (see e.g. \cite{ElKaffas:2006gdt})
\begin{eqnarray} \label{PertStab}
\lambda_1, \lambda_2  &>& 0\,, \nonumber \\   
\lambda_3  &>&  - \sqrt{ \lambda_1  \lambda_2}\,,  \\
\lambda_3 +\lambda_4-  |\lambda_5 |&>&  - \sqrt{ \lambda_1  \lambda_2}\,. \nonumber 
\end{eqnarray}  
We have also imposed  constraints on the size of the various couplings from perturbativity and the requirement  that unitarity is not violated in 2HDM scalar scattering processes, along the lines of Refs~\cite{Akeroyd:2000wc,Ginzburg:2005dt,Bhattacharyya:2015nca}.

As long as the Higgs mass matrix possesses at least one negative eigenvalue, the scalar fields develop non-zero vacuum expectation values (VEVs):
\beq\label{potmin}
\langle \Phi_1 \rangle=\frac{1}{\sqrt{2}} \left(
\begin{array}{c} 0\\ v_1\end{array}\right), \qquad \langle
\Phi_2\rangle=
\frac{1}{\sqrt{2}} \left(\begin{array}{c}0\\ v_2
\end{array}\right)\,,
\eeq
which must satisfy 
$v_1^2+v_2^2=v^2\simeq(246\ \GeV)^2$. As usual, we define the $\beta$ angle such that
\beq
v_1 = v \cos \beta = v \, c_\beta, \ \ \ \   v_2 = v \sin \beta= v \, s_\beta\ .
\eeq

The vacuum is CP-conserving provided
$|m_{12}^2|\geq\lam_5|v_1||v_2|$ \cite{Gunion:2002zf}. Besides, the fields can be redefined so that $v_1, v_2\geq 0$.
 
 The minimization conditions $\partial_{\Phi_i} {V}|_{(v_1,v_2)}=0$ read
\begin{eqnarray} \label{eqminv12}
2 m_{11}^2 v_1 - 2 m_{12}^2  v_2 +  v_1^3 \lambda_1 + v_1  v_2^2  \lambda_{345}   & = & 0\,, \\
- 2 m_{12}^2 v_1 + 2 m_{22}^2  v_2 +  v_2^3 \lambda_2 + v_1^2  v_2  \lambda_{345}  & = & 0\,,
\label{eqminv2}
\end{eqnarray} 
where $ \lambda_{345} = \lambda_3 + \lambda_4 + \lambda_5$. 

A special instance takes place when either $v_1=0$ or $v_2=0$, i.e. $t_\beta=\tan \beta=0,\infty$. This requires $m_{12}^2=0$ and corresponds to the so-called inert-doublet model \cite{Barbieri:2006dq}. This case is protected by the symmetry $\Phi_1\rightarrow \Phi_1$, $\Phi_2\rightarrow -\Phi_2$, and represents a somehow trivial situation where one of the two Higgs fields plays exactly the SM-Higgs role while the other one does not couple to the rest of the SM fields.

For the non-trivial cases ($t_\beta\neq 0,\infty$), we can write the previous conditions in terms of $v$ and $t_\beta$, 
\begin{eqnarray} \label{dV1}
\mathscr{P}_1&\equiv& 2 m_{11}^2 (1+  t_\beta^2)  - 2 m_{12}^2 ( t_\beta + t_\beta^3) + v^2 \left(  \lambda_1 + t_\beta^2  \lambda_{345}   \right)  =  0\,, \\
\mathscr{P}_2&\equiv& - 2 m_{12}^2 (1+  t_\beta^2)  + 2 m_{22}^2 ( t_\beta + t_\beta^3) +  v^2  \left( t_\beta ^3  \lambda_2 +t_\beta  \lambda_{345}   \right)  =  0\,. \label{dV2}
\end{eqnarray}

Eliminating $v$ from Eqs.(\ref{dV1}) and (\ref{dV2}), we get a quartic equation for $t_\be$:
\begin{equation}\label{eqbeta}
\mathscr{P}_\beta \ \  \equiv\ \  m_{12}^2 (-\lambda_1+  t_\beta^4 \lambda_2)  -
   m_{11}^2 \left( t_\beta ^3  \lambda_2 +  t_\beta  \lambda_{345} \right) +
    m_{22}^2  \left( t_\beta   \lambda_1 +t_\beta ^3  \lambda_{345}  \right) = 0,
\end{equation} 
which allows the replacement of either $\mathscr{P}_1=0$ or $\mathscr{P}_2=0$ by $\mathscr{P}_\beta=0$ in the couple of  minimization equations (\ref{dV1}, \ref{dV2}).

The  $\be$ angle introduced before  relates the initial basis (in which the $\ZZ_2$-symmetry is apparent) to the so-called Higgs-basis
\beq\label{HiggsBasis} 
H_1=\left(\ba{c} H_1^+\\ \ecart\dis H_1^0 \ea\right)\equiv\Phi_1 c_\be +\Phi_2 s_\be\,,\quad H_2=\left(\ba{c} H_2^+\\ \ecart\dis H_2^0 \ea\right)\equiv -\Phi_1 s_\be +\Phi_2 c_\be\,,
\eeq
which satisfies $\langle H_1^0\rangle=v/\sqrt{2}$, $\langle H_2^0\rangle=0$.
The associated quadratic (quartic) parameters of the Higgs scalar potential in this basis are denoted by $Y_i$ ($Z_j$), where $i\in\{1,2,3\}$ ($j\in\{1,\dots, 7\}$) (expressions for $Y_i, Z_j$ in terms of the initial parameters, $m_{ij}$, $\lambda_j$ can be found in ref.~\cite{Bernon:2015qea}).

From the original eight scalar degrees of freedom in the two Higgs  doublets, three Goldstone bosons are absorbed by the electroweak bosons, $W^\pm$ and $Z$, and the remaining degrees of freedom correspond to five physical Higgs particles: two CP-even scalars ($h$ and $H$,  with the convention $m_h\leq m_H$), one CP-odd scalar ($A$) and one pair of charged Higgses ($H^\pm$).

The $A$ and $H^\pm$ fields stem directly from the above $H_2$ doublet, namely $A=\sqrt{2}\ {\rm Im}(H_2^0)$, $H^+=H_2^+$, $H^-=(H_2^+)^\dagger=H_2^-$, with masses given by
\begin{eqnarray}
\label{AHcMasses}
&&m_A^2=m_{12}^2\dfrac{1+t_\be^2}{t_\be}-v^2 \lam_5, 
\\
&&m_{H^\pm}^2=m_A^2+\dfrac{1}{2}v^2 (\lam_5-\lam_4)\,.
\label{AHcMasses2}
\end{eqnarray}
On the other hand, in the initial basis the CP-even neutral Higgs fields, $(\sqrt{2} {\rm Re}\Phi_1^0-v_1), (\sqrt{2} {\rm Re}\Phi_2^0-v_2)$ mix through the squared-mass matrix
\begin{equation}
\M^2=\left(
\begin{array}{cc}
m_{11}^2 +\frac{1}{2}v^2(3\lambda_1c_\beta^2 + \lambda_{345}s_\beta^2)
& m_{12}^2+\lambda_{345}v^2s_\beta c_\beta
\\
 m_{12}^2+\lambda_{345}v^2s_\beta c_\beta & m_{22}^2 +\frac{1}{2}v^2(3\lambda_2s_\beta^2 + \lambda_{345}c_\beta^2) \end{array}
\right)\ ,
\label{massmatrix}
\end{equation}
which is diagonalized by 
\begin{equation} 
\label{masseigenvalues}
\left(\begin{array}{cc} m_H^2 & 0 \cr 0 &m_h^2\end{array}\right)=
\left(\begin{array}{cc} c_\alpha & s_\alpha  \cr -s_\alpha  & c_\alpha  \end{array}\right)
\left(\begin{array}{cc} \M_{11}^2 & \M_{12}^2 \\
\M_{12}^2 &\M_{22}^2\end{array}\right)
\left(\begin{array}{cc}   c_\alpha & -s_\alpha  \cr s_\alpha  & c_\alpha \end{array}\right)
\end{equation} 
(with the usual notation, $c_\alpha=\cos \al$, $s_\al=\sin \al$).

Obviously, for the trivial cases $s_\beta=0$ ($c_\beta=0$), i.e. $t_\beta=0$ ($\infty$), the mass matrix (\ref{massmatrix}) is diagonal from the beginning, since $m_{12}^2=0$. Then the SM-Higgs corresponds to $\Phi_1$ $(\Phi_2)$  and $\alpha=-\pi/2$ $(0)$
. Otherwise,
$\alpha$ is determined by the equation
\begin{equation}   \label{eqalpha}
\mathscr{P}_\alpha\ \equiv \ 
    m_{12}^2  (1+ t_\beta^2) ( t_\alpha -   t_\beta)(1 + t_\alpha t_\beta) 
    + v^2  t_\beta \left(
                               t_\alpha ( - \lambda_1  + t_\beta ^2  \lambda_2) + t_\beta (1-  t_\alpha^2 ) 
                                 \lambda_{345} 
                             \right)=0\,,
\end{equation} 
where we have used Eqs.(\ref{dV1}, \ref{dV2}) to eliminate $m_{11}^2, m_{22}^2$.
The $\al$ angle is defined modulo $\pi$ and we have chosen the convention $-\pi/2\leq \alpha\leq \pi/2$, thus $c_\alpha\geq 0$.
The mass eigenvalues (\ref{masseigenvalues}) for the light and heavy neutral Higgses are then given by

\begin{eqnarray}\label{eqmH}
m_H^2&=&\frac{m_{12}^2 (t_\al - t_\be)^2 (1 + t_\be^2) + t_\be v^2 (\lam_1 + t_\al t_\be (t_\al t_\be \lam_2+ 2 \lam_{345}))}{(1 + t_\al^2)t_\be (1 + t_\be^2)}\,,  \\
m_h^2&=&\frac{m_{12}^2(1 +t_\al t_\be)^2 (1 +t_\be^2) + t_\be v^2 (t_\al^2 \lam_1+ t_\be^2 \lam_2 - 2 t_\al t_\be \lam_{345})}{(1 + t_\al^2)t_\be (1 + t_\be^2)}\,,\label{eqmh}
\end{eqnarray}
corresponding to the physical mass-eigenstates 
\begin{eqnarray}  \label{PhysBasis}
H = (\sqrt{2} Re\ \Phi_1^0-v_1) c_\al+(\sqrt{2} Re\ \Phi_2^0-v_2) s_\al\,, \\
h = -(\sqrt{2} Re\ \Phi_1^0-v_1) s_\al+(\sqrt{2} Re\ \Phi_2^0-v_2) c_\al\,.
\end{eqnarray} 
This is usually called the physical basis, which is related to the Higgs-basis (\ref{HiggsBasis}) by a $\be-\al$ rotation:
\begin{eqnarray} 
H &=& c_{\be-\al}(\sqrt{2} Re\ H_1^0-v) -s_{\be-\al}(\sqrt{2} Re\ H_2^0) \,, \\
h &=& s_{\be-\al}(\sqrt{2} Re\ H_1^0-v) +c_{\be-\al}(\sqrt{2} Re\ H_2^0) \,.
 \label{PhysBasis2}
\end{eqnarray}

The  couplings of these Higgses to the gauge bosons, $V=W^{\pm},\,Z$, are usually parametrized by the coefficients $C_V^h=C_V^H=\sin (\beta-\alpha)$, which relate the former to the SM Higgs couplings
\begin{equation}
g_{hVV}=C_V^h\ g_{h_{SM} VV}\,,\quad g_{HVV}=C_V^H\  g_{h_{SM} VV}\,,
\end{equation}
where  $h_{SM}$ is the SM Higgs. 

Similarly, the couplings of these Higgses to the fermions are parametrized  by the $C_F^h,\, C_F^H$ coefficients, which in turn depend on the initial couplings of the two $\Phi_1, \Phi_2$ doublets to the fermions. The latter are strongly restricted by the requirement of the absence of dangerous FCNCs. As mentioned above, this is  guaranteed by imposing a  softly-broken $\ZZ_2$ symmetry, which affects both the Higgs doublets and the fermion fields. The possible $\ZZ_2$ charge assignments lead to the well-known four types of 2HDM, shown
 in Table \ref{table:2HDM_Types} \cite{Aoki:2009ha}. 

\begin{table}[h]
    \centering
    \resizebox{14cm}{!}{
    \begin{tabular}{|c||c|c|c|c|c|c|}
    \hline
      & $\Phi_1$&
      $\Phi_2$ & $u_R$ &  $d_R$ & $e_R$ & $u_L,\, d_L,\, \nu_L,\, e_L$ \\
      \hline
      Type I                    & $+$ & $-$ & $-$ & $-$ & $-$ & $+$\\
      Type II                   & $+$ & $-$ & $-$ & $+$ & $+$ & $+$\\
      Type X (lepton specific)  & $+$ & $-$ & $-$ & $-$ & $+$ & $+$\\
      Type Y (flipped)          & $+$ & $-$ & $-$ & $+$ & $-$ & $+$\\
      \hline  
    \end{tabular}
    }
    \caption{$\ZZ_2$ charge assignments that forbid tree-level Higgs-mediated FCNC effects.}
    \label{table:2HDM_Types}
\end{table}

Then, the corresponding $C_V$ and $C_F$ coefficients are given in Table \ref{table:couplings} for the four types of 2HDM.
\begin{table}[h]
    \centering
    \resizebox{17cm}{!}{
    \begin{tabular}{|c||c|c|c|c|c|c|c|c|}
    \hline
      & \multicolumn{2}{c|}{All types}& Type I &  Type II & \multicolumn{2}{c|}{Type X}& \multicolumn{2}{c|}{Type Y} \\
      \hline
      \multirow{2}{*}{}                    & \multirow{2}{*}{$VV$} & \multirow{2}{*}{$u$-quarks} & $d$-quarks,  & $d$-quarks, & \multirow{2}{*}{$d$-quarks} & \multirow{2}{*}{leptons}&\multirow{2}{*}{$d$-quarks}&\multirow{2}{*}{leptons}\\
       & & &  leptons & leptons & & & &\\
      \hline
      $h$                   & $s_{\beta-\alpha}$ & $c_\alpha/s_\beta$ & $c_\alpha/s_\beta$ & $-s_\alpha/c_\beta$ & $c_\alpha/s_\beta$ & $-s_\alpha/c_\beta$&$-s_\alpha/c_\beta$&$c_\alpha/s_\beta$\\
      $H$  & $c_{\beta-\alpha}$ & $s_\alpha/s_\beta$ & $s_\alpha/s_\beta$ & $c_\alpha/c_\beta$ & $s_\alpha/s_\beta$ & $c_\alpha/c_\beta$ & $c_\alpha/c_\beta$ & $s_\alpha/s_\beta$\\
      $A$
      & $0$ & $ 1/t_\beta$ & $-1/t_\beta$ & $t_\beta$ & $-1/t_\beta$ & $t_\beta$&$t_\beta$&$-1/t_\beta$\\
      \hline  
    \end{tabular}
    }
    \caption{Tree-level vector boson couplings, $C_V$, and fermionic couplings,  $C_F$, normalised to the SM-Higgs couplings for the  $h,  H, A$ fields in the four 2HDM types; $s_a, c_a$ stand for $\sin a, \cos a$. }
    \label{table:couplings}
\end{table}

\subsubsection*{The Alignment Limit}\label{TheAlignLim}

\quad This limit occurs when the light CP-even Higgs, $h$, behaves as the SM Higgs, i.e. it presents the same couplings as the latter to all the SM fields. From the couplings of $h$ to vector bosons and fermions shown in Table \ref{table:couplings}, it is clear that this is achieved for
\begin{equation}
\label{allimit}
\hspace{-2cm}\text{Alignment limit:}\hspace{1cm}c_{\be-\al}\to0\ . 
\end{equation}
Since $0\leq \beta\leq \pi/2$, $-\pi/2\leq \alpha\leq \pi/2$, this is accomplished for $\alpha=\beta-\pi/2$.
In this limit the Higgs-basis and the physical basis are equivalent. Namely, from Eq.(\ref{PhysBasis2}), $h$ becomes aligned with $(\sqrt{2} Re\ H_1^0-v)$ and $H_1$ plays the role of the ordinary SM doublet. Note also that in this limit the first term in the numerator of Eq.(\ref{eqmh}) vanishes, so $m_h^2={\cal O}(\lambda v^2)$, as in the SM.

Up to now, the Higgs boson observed in the LHC looks remarkably similar to the SM one \cite{ParticleDataGroup:2020ssz}.
This is why the 2HDM is usually considered at the (either exact or approximate) alignment limit \cite{Craig:2013hca}. 
More precisely, 
ATLAS~\cite{ATLAS:2019nkf} and CMS~\cite{CMS:2018uag} Run 2 Higgs data allow to determine $(C_V^h, C_F^h)$ within a $\sim 10\%$ error (reduced to $\sim 7\%$ when combining both analysis). 
These bounds are more restrictive when evaluated in a given model since $C_F$'s are in general non-universal and  correlated with $C_V$. In the 2HDM, these correlations are given in Table \ref{table:couplings}. In particular, for type I, deviations from $C_V^h=1$ are $\sim 1\%$  and \cite{Bertrand:2020lyb,Arco:2020ucn} 
\beq
|c_{\be-\al}| \lesssim 0.15\ .
\eeq
The range is even  more reduced in type II.

It is worth commenting that  there are symmetries of the potential\cite{Ferreira:2009wh, Ferreira:2010bm,Ferreira:2010yh, Battye:2011jj}
 that lead to an exact alignment, i.e. $c_{\be-\al}=0$. A rather trivial instance is the above-mentioned inert-doublet scenario, where $t_\beta=0,\infty$ and there is an exact $\Phi_1\rightarrow \Phi_1$, $\Phi_2\rightarrow -\Phi_2$ symmetry, so that one of the Higgs doublets precisely corresponds to the SM-Higgs, coupled to all fermions, while the other one is completely decoupled. 
  
  Beside this special case, the existence of symmetries that lead to alignment can be explored  by eliminating $t_\beta$ in 
$\{ \mathscr{P}_\alpha\ , \mathscr{P}_\beta    \}=0 $ evaluated at  the alignment limit, $t_\alpha =  - 1/{t_\beta} $; namely
\beq
t_\beta= \sqrt{\frac{\lambda_1-\lambda_{345}}{\lambda_2-\lambda_{345}}}\ ,
\label{tbetaalign}
\eeq
 subject to the consistency condition
\begin{equation}
\left(\lambda_1 \lambda_2 -\lambda_{345}^2 \right) \left( m_{12}^2 ( \lambda_1 - \lambda_2) +
   (m_{11}^2 -m_{22}^2)
    \sqrt{(\lambda_1  -  \lambda_{345} )
    (\lambda_2  -  \lambda_{345}  )} \right)
  = 0\,.    
  \label{eq:constraint}
\end{equation}
Whenever  Eq.(\ref{eq:constraint}) is fulfilled by the initial parameters of the potential, there exists a minimum with $t_\beta$ given by Eq.(\ref{tbetaalign}) where the alignment is exact.
 Notice, in particular, that Eq~(\ref{eq:constraint}) is satisfied for some simple relations between the potential parameters, such as
$\{ \lambda_1=\lambda_2 \; \& \; m_{11}^2 =m^2_{22} \} $, which  can be achieved by imposing certain global symmetries in the potential\cite{BhupalDev:2014bir,BhupalDev:2017txh,
Draper:2020tyq,Haber:2021zva}. However, for phenomenological reasons these symmetries cannot be exact. Still, an approximate alignment can arise if the symmetry is softly broken in the scalar sector. Nevertheless, as 
 pointed out in ref. \cite{Draper:2020tyq}, this requires to extend the Yukawa sector, e.g. with vector-like top quark partners coupled to the Higgses, which goes beyond the scope of this paper.
 
 In any case, even if the parameters are in a suitable combination to get exact or approximate alignment, if the underlying symmetry is not exact one must vary the parameters in a free way to evaluate the fine-tuning. This is exactly what we have done in the present analysis, which we expose below, so these possibilities are taken into account.

\section{Fine-tuning in the 2HDM. Decoupling and non-decoupling regimes}\label{FT2HDM}

\quad A theoretical  model  presents  fine-tuning  (or,  equivalently,  absence  of  naturalness)  when  some  observable quantity depends critically on a fine adjustment (or ``conspiracy") of the fundamental parameters.
Such adjustment in the parameter space is conceptually problematic, as it is implausible unless it can be explained from the theory itself.

Regarding the 2HDM, we distinguish two potential fine-tunings. The first one is related to the EW breaking, when the magnitude of the vacuum expectation value, $v^2$, is much smaller than  the other squared-mass terms entering the theory. Namely, from Eq.(\ref{dV1}) or Eq.(\ref{dV2}):
\beq
v^2\backsim \frac{\sum {\cal O} (m_{ij}^2)\hbox{-terms}}{{\cal O} (\lam_k)}\,.
\eeq
 Since $\lam_i\simlt 1$, if some $m_{ij}^2\gg v^2$ (as required for $m_H^2\gg v^2$), then $v^2$ is likely to be fine-tuned. This is sometimes called the ``little hierarchy problem"~\cite{Barbieri:2000gf}. Given that in the 2HDM there are two expectations values, $v_1$ and $v_2$, or equivalently $v^2$, $\tan \beta$; one can wonder if there are two independent fine-tunings, i.e. if  $\tan \beta$ might also be a fine-tuned parameter. We will examine this issue throughout the paper. 
 Let us note that 
the mass of the SM-like Higgs, $m_h$, has electroweak size, however it is not a fine-tuned parameter since, as mentioned above, $m_h^2={\cal O}(\lambda v^2)$, as in the SM. Therefore, {\em once} the value of $v^2$ has been set, $m_h^2$, similarly to $M_W^2, M_Z^2$, is naturally of electroweak size.

The second potential fine-tuning is related to the magnitude of $c_{\be-\al}$ in the alignment regime, i.e. $|c_{\be-\al}|\ll 1$,  since, in principle, there is no reason why the initial parameters should yield such small value.
To see this more closely, let us use the expression given in ref.~\cite{Bernon:2015qea} that links $c_{\be-\al}$ with  $m_h^2$, $m_H^2$ and the   $Z_{1,6}$ parameters:
\beq\label{cosZ6}
c_{\be-\al}=\frac{-Z_6 v^2}{\sqrt{(m_H^2-m_h^2)(m_H^2-Z_1 v^2)}},
\eeq
where
\beq 
\label{Z6Z1}
Z_6  = 
-\frac{1}{2}\ , s_{2\beta} \left[ 
\lambda_1 c_{\beta}^2  -\lambda_2 s_{\beta}^2 - \lambda_{345} c_{2\beta}
\right],\quad Z_1= \lam_1 c_\be^4 + \lam_2 s_\be^4 +\frac{1}{2}\ ,\lam_{345}s_{2\be}^2\,,
\eeq
Demanding an approximate (exact) alignment limit is equivalent to demand a small (vanishing) value of $|c_{\be-\al}|$. This can be achieved in two ways:
\begin{enumerate}
\item  
 $m_H^2\gg v^2$. Since 
 $Z_{1,6}=\sum_i {\cal O} (\lam_i)
 \simlt 1$, the denominator of Eq.(\ref{cosZ6}) becomes in this case much larger than the numerator, leading to a small $|c_{\be-\al}|$.  This instance is called in the literature \emph{alignment by decoupling}\cite{Gunion:2002zf,Asner:2013psa,Haber:2013mia}. It is worth-noticing that  this limit requires large $m_{12}^2\gg v^2$ (see Eq.(\ref{eqmH})). Then, from Eqs.(\ref{AHcMasses}, \ref{AHcMasses2}, \ref{eqmH}),  all the extra Higgs states get similar masses:
 \beq
 \label{similar masses}
 m_H^2\simeq m_A^2\simeq m_{H^\pm}^2\simeq \dfrac{1+t_\be^2}{t_\be}m_{12}^2\ .
 \eeq
 Actually, in this regime one can integrate out the heavy Higgs states  and the resultant effective theory is essentially the SM (more precisely, an SMEFT), with $h$ playing the role of the SM-like Higgs boson.
This may seem quite natural, but, as discussed above, it implies a potential fine-tuning related to the EW breaking.

\item $Z_6\ll1$. Then, $|c_{\be-\al}|\ll1$ without requiring large Higgs masses. 
This regime is commonly called \emph{alignment without decoupling}\cite{Craig:2013hca,Carena:2013ooa, Carena:2001bg}.
However, as it is clear from Eq.(\ref{Z6Z1}), this typically requires
a precise cancellation inside 
$Z_{6}=\sum_i {\cal O} (\lam_i)
$. A possible exception, however, occurs when $s_{2\beta}\simeq 0$, i.e. for $t_\beta\gg 1$. Note here that the  $t_\be\ll1$ case is excluded because it leads to a non-perturbative Yukawa coupling for the top quark (which, by definition, is coupled to the $\Phi_2$ doublet).

\end{enumerate}

Hence, the alignment generically requires fine-tuning, whether it is achieved by decoupling or not.
Thus, evaluating both potential fine-tunings in detail may lead to a better understanding of the parameter space, and to find regions in which none of them is too large. This is the main goal of this paper.

In order to quantify the fine-tuning associated with a generic observable, $\Om$, with respect to an initial parameter, $\theta_i$, we will use the somewhat standard criterion proposed by by Ellis et al. \cite{Ellis:1986yg} and Barbieri and Giudice \cite{Barbieri:1987fn}. Namely, we define the fine-tuning parameters as
\begin{equation}\label{FTDef}
\De_{\th_i}\Om=\frac{d \ln \Omega}{d \ln \theta_i }=\frac{\th_i }{\Om}\frac{d  \Omega}{d \theta_i },\quad  \Delta\Om \equiv \max |\Delta_{\theta_i}\Om|.
\end{equation}
Then $\De \Om \sim 10$ $(100)$ denotes a fine-tuning of about $10\%$ $(1\%)$, etc~\cite{Barbieri:1987fn}.

As discussed above, the set of (potentially fine-tuned) observables that we will consider is
\begin{eqnarray}  \label{dV}
\Omega = \{v^2,t_\be, c_{\beta-\alpha} \}\,, 
\end{eqnarray} 
while the set of initial parameters is 
\begin{eqnarray}  \label{theta}
\theta_i =  \{ m_{11}^2 , m_{22}^2 , m_{12}^2 , \lambda_1,  \lambda_2,  \lambda_3,  \lambda_4,  \lambda_5  \}\,.
\end{eqnarray} 
Our aim is to find analytical expressions for $\De_{\theta_i} v^2$, $\De_{\theta_i} t_\beta$ and $\De_{\theta_i} c_{\beta-\alpha}$ in terms of the initial parameters or (most conveniently) the physical parameters, and discuss their magnitude in different regimes.



\section{Analytical expressions for the Fine-tuning}\label{EvFT}

\subsection{Fine-tuning in terms of the initial parameters}\label{sec:inpar}

The most straightforward way to get analytical expressions for the fine-tuning on any observable, $\Omega$, is to start with its explicit dependence on the initial parameters,
 $\Om(\th_i)$, and evaluate the derivatives $d  \Omega/{d \theta_i }$ involved in Eq.(\ref{FTDef}).
 However, in our case these kinds of expressions are cumbersome; e.g. $t_\be$ is given by the quartic equation (\ref{eqbeta}). Fortunately, we can still extract analytical expressions for the fine-tuning using the constraints
 $\mathscr{P}_{1,2,\beta,\alpha}=0$ (see Eqs.(\ref{dV1}, \ref{dV2}, \ref{eqbeta}, \ref{eqalpha})),
  together with the Implicit Function Theorem.
  
  Next we do this for the three potentially fine-tuned observables, $v^2$, $t_\beta$, $c_{\beta-\alpha}$.
  
  \subsubsection*{ \underline{Fine-tuning in $t_\beta$}}
 \vspace{0.2cm}
 
For convenience, we start with the fine-tuning in $t_\be$. Taking derivatives in Eq.(\ref{eqbeta}), 
\begin{equation}
\frac{d \mathscr{P}_\be }{d\th_i} = 0\implies \frac{\partial \mathscr{P}_\be }{\partial \th_i }+\frac{\partial \mathscr{P}_\be}{\partial t_\be}\frac{\partial t_\be }{\partial \th_i }=0,
\end{equation}
so
\begin{equation}
\label{Dtp}
\De_{\th_i}t_\be =-\frac{\th_i}{t_\be}\frac{\partial \mathscr{P}_\be}{\partial \th_i}/\frac{\partial \mathscr{P}_\be}{\partial t_\be} \,.
\end{equation}
Hence, replacing here the explicit expression (\ref{eqbeta}) for $\mathscr{P}_\be$ 
 we get analytical formulas for all $\De_{\th_i}t_\be$.
 
 The above expression is enough to see that, in general, $\De_{\th_i}t_\be\simlt{\cal O} (1)$, and thus
 $t_\beta$ is not a fine-tuned parameter. To check this, note that the size of the denominator
 \begin{equation}
     \frac{\partial \mathscr{P}_\be}{\partial t_\be}=m_{12}^2(4t_\beta^3\lambda_2) - m_{11}^2(3t_\beta^2\lambda_2+\lambda_{345}) + m_{22}^2(\lambda_1 + 3t_\beta^2\lambda_{345}) 
 \end{equation}
 is typically $\simgt{\cal O}(m_{ij}^2)$, barring accidental cancellations; and this is also the expected size of the numerator for any $\theta_i$. E.g. let $\theta_i$ be one of the initial mass-parameters (the ones more directly involved in the electroweak fine-tuning), say
 $\theta=m_{12}^2$. Then Eq.(\ref{Dtp}) reads
 \begin{equation}
\label{Dm12tb}
\De_{m_{12}^2}t_\be =-\frac{m_{12}^2(-\lambda_1+t_\beta^4\lambda_2)}{m_{12}^2(4t_\beta^4\lambda_2) - m_{11}^2(3t_\beta^3\lambda_2+t_\beta\lambda_{345}) + m_{22}^2(t_\beta\lambda_1 + 3t_\beta^3\lambda_{345})} 
\,.
\end{equation}
which is typically $\simlt{\cal O}(1)$. 
This is obviously the case for $\lambda_i, t_\beta = {\cal O}(1)$, and also for $t_\beta\gg 1$. (In the last instance $|\De_{m_{12}^2}t_\be| \sim 1/4$.)

The fact that $t_\beta$ is not fine-tuned in most cases can be traced back to the origin of this observable.
If $m_{11}^2>0$, the VEV of $\Phi_1$ is triggered by that of  $\Phi_2$ through the linear term in the potential (\ref{pot}),
\beq
\label{v1v2}
v_1 \sim \frac{m_{12}^2}{m_{11}^2}v_2\ .
\eeq
 Hence, $v_1$ can have a similar size as $v_2$ with no need of tunings or cancellations (whether or not $v_2$ has electroweak size); thus $t_\beta$ can be 
 ${\cal O}(1)$, or actually any size, with no fine-tuning. Note that a
 very large (or small) $t_\beta$ implies a strong hierarchy between $m_{11}^2$ and $m_{12}^2$ . This may be considered as an odd fact (as it is e.g. the hierarchy of masses of the fermionic generations), but it is not a fine-tuning.
 
 \subsubsection*{ \underline{Fine-tuning in $v^2$}}
  \vspace{0.2cm}

For the fine-tuning in $v^2$, we proceed in a similar way. From Eqs.(\ref{dV1}, \ref{dV2}), we get
\begin{equation}
\dfrac{d \mathscr{P}_{1, 2} }{d\th_i} = 0\implies \dfrac{\partial \mathscr{P}_{1, 2} }{\partial \th_i }+\dfrac{\partial \mathscr{P}_{1, 2}}{\partial t_\be}\dfrac{\partial t_\be }{\partial \th_i } +\dfrac{\partial \mathscr{P}_{1, 2}}{\partial v^2}\dfrac{\partial v^2 }{\partial \th_i }=0\,,
\end{equation}
which leads to:
\begin{equation}
\label{Dvp}
\De_{\th_i}v^2 =-\frac{\th_i}{v^2}\left(\frac{\partial \mathscr{P}_{1, 2}}{\partial \th_i}+ \frac{t_\be}{\th_i}\frac{\partial \mathscr{P}_{1, 2}}{\partial t_\be }\De_{\th_i}t_\be \right)/\frac{\partial \mathscr{P}_{1, 2}}{\partial v^2}\,.
\end{equation}
Thus, replacing the explicit expressions
(\ref{dV1}, \ref{dV2}) for $\mathscr{P}_{1, 2}$, we get two alternative but equivalent analytical forms for each $\De_{\th_i}v^2$.

Now it is easy to see that $v^2$ is generically fine-tuned with respect to the initial mass-parameters. E.g. for $\theta_i=m_{11}$, we replace
\begin{equation}
     \frac{\partial \mathscr{P}_1}{\partial t_\be}=4t_\beta m_{11}^2
     -2(1+3t_\beta^2)m_{12}^2 +2t_\beta  \lambda_{345}v^2  \ ,
 \end{equation}
\begin{equation}
     \frac{\partial \mathscr{P}_1}{\partial v^2}=\lambda_1 +t_\beta^2\lambda_{345} \ ,
 \end{equation}
 \begin{equation}
     \frac{\partial \mathscr{P}_1}{\partial m_{11}^2}=2(1+t_\beta^2) \ ,
 \end{equation}
into Eq.(\ref{Dvp}), getting
\begin{equation}
\label{Dm11v2}
\De_{m_{11}^2}v^2 =-\frac{m_{12}^2}{v^2}
\frac{1}{\lambda_1+t_\beta^2 \lambda_{345}}
\left[
2(1+t_\beta^2)
+\frac{t_\beta}{m_{11}^2}\left(4t_\beta m_{11}^2-2
(1+3t_\beta^2)m_{12}^2 +2t_\beta\lambda_{345}v^2\right)\Delta_{m_{11}^2}t_\beta\right]
\,,
\end{equation}
which is typically $\simgt{\cal O}(m_{ij}^2/v^2)$. Hence, $\De_{\theta}v^2$ captures the whole fine-tuning associated to the scale of the EW breaking (the little hierarchy problem).

 \subsubsection*{ \underline{Fine-tuning in $c_{\beta-\alpha}$}}
 \vspace{0.2cm}
 
Let us finally consider the fine-tuning in the alignment, i.e. $\De_{\th_i} c_{\be-\al}$.  Using the trigonometric identity
\begin{equation}
c_{\be-\al}=\frac{1+t_\al t_\be}{\sqrt{(1+t_\be^2)(1+t_\al^2)}}
\end{equation}
in Eq.(\ref{FTDef}) (for $\Omega = c_{\be-\al}$)
we can express the fine-tuning in $c_{\be-\al}
$ as
\begin{equation}
 \De_{\th_i} c_{\be-\al} = \frac{1}{c_{\be-\al}}\left(t_\al \frac{\partial c_{\be-\al}}{\partial t_\al }\De_{\th_i}t_\al+t_\be \frac{\partial c_{\be-\al}}{\partial t_\be }\De_{\th_i}t_\be \right)\,,
\end{equation}
where $\De_{\th_i}t_\al\equiv\dfrac{\th_i }{t_\al}\dfrac{d  t_\al}{d \theta_i }$ is the ``fine-tuning in $\alpha$",
which can be obtained from Eq.(\ref{eqalpha}) following similar steps as for the other fine-tunings:
\beq
\dfrac{d \mathscr{P}_\al }{d\th_i} = 0\implies \dfrac{\partial \mathscr{P}_\al }{\partial \th_i }+\dfrac{\partial \mathscr{P}_\al}{\partial t_\be}\dfrac{\partial t_\be }{\partial \th_i }+\dfrac{\partial \mathscr{P}_\al}{\partial v^2}\dfrac{\partial v^2}{\partial \th_i }+\dfrac{\partial \mathscr{P}_\al}{\partial t_\al}\dfrac{\partial t_\al }{\partial \th_i }=0\,.
\eeq
Hence,
\beq\label{FTtalfa}
\De_{\th_i} t_\al =- \frac{\th_i}{t_\al}\left(\frac{\partial\mathscr{P}_\al}{\partial \th_i}+ \frac{t_\be}{\th_i}\frac{\partial \mathscr{P}_\al}{\partial t_\be }\De_{\th_i}t_\be + \frac{v^2}{\th_i}\frac{\partial \mathscr{P}_\al}{\partial v^2 }\De_{\th_i}v^2\right) /\frac{\partial \mathscr{P}_\al}{\partial t_\al}\,,
\eeq
Replacing here the explicit expression (\ref{eqalpha}) for $\mathscr{P}_\al$, beside Eqs.(\ref{Dtp}, \ref{Dvp}),
 we get analytical formulas for all $\De_{\th_i}t_\al$.
 Finally, the alignment fine-tuning reads
\begin{equation} \label{FTcosAnaly}
\De_{\th_i} c_{\be-\al}=-\frac{t_\al-t_\be}{1+t_\al t_\be}\left(\frac{t_\alpha}{1+t_\al^2}\De_{\th_i}t_\al-\frac{t_\be}{1+t_\be^2}\De_{\th_i}t_\be\right)\,,
\end{equation}
which, upon replacement of Eqs.(\ref{Dtp}, \ref{FTtalfa}), has an explicit analytical expression for all ${\th_i}$.
Note that, in the alignment limit,  $1+t_\al t_\be\rightarrow 0$, thus we expect
$c_{\be-\al}$ to be generically fine-tuned.

It is worth-noticing that the term proportional to $\De_{\th_i}t_\al$ has ``inside" the electroweak fine-tuning, $\De_{\th_i}v^2$,  through Eq.(\ref{FTtalfa}). 
Actually, it can be checked from the previous expressions that for $m_{ij}^2\gg v^2$ (decoupling limit) and $1+t_\al t_\be\rightarrow 0$ (alignment limit) the coefficient of $\Delta_{\theta_i} v^2$ is 1:
 \begin{eqnarray}
\label{XX}
\De_{\th_i} c_{\be-\al}\ \supset\
\De_{\th_i}v^2
\ . 
\end{eqnarray}
This remarkably simple result can be understood as follows.  Let us recall from Eq.(\ref{cosZ6}) that 
\beq
\label{cosZ62}
c_{\be-\al}=\frac{-Z_6 v^2}{\sqrt{(m_H^2-m_h^2)(m_H^2-Z_1 v^2)}},
\eeq
Hence, in absolute value,
\begin{eqnarray}
\label{cosZ6FT}
\Delta_{\theta_i}c_{\be-\al}&=&
\frac{d\ \log c_{\be-\al}}{d\ \log \theta_i}
\nonumber \\
&=&\frac{d} {d\ \log \theta_i}\left[
\log v^2
+ \log Z_6
- \frac{1}{2}\log ((m_H^2-m_h^2)(m_H^2-Z_1 v^2))\right]
\end{eqnarray}
The first term is exactly $\De_{\th_i} v^2$, the second one represents the fine-tuning in $c_{\be-\al}$ when there is no decoupling and the third term is ${\cal O}(1)$ since $(m_H^2-m_h^2)(m_H^2-Z_1 v^2) \simeq  m_H^4$, which is not a fine-tuned quantity.

Eqs.(\ref{cosZ62}, \ref{cosZ6FT}) show in a transparent way why the electroweak fine-tuning, $\De v^2$, appears involved in the  the alignment one.  The definition of fine-tuning takes into account {\em all} the adjustments required to get the desired value of the observable under consideration. In our case, if the alignment is achieved through decoupling, there is a fine-tuning price, namely the one required to get $v^2$ much smaller than the mass parameters of the potential.

From the above discussion it is clear that the fine-tuning in $v^2$ and $c_{\beta-\alpha}$ are {\em not} independent. Thus, it would be incorrect to multiply them to get the ``total" fine-tuning. In particular, in the decoupling limit, once the initial parameters have been adjusted to get the right size of $v^2$, there is no need of extra adjustments to get alignment.
Hence, a practical way to remove the ``double counting" in the evaluation of the $c_{\beta-\alpha}$ fine-tuning is to discard variations in the $\theta_i$ parameters that change the value of $v^2$. In the $\{\log\theta_i\}$ space, this is equivalent to project the gradient $\Delta_{\theta_i}  c_{\beta-\alpha}=\partial \log c_{\beta-\alpha}/\partial \log\theta_i$ onto the constant-$v^2$ hypersurface~\cite{Casas:2005ev}: 
\begin{equation} 
\label{Projection}
\Delta_{\th_i}  c_{\beta-\alpha}
\rightarrow
\Delta_{\th_i}  c_{\beta-\alpha}-\left(\sum_j \Delta_{\th_j}  c_{\beta-\alpha}\cdot 
n_j\right) n_i
\end{equation}
where $n_i=\Delta_{\th_i} v^2/\sqrt{\sum_j(\Delta_{\th_j} v^2)^2}$ is the unitary vector normal to the
constant-$v^2$ hypersurface. Obviously, in doing so, all the $\Delta_{\th_i} v^2$ pieces in Eq.(\ref{FTcosAnaly}) are removed.

\subsection{Fine-tuning in terms of the physical parameters}
 
Very often it is convenient and more meaningful to express the fine-tuning in terms of the physical parameters of the model, rather than in terms of the initial ones. To this end, we have to trade the eight initial parameters, $\theta_i$, of Eq.(\ref{theta}) for eight physical observables, $\chi_i$. Specifically, we have chosen 
\begin{equation} 
\label{PhysPar}
\chi_i = \{v, t_\beta, t_\alpha, m_h, m_H, m_A, m_{H^\pm} ,  \lambda_5\}\,.
\end{equation}
Although $\lam_5$ is not strictly an observable, it is related to the triple scalar self-couplings, 
e.g.
\beq 
\label{eq:Hhhcoup}
g_{Hhh}=\frac{m_H^2+2m_h^2-4(m_A^2+v^2\lam_5)}{v}\ c_{\be-\al}+{\cal O} (c_{\be-\al}^2)\,. \eeq

Now the fine-tunings $\De_{\th_i} t_\beta, \De_{\th_i} v^2,
\De_{\th_i} c_{\be-\al}$, given by Eqs.(\ref{Dtp}, \ref{Dvp}, \ref{FTcosAnaly}), can be written as functions of the physical parameters, $\chi_i$, by replacing the initial parameters in terms of the latter, i.e. $\th_i(\chi_j)$. This can be done with the help of the relations obtained in Sect.\ref{2HDM}. Namely, from Eqs.(\ref{AHcMasses}, \ref{AHcMasses2}) we derive $m_{12}^2(\chi_i)$, $\lambda_4(\chi_i)$. Then, from Eqs.(\ref{eqalpha}, \ref{eqmH}, \ref{eqmh}) we derive $\lambda_{1,2,3}(\chi_i)$; and finally, from
Eqs.(\ref{dV1}, \ref{dV2}) we get $m_{11}^2(\chi_i)$, $m_{22}^2(\chi_i)$. Altogether, the explicit expressions for $\th_i(\chi_j)$ read
\begin{eqnarray} \label{eq:cambio}
 m_{11}^2  & = & 
 \frac{m_h^2t_\alpha (t_\beta - t_\alpha)}{2(1 +  t_\alpha^2)} 
- \frac{m_H^2(1+  t_\alpha t_\beta  )}{2(1 +  t_\alpha^2)} 
+ \frac{t_\beta^2}{1 +  t_\beta^2} (m_A^2 + v^2 \lambda_5)\,, \nonumber \\
 m_{22}^2  & = & 
 \frac{m_h^2(t_\alpha-t_\beta  )}{2 t_\beta (1 +  t_\alpha^2)} 
- \frac{m_H^2t_\alpha (1+  t_\alpha t_\beta  )}{2t_\beta (1 +  t_\alpha^2)} 
+ \frac{1}{1 +  t_\beta^2} (m_A^2 + v^2 \lambda_5)\,, \nonumber \\
 m_{12}^2  & = & \frac{t_\beta}{1 +  t_\beta^2} (m_A^2 + v^2 \lambda_5)\,, \nonumber \\ 
 \lambda_1 & = & \frac{m_h^2t_\alpha^2+m_H^2}{v^2}  \frac{1+t_\beta^2}{1 +  t_\alpha^2} - t_\beta^2 \frac{m_A^2 + v^2 \lambda_5}{v^2}\,,
 \\
  \lambda_2 & = & \frac{m_h^2+ m_H^2 t_\alpha^2 }{v^2}  \frac{1+t_\beta^2}{(1 +  t_\alpha^2)t_\beta^2} - \frac{m_A^2 + v^2 \lambda_5}{t_\beta^{2} v^2}\,,  \nonumber \\
 \lambda_3 & = & \frac{2 m_{H^\pm}^2 -  m_A^2}{v^2} 
  - \frac{m_h^2- m_H^2}{v^2}   
   \frac{ t_\alpha(1+  t_\beta^2  )}{t_\beta(1 +  t_\alpha^2)}  - \lambda_5\,, \nonumber \\
    \lambda_4 & = & \frac{2 ( m_A^2 - m_{H^\pm}^2)}  {v^2} 
    + \lambda_5\,, \nonumber \\
     \lambda_5  & = &  \lambda_5\,. \nonumber
\end{eqnarray} 
Incidentally, the change of variables from the initial parameters to the physical observables corresponds to a remarkably simple Jacobian, 
\begin{equation} \label{Jacobian}
J = \displaystyle{\frac{32 m_h^3 m_H^3 (m_H^2 -m_h^2 )m_A  m_{H^\pm} (1 + t_\beta^2)^2}
{v^9 t_\beta^3 (1 + t_\alpha^2)}}.
\end{equation}
We present the final analytical expressions for all the fine-tunings in the Appendix. This is one of the main results of the present work.

\subsection*{ Results in $\epsilon-$expansion}\label{epsilon_exp}

Due to the smallness of $c_{\be-\al}$ in the alignment regime,  we can rewrite all the fine-tuning expressions of the Appendix as power series in $c_{\be-\al}=\ep$, using
\begin{equation} \label{AlignLim}
 t_\alpha =  - \frac{1}{t_\beta}   + \frac{1+t_\be^2}{t_\beta^2}\epsilon  +{\cal O} (\epsilon^3 )\,.
\end{equation}
This expansion yields simpler formulas, which capture the general behaviour of the fine-tuning and allow for a closer examination.

Next we gather and discuss the expressions of the various fine-tunings at the lowest order in $\epsilon$.
We will focus the discussion on the two instances that, as argued in Sect.\ref{FT2HDM},
naturally lead to alignment, namely the decoupling limit and the $t_\beta\gg 1$ regime.

 \subsubsection*{ \underline{Fine-tuning in $v^2$}}

\begin{eqnarray}\nonumber
\label{v-lowestorder}
 \Delta_{m^2_{11}}  v^2   & = &
 \displaystyle \frac{(1+t_\beta^2) m_h^2 - 2 t_\beta^2 (m_A^2 + v^2 \lambda_5)}{(1+t_\beta^2)^2 m_h^2}  
 + {\cal O} (\epsilon)\,,    \\[5mm]\nonumber
 \Delta_{m^2_{22}}   v^2  & = &
 \displaystyle \frac{  t_\beta^2 (1+t_\beta^2) m_h^2 - 2 t_\beta^2 (m_A^2 + v^2 \lambda_5)}{(1+t_\beta^2)^2 m_h^2}  
 + {\cal O} (\epsilon)\,,   \\[5mm]\nonumber
    \Delta_{m^2_{12}}  v^2  & = &
 \displaystyle \frac{ 4  t_\beta^2 (m_A^2 + v^2 \lambda_5)}{(1+t_\beta^2)^2 m_h^2}  
 + {\cal O} (\epsilon)\,,  \\[5mm]
  \Delta_{\lambda_1}   v^2   & = &
 \displaystyle \frac{ -m_h^2 +  t_\beta^2 (m_A^2 + v^2 \lambda_5-m_H^2)}{(1+t_\beta^2)^2 m_h^2}  
 + {\cal O} (\epsilon)\,,  \\[5mm]\nonumber
  \Delta_{\lambda_2}  v^2   & = &
 \displaystyle \frac{t_\beta^2 (m_A^2 + v^2 \lambda_5-m_H^2 -m_h^2t_\beta^2  )}{(1+t_\beta^2)^2 m_h^2}  
 + {\cal O} (\epsilon)\,,   \\[5mm]\nonumber
  \Delta_{\lambda_3}   v^2   & = &
 \displaystyle \frac{2 t_\beta^2 (m_A^2 + v^2 \lambda_5+m_H^2 -m_h^2 -2 m_{H_\pm}^2 )}{(1+t_\beta^2)^2 m_h^2}  
 + {\cal O} (\epsilon)\,,  \\[5mm]\nonumber
  \Delta_{\lambda_4}   v^2   & = &
 -\displaystyle \frac{2 t_\beta^2 (2 m_A^2 + v^2 \lambda_5-2 m_{H_\pm}^2 )}{(1+t_\beta^2)^2 m_h^2}  
 + {\cal O} (\epsilon)\,,   \\[5mm]\nonumber
  \Delta_{\lambda_5}   v^2   & = &
 -\displaystyle \frac{2 t_\beta^2  v^2 \lambda_5}{(1+t_\beta^2)^2 m_h^2}  
 + {\cal O} (\epsilon)\,.   
 \end{eqnarray} 

In the decoupling regime, $m_H^2\simeq 
m_A^2\simeq m_{H^\pm}^2
\gg v^2$,  so $\De_{m_{ij}^2} v^2={\cal O}(\frac{m_H^2}{v^2})+{\cal O} (\ep)$, indicating that generically the EW fine-tuning diverges as the masses of the extra Higgses increase, as expected from the discussion in section \ref{sec:inpar}. 
More precisely, in that regime 
\beq
\label{aproxDv2}
\frac{1}{2}\De_{m_{12}^2}v^2\simeq -\De_{m_{11}^2}v^2\simeq
-\De_{m_{22}^2}v^2\simeq 2\left(\frac{t_\beta}{1+t_\beta^2}\right)^2\frac{m_A^2}{m_h^2} \ .
\eeq
In contrast $\De_{\lambda_{i}} v^2$ remain ${\cal O}(1)$.

Let us consider now the $t_\beta\gg 1$ regime. It can be checked from Eqs. (\ref{v-lowestorder}) that in this case all $\Delta_{\theta_i}v^2$ fine-tunings tend to zero, except $\Delta_{m_{22}^2}v^2$, $\Delta_{\lambda_2}v^2$, which become ${\cal O}(1)$, i.e. irrelevant.
This remarkable absence of EW fine-tuning comes from the following. In the $t_\be\to\infty$ limit
\beq
v^2\simeq v^2_2 \simeq\dfrac{- 2m_{22}^2}{\lam_2}\,,
\eeq
(this can be seen, e.g. from Eqs.(\ref{eqminv2}) or (\ref{dV2})).
Since $\lambda_2\leq {\cal O}(1)$, this situation requires $m_{22}^2 = {\cal O} (m_h^2)$, which by itself does not entail any fine-tuning. Note that the extra Higgses can still be heavy provided $m_{11}^2$ is large. So for $t_\beta\gg1$
there is a hierarchy between $m_{ij}^2$ mass terms; namely, from Eqs.(\ref{eq:cambio}):
\beq
m_{11}^2\simeq m_A^2,\quad  |m_{22}^2|\simeq m_h^2/2,\quad m_{12}^2\simeq m_A^2/ t_\be\ ,
\eeq
so that
\beq 
m_{11}^2\gg |m_{22}^2|,m_{12}^2,
\eeq
while
\begin{equation}
|m_{22}^2|\simeq \left(
\frac{m_h^2}{2m_A^2}
\right)m_{11}^2
\end{equation}

As already mentioned, a hierarchy of mass terms may be or may be not seen as an odd fact, but it does not imply a fine-tuning between parameters.

This situation contrasts with what happens in the supersymmetric case. In the minimal supersymetric Standard Model (MSSM), the $t_\beta\gg 1$ regime (which is desirable in order to reproduce the ordinary Higgs mass) implies a notorious fine-tuning. This occurs because $m_{22}^2$ at low energy contains several contributions proportional to the soft supersymmetric masses, e.g. a large and negative contribution proportional to the mass-squared of the gluinos, and a positive contribution equal to the $\mu-$term (squared). Then, the smallness of $|m_{22}^2|$ can only be achieved by a cancellation between all these large contributions, and this does imply a fine-tuning. However, this does not need to be the case in a generic 2HDM. 

Of course, the previous discussion has nothing to do with the famous hierarchy problem~\cite{Maiani:1979cx,tHooft:1980xss}, which is related to the stability of the scalar mass-parameters under quadratic radiative corrections.

In summary, from Eq.(\ref{aproxDv2}) the electroweak fine-tuning in the 2HDM has the approximate size
\beq
\label{aproxDv2_2}
\De v^2\sim  \frac{4}{t_\beta^2}\frac{m_A^2}{m_h^2}
\eeq
which shows that it increases with the masses of the extra Higgses (as expected), but it can be compensated by a large $t_\beta$.

 \subsubsection*{\underline{Fine-tuning in $t_\beta$}}

\begin{eqnarray} \nonumber
 \Delta_{m^2_{11}} t_\be   & = &
  -\displaystyle \frac{(1+t_\beta^2) m_h^2 - 2 t_\beta^2 (m_A^2 + v^2 \lambda_5)}{2 m_H^2(1+t_\beta^2) }  
 + {\cal O} (\epsilon)\,,  \\[5mm]\nonumber
  \Delta_{m^2_{22}} t_\be  & = & 
 \displaystyle \frac{(1+t_\beta^2) m_h^2-2 (m_A^2 + v^2 \lambda_5)}
 {2 m_H^2(1+t_\beta^2)}  
 + {\cal O} (\epsilon)\,,  \\[5mm]\nonumber
   \Delta_{m^2_{12}} t_\be   & = & -  
 \displaystyle \frac{(t_\beta^2-1)(m_A^2 + v^2 \lambda_5)}
 {m_H^2(1+t_\beta^2)} 
 + {\cal O} (\epsilon)\,,  \\[5mm]
    \Delta_{\lambda_1} t_\be   & = & 
 \displaystyle \frac{ m_h^2 - t_\beta^2(m_A^2-m_H^2 + v^2 \lambda_5)  } {2 m_H^2(1+t_\beta^2)} 
 + {\cal O} (\epsilon)\,, \\[5mm]\nonumber
     \Delta_{\lambda_2} t_\be   & = & 
 \displaystyle \frac{m_A^2 -m_H^2  -m_h^2 t_\beta^2 + v^2 \lambda_5}{2 m_H^2(1+t_\beta^2)}
 + {\cal O} (\epsilon)\,,   \\[5mm]\nonumber
\Delta_{\lambda_3} t_\be  & = & - \displaystyle \frac{ (t_\beta^2 -1) (m_A^2 - m_h^2 + m_H^2 - 2 m_{H_\pm}^2 + v^2 \lambda_5) }{2 m_H^2(1+t_\beta^2)}
 + {\cal O} (\epsilon)\,,   \\[5mm]\nonumber
 \Delta_{\lambda_4} t_\be  & = &  
 \displaystyle \frac{(t_\beta^2 -1)(2 m_A^2 - 2 m_{H_\pm}^2 + v^2 \lambda_5)}{2 m_H^2(1+t_\beta^2)}
 + {\cal O} (\epsilon)\,,   \\[5mm]\nonumber
  \Delta_{\lambda_5} t_\be  & = &  
 \displaystyle \frac{(t_\beta^2 -1) v^2 \lambda_5}{2 m_H^2(1+t_\beta^2)} 
 + {\cal O} (\epsilon)\,.  
 \end{eqnarray} 
 
 It is clear from the previous expressions that $t_\beta$ is not a fine-tuned parameter in any instance, as anticipated in section \ref{sec:inpar}. In particular, in the decoupling regime, $m_H^2\simeq 
m_A^2\simeq m_{H^\pm}^2
\gg v^2$,  all 
$\De_{\lambda_i} t_\beta\rightarrow {\cal O} (\epsilon)$, whereas
$\De_{m_{11}^2} t_\beta, \De_{m_{12}^2} t_\beta = {\cal O} (1)$ and $\De_{m_{22}^2} t_\beta,  \sim 1/(1+t_\beta^2)$.

Something similar occurs in the large $t_\beta$ regime, where all the fine-tuning parameters are smaller than 1, except $\De_{m_{11}^2} t_\beta \sim \De_{m_{12}^2} t_\beta \sim 1$. This comes from the fact that if $m_{11}^2>0$, then $t_\beta \sim m_{11}^2/m_{12}^2$, as discussed around Eq.(\ref{v1v2}).

In summary, $t_\beta$ is not a fine-tuned parameter and typically
\beq
\Delta t_\beta \simlt {\cal O}(1)
\eeq

 \subsubsection*{\underline{Fine-tuning in $c_{\beta-\alpha}$}}
  \vspace{0.2cm}

In this case the lowest order in the expansion is always $1/\ep$, reflecting the fact that we are evaluating the fine-tuning in $c_{\beta-\alpha}=\epsilon$ itself, which is logically more severe as $\epsilon$ shrinks:
\begin{eqnarray} \nonumber
\label{cba-lowestorder}
 \Delta_{m^2_{11}} c_{\beta - \alpha}  & =&-  \frac{1}{\epsilon}\;
 \displaystyle \frac{t_\beta(m_A^2 -m_H^2 + v^2 \lambda_5)(m_h^2(1+t_\beta^2)  - 2 t_\beta^2(m_A^2 +v^2 \lambda_5))  }{m_H^2 (m_H^2-m_h^2)(1+t_\beta^2)^2 } 
+ {\cal O} (\epsilon^0)\,, \\[5mm]\nonumber
  \Delta_{m^2_{22}} c_{\beta - \alpha}   & = &  \phantom{-}  \frac{1}{\epsilon}\;
 \displaystyle \frac{t_\beta(m_A^2 -m_H^2 + v^2 \lambda_5)
 (-2 m_A^2+ m_h^2(1+t_\beta^2)  - 2 v^2 \lambda_5)  }
 {m_H^2 (m_H^2-m_h^2)(1+t_\beta^2)^2 } 
+ {\cal O} (\epsilon^0)\,, \\[5mm]\nonumber
   \Delta_{m^2_{12}} c_{\beta - \alpha}   & = &  \frac{1}{\epsilon} \;
 \displaystyle \frac{2 t_\beta (t_\beta^2 -1) (m_A^2 -m_H^2 + v^2 \lambda_5)
 (m_A^2+ v^2 \lambda_5)  }
 {m_H^2 (m_H^2-m_h^2)(1+t_\beta^2)^2 }
+ {\cal O} (\epsilon^0)\,,  \\[5mm]
 \Delta_{\lambda_1} c_{\beta - \alpha}  & = & -  \frac{1}{\epsilon} \;
 \displaystyle \frac{ t_\beta
 (m_A^2+ v^2 \lambda_5)
 (-m_h^2 + t_\beta^2(m_A^2-m_H^2 + v^2 \lambda_5) )  }
 {m_H^2 (m_H^2-m_h^2)(1+t_\beta^2)^2 } 
+ {\cal O} (\epsilon^0)\,,  \\[5mm]\nonumber
     \Delta_{\lambda_2} c_{\beta - \alpha}   & = & \phantom{-}   \frac{1}{\epsilon} \;
 \displaystyle \frac{ t_\beta(m_A^2+ v^2 \lambda_5)(m_A^2 -m_H^2  -m_h^2 t_\beta^2 + v^2 \lambda_5)}
 {m_H^2 (m_H^2-m_h^2)(1+t_\beta^2)^2 } 
+ {\cal O} (\epsilon^0)\,,   \\[5mm]\nonumber
\Delta_{\lambda_3} c_{\beta - \alpha}   & = & - \frac{1}{\epsilon} \;
 \displaystyle \frac{ t_\beta  (t_\beta^2 -1)(m_A^2+ v^2 \lambda_5)
 (m_A^2 - m_h^2 + m_H^2 - 2 m_{H_\pm}^2 + v^2 \lambda_5)}{m_H^2 (m_H^2-m_h^2)(1+t_\beta^2)^2 }
+ {\cal O} (\epsilon^0)\,,  \\[5mm]\nonumber
 \Delta_{\lambda_4} c_{\beta - \alpha}  & = &  \phantom{-}   \frac{1}{\epsilon} \;
 \displaystyle \frac{ t_\beta  (t_\beta^2 -1) (m_A^2+ v^2 \lambda_5)
 (2 m_A^2 - 2 m_{H_\pm}^2 + v^2 \lambda_5)} {m_H^2 (m_H^2-m_h^2)(1+t_\beta^2)^2 } 
+ {\cal O} (\epsilon^0)\,,  \\[5mm]\nonumber
  \Delta_{\lambda_5} c_{\beta - \alpha}   & = & \phantom{-}   \frac{1}{\epsilon} \;
 \displaystyle \frac{ t_\beta  (t_\beta^2 -1) v^2 \lambda_5 (m_A^2+ v^2 \lambda_5) }{m_H^2 (m_H^2-m_h^2)(1+t_\beta^2)^2 }
 + {\cal O} (\epsilon^0)\,.
\end{eqnarray}
Let us note that  the ${\cal O} (\ep^0)$ term, that has been dropped above, precisely contains the $\Delta_{\theta_i}v^2$ contribution to  $\Delta_{\theta_i}c_{\beta-\alpha}$, see the discussion around Eq.(\ref{Projection}). Actually, Eqs.(\ref{cba-lowestorder}) are a good approximation of the {\em independent} $\Delta_{\theta_i}c_{\beta-\alpha}$ fine-tuning, i.e. once it is projected onto the constant-$v^2$ hypersurface, according to Eq.(\ref{Projection}). Consistently with that, we note that at this order in the $\epsilon-$expansion, $\De_{\th_i} c_{\be-\al}={\cal O}\left( \frac{v^2}{m_H^2}\frac{1}{\ep}\right)$, 
thus the alignment becomes natural in the decoupling limit. 

Likewise, if $t_\beta\gg 1$, then $\De_{\th_i} c_{\be-\al}\sim 1/t_\beta$, showing that the alignment becomes non-fine-tuned in that regime. The origin of this remarkable fact was already mentioned in Sect.\ref{FT2HDM}, when discussing Eq.(\ref{cosZ6}).

In summary,
the alignment fine-tuning in the 2HDM is of the order
\beq
\label{aproxDcba}
\De c_{\beta-\alpha}\sim  \frac{1}{c_{\beta-\alpha}}\frac{1}{t_\beta}\frac{v^2}{m_H^2}
\eeq
which shows that it is mitigated in the decoupling limit and for large $t_\beta$.

\section{Numerical Analysis} \label{NumAn}
\quad 
We perform in this section a numerical exploration of the the fine-tuning(s) in the 2HDM parameter-space. We will focus on their dependence on the masses of the extra Higgses and on $t_\beta$; which, as  discussed above, are the main determinants of the size of the fine-tuning.
We will also consider two possible circumstances: that the alignment is sharp, i.e. quite smaller than the present experimental upper bounds, or close to them.

For these purposes we have designed five scenarios which capture all the relevant features of the problem, see Table \ref{table:scenarios}. We examine them in order.
\begin{table}[h]
    \centering
    \begin{tabular}{|c||c|c|c|c|c|}
    \hline
      Scenario & $\cos{(\beta-\alpha)}$&
      $m_H (\GeV)$ & $m_A,\ m_{H^\pm}(\GeV)$ &  $\tan\beta$ & $\lambda_5$ \\
      \hline
      1 &	$0.01$ & $[200, 2000]$ & $=m_H$ & $1.1$ & $-5.5$\\
      2 &	$0.1$ & $[200, 900]$ & $=m_H$ & $1.1$ & $-5.5$\\
      3 &	$0.01$ & $240$ & $280$ & $[1,15]$ & $-0.4$\\
      4 &	$0.001$ & $1498$ & $1502$ & $[1,15]$ & $-0.25$\\
      5 &	$0.1$ & $260$ & $320$ & $[1,15]$ & $-0.625$\\
      \hline  
    \end{tabular}
    \caption{Benchmark Scenarios used in the numerical analysis to explore the electroweak and alignment fine-tunings in the 2HDM. The fact that $m_A=m_{H^\pm}$ implies $\lambda_4=\lambda_5$ in all cases.}
    \label{table:scenarios}
\end{table}

\subsection*{Scenario 1}

This scenario is chosen to show the dependence of the fine-tuning on the extra-Higgs masses in the case of sharp alignment,  $\epsilon=c_{\beta-\alpha}=0.01$ and low $t_\beta$. As discussed above, in the alignment limit the larger the extra-Higgs masses, the closer each other become. This feature is trivially fulfilled here as we have set 
$m_H=m_A= m_{H^\pm}\equiv m_0$. 
On the other hand, the value of $\lambda_5$ has been chosen to enhance the range of $m_0$ that preserves perturbativity and stability, but it does not play an important role in the discussion.

\begin{figure} [h!]
    \centering
    \subfigure{\includegraphics[height=4.5cm]{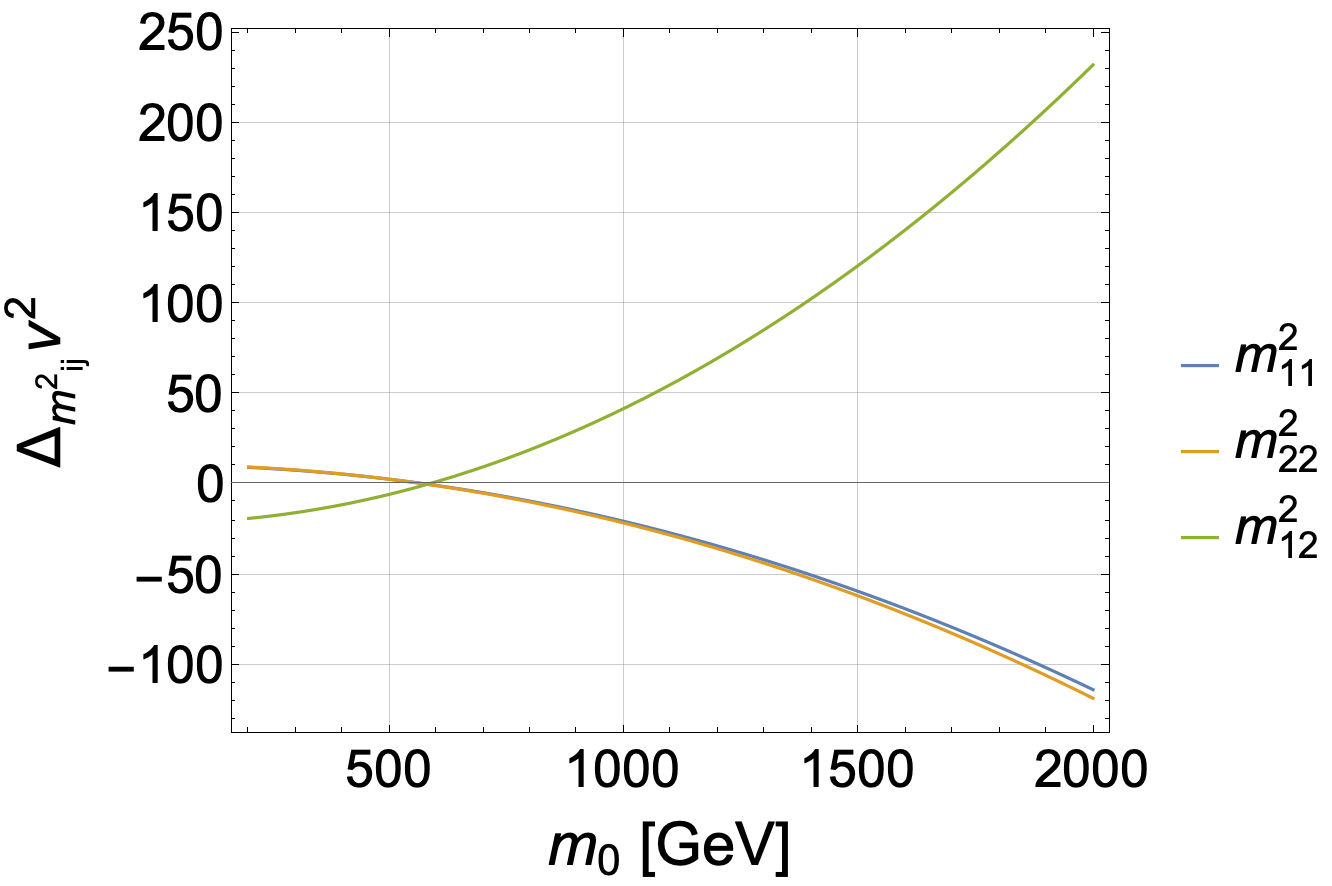}}\hspace{5mm}
    \subfigure{\includegraphics[height=4.4cm]{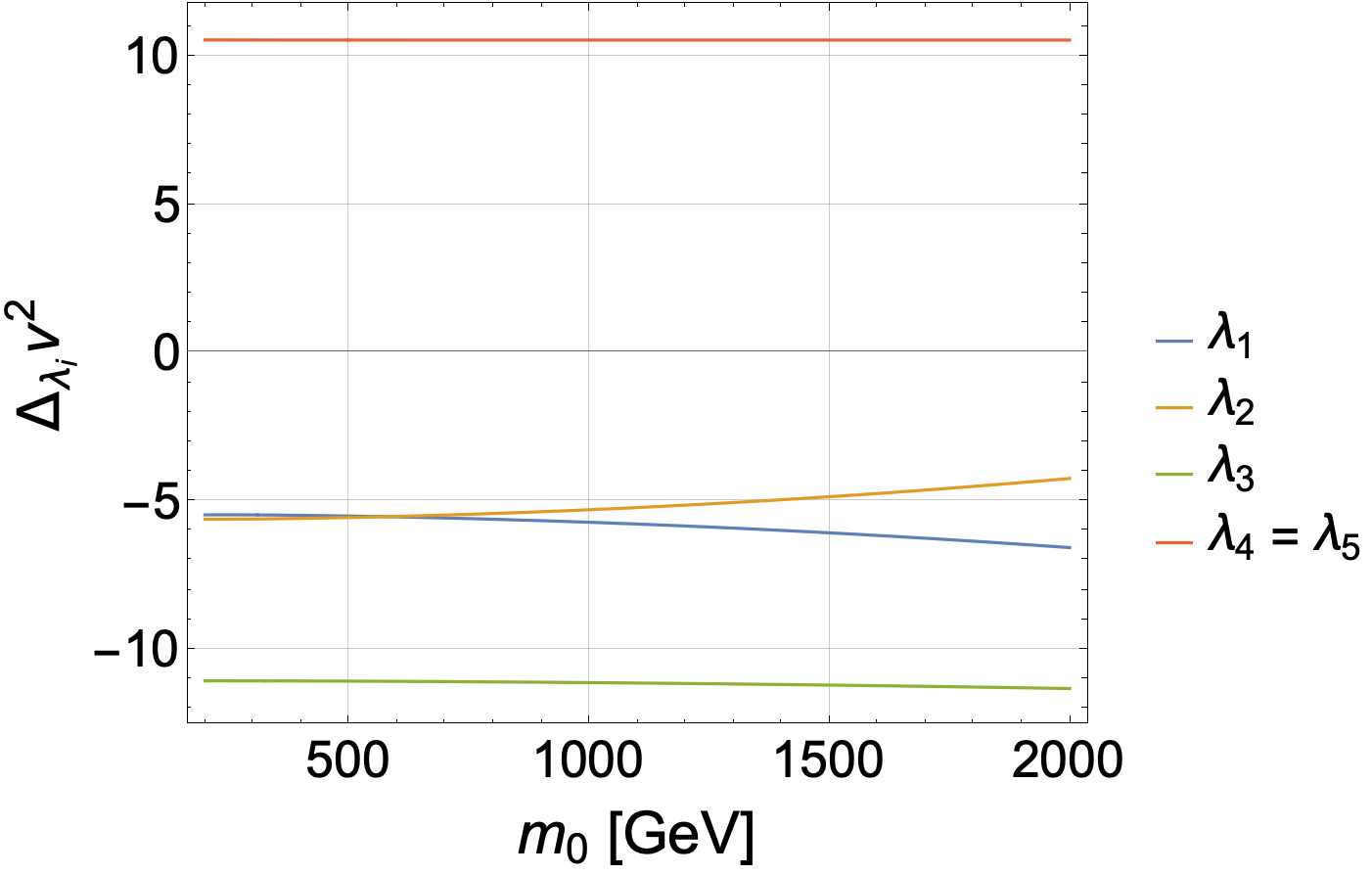}}
    \caption{Dependence of the electroweak fine-tuning, $\Delta_{\theta_i}v^2$,  on the common mass of the extra Higgses, $m_0$ for Scenario 1.
    }
    \label{Fig:Van1_FTv}
\end{figure}

The electroweak fine-tuning, $\Delta_{\theta_i}v^2$ vs. $m_0$ is shown in Fig.~\ref{Fig:Van1_FTv} for the various  $\theta_i$ parameters.
As expected, the only important fine-tunings are those associated with mass parameters, $m_{ij}^2$, which, for large $m_0$ increase according to the trend of Eq.(\ref{aproxDv2}), becoming as large as 250 for $m_0 \sim 2$ TeV. In contrast, for $m_0\simlt 700$ GeV the electroweak fine-tuning is quite mild.

\begin{figure} [h!]
    \centering
    \subfigure{\includegraphics[height=4.5cm]{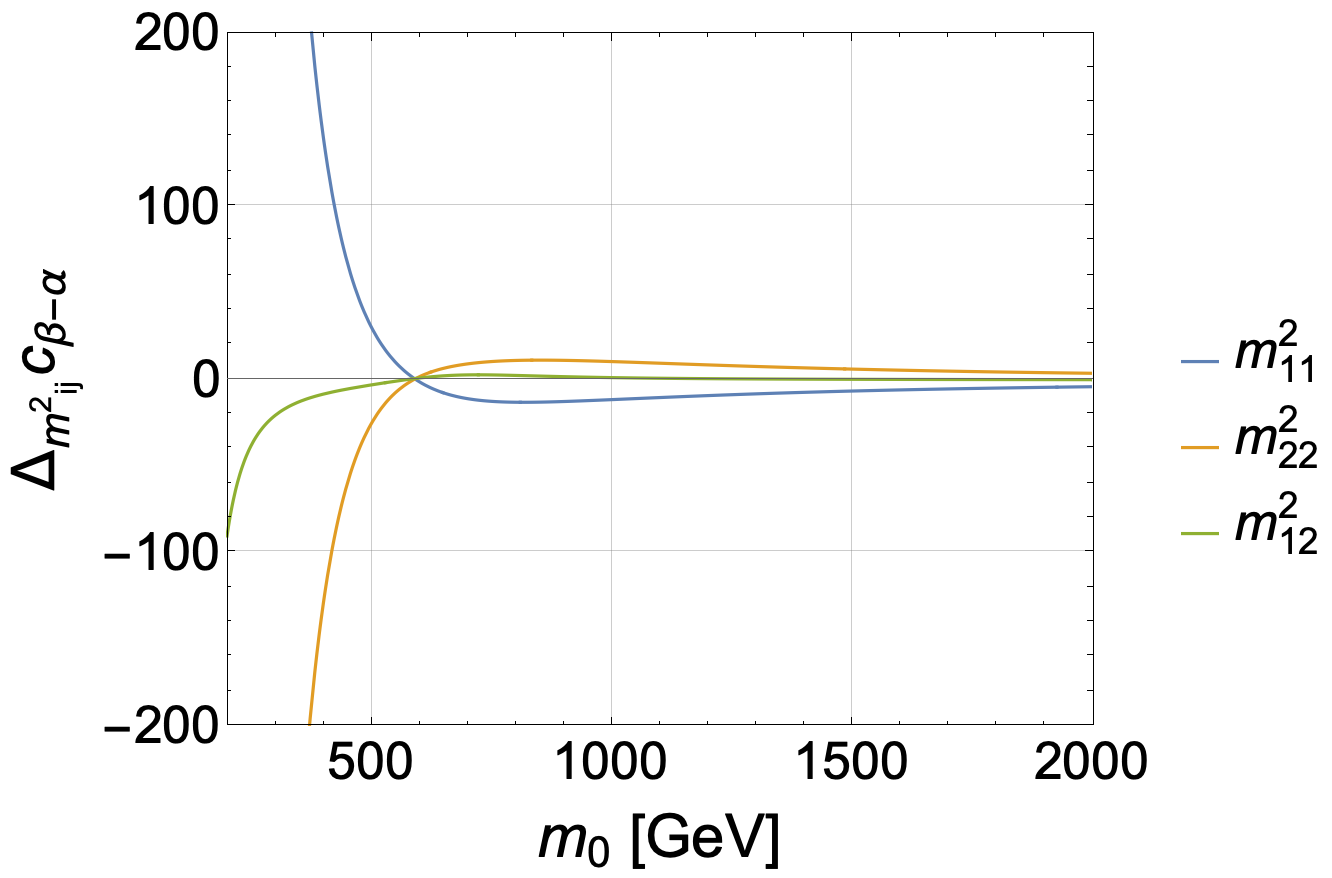}}\hspace{5mm}
    \subfigure{\includegraphics[height=4.5cm]{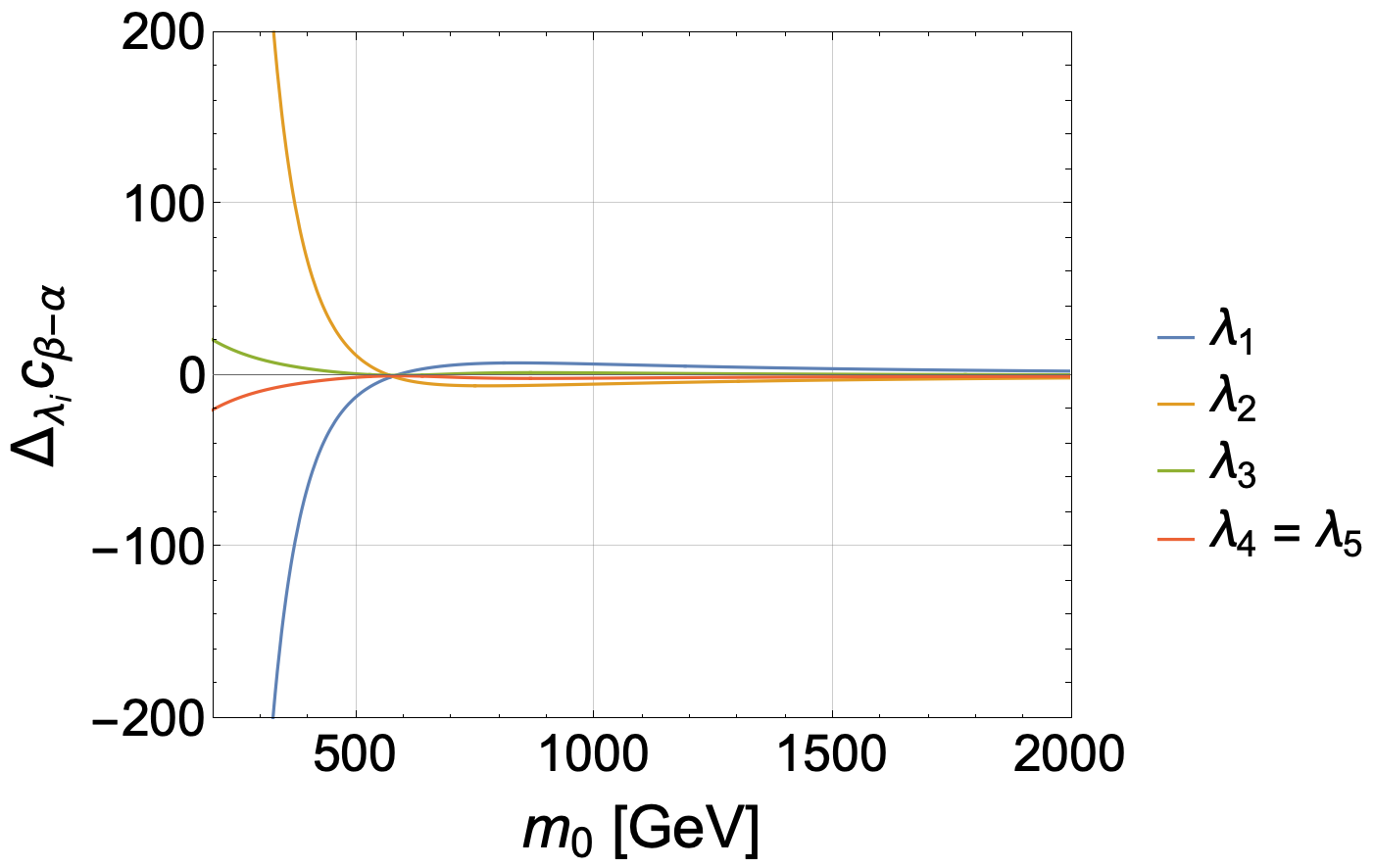}}
    \caption{The same as Fig.~\ref{Fig:Van1_FTv} for  the independent $\Delta_{\theta_i} c_{\be-\al}$ fine-tuning, i.e. projected onto the once the $v^2={\rm const.}$ hypersurface.}
    \label{Fig:Van1_FTcos}
\end{figure}

Concerning the alignment fine-tuning, we have projected the complete expressions for $\Delta_{\theta_i}c_{\beta-\alpha}$ given in the Appendix onto the constant-$v^2$ surface, according to Eq.(\ref{Projection}), in order to show only the independent fine-tuning. As commented at the end of the previous section, 
this is essentially equivalent to use the lowest-order expressions for $\Delta_{\theta_i}c_{\beta-\alpha}$, Eqs.(\ref{cba-lowestorder}).
 The corresponding results are shown in Fig. \ref{Fig:Van1_FTcos}. As expected, the alignment fine-tuning increases for decreasing masses of the extra Higgses, becoming as large as $\sim 1000$ at $m_0=200$ GeV. 

By comparing Figs.~\ref{Fig:Van1_FTv} and \ref{Fig:Van1_FTcos}, we see that there is an intermediate region, $550\  {\rm GeV}\simlt m_0\simlt 700\ {\rm GeV}$, where both the EW and the alignment fine-tunings remain at acceptable values, $\simlt {\cal O}(10)$.

\begin{figure} [h!]
    \centering
    \subfigure{\includegraphics[height=4.5cm]{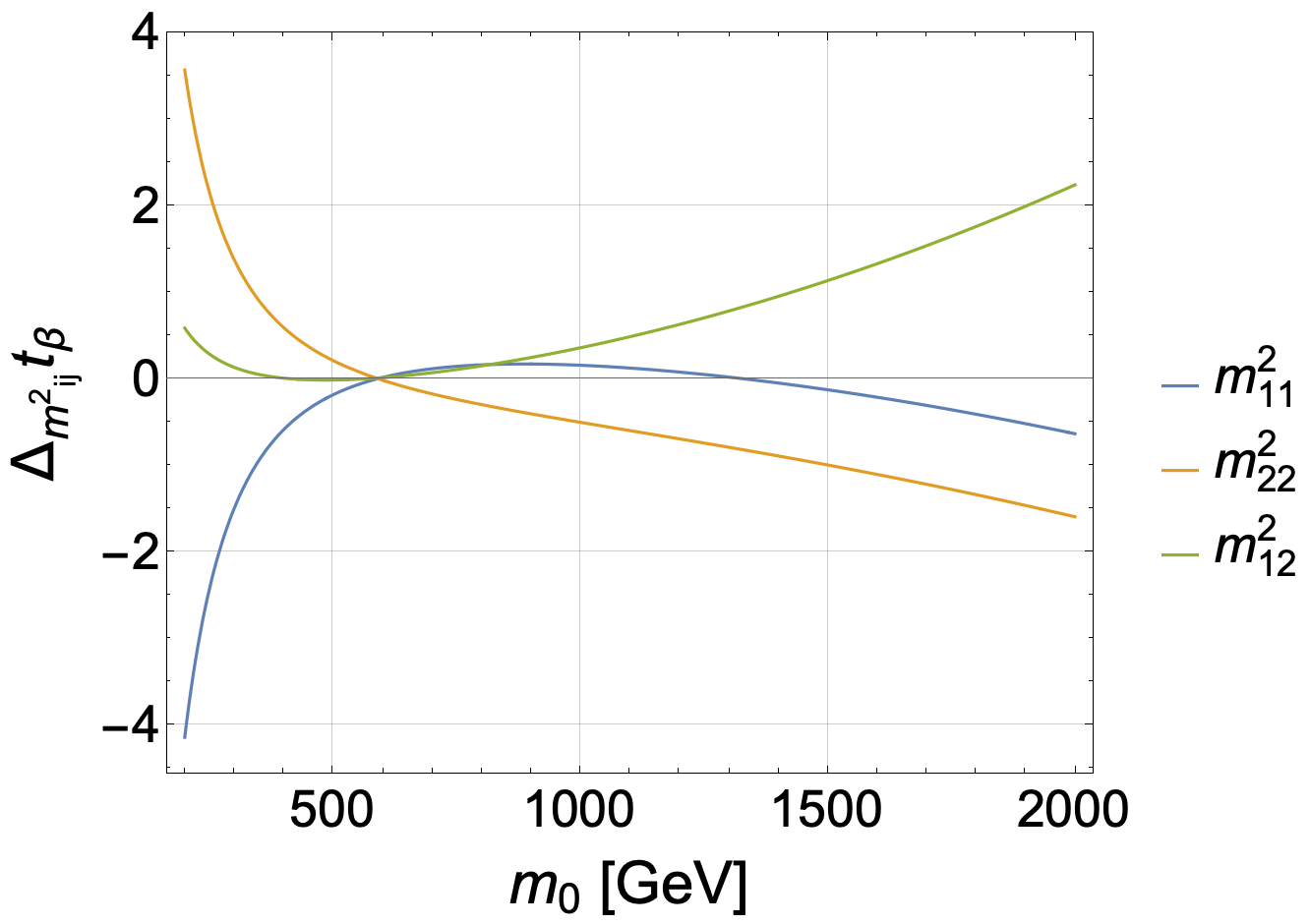}}\hspace{5mm}
    \subfigure{\includegraphics[height=4.4cm]{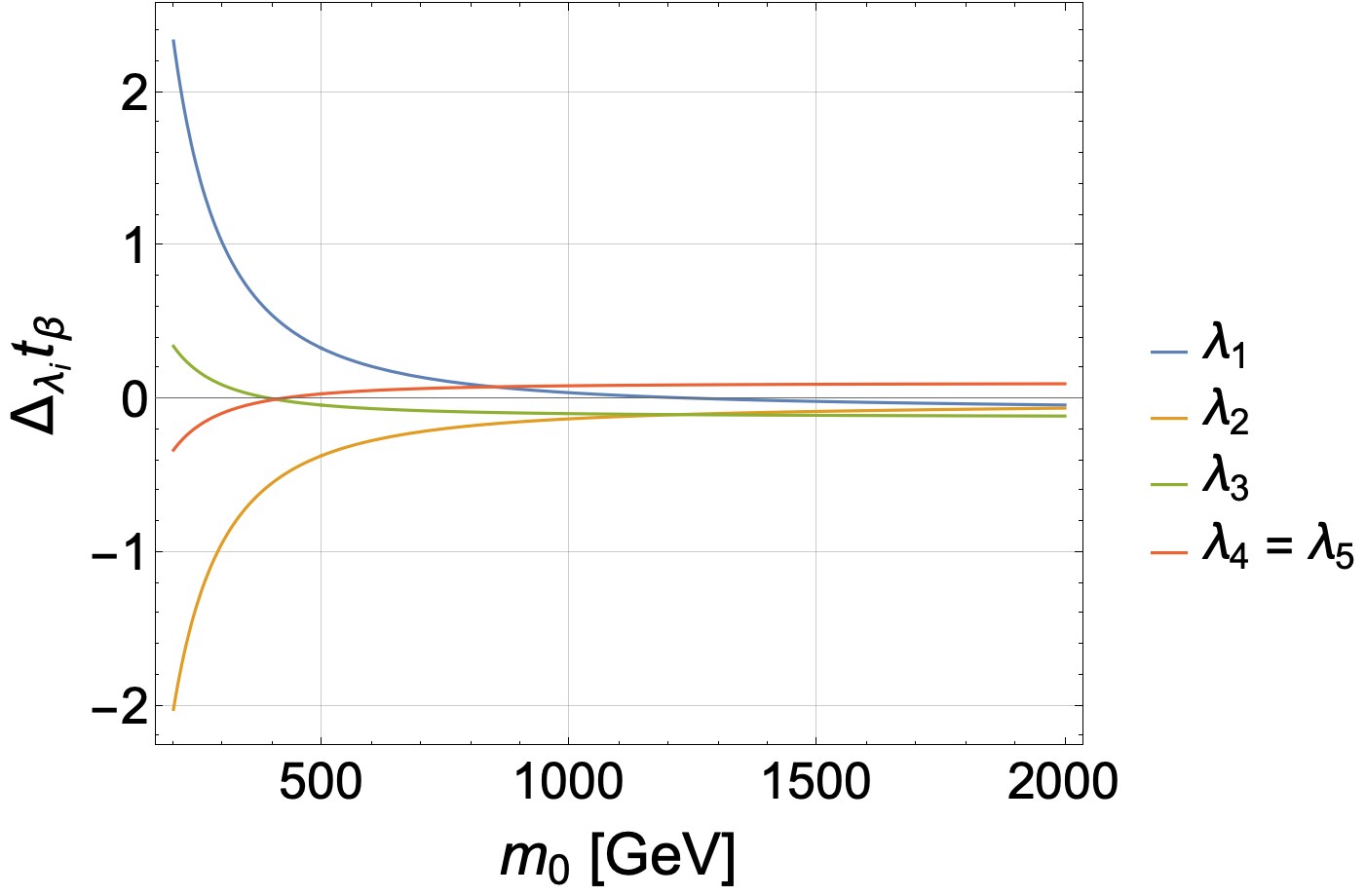}}
    \caption{The same as Fig.~\ref{Fig:Van1_FTv} for  the $\Delta_{\theta_i} t_\beta$ fine-tuning.}
    \label{Fig:Van1_FTtan}
\end{figure}

Finally, the fine-tuning in $t_\be$ is shown in Fig. \ref{Fig:Van1_FTtan}. As expected, 
$t_\be$ is not a fine-tuned quantity, i.e. the associated fine-tuning parameters, $\Delta_{\theta_i}t_\beta$ are ${\cal O}(1)$ in all cases.

\subsection*{Scenario 2}

This is similar to Scenario 1, but with a weak alignment; namely, $\epsilon=c_{\beta-\alpha}=0.1$, about as large as allowed by experimental constraints. An important consequence of the mild alignment is that
now $m_0$ cannot be too large. The reason is that for very large masses of the extra-Higgses we enter the decoupling regime, which automatically leads to alignment. Under such circumstances, it is very difficult to keep a certain misalignment (typically it requires non-perturbative $\lambda-$couplings). In consequence the available range for $m_0$ lies below $\sim 1$ TeV. As we will see later, for higher $t_\beta$, this upper bound becomes stronger.

\begin{figure} [h!]
    \centering
    \subfigure
    {\includegraphics[height=4.5cm]{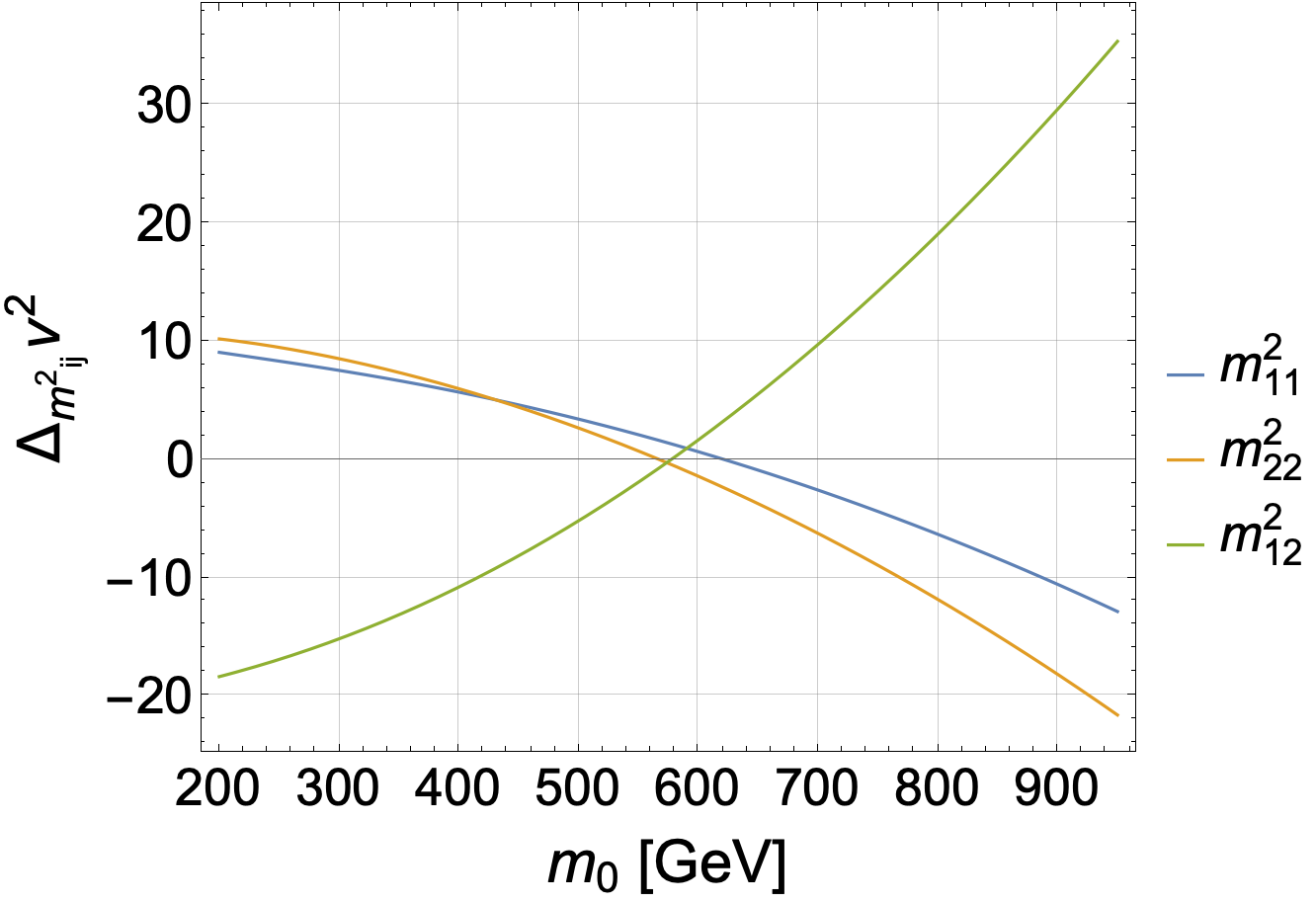}}\hspace{5mm}
    \subfigure
    {\includegraphics[height=4.5cm]{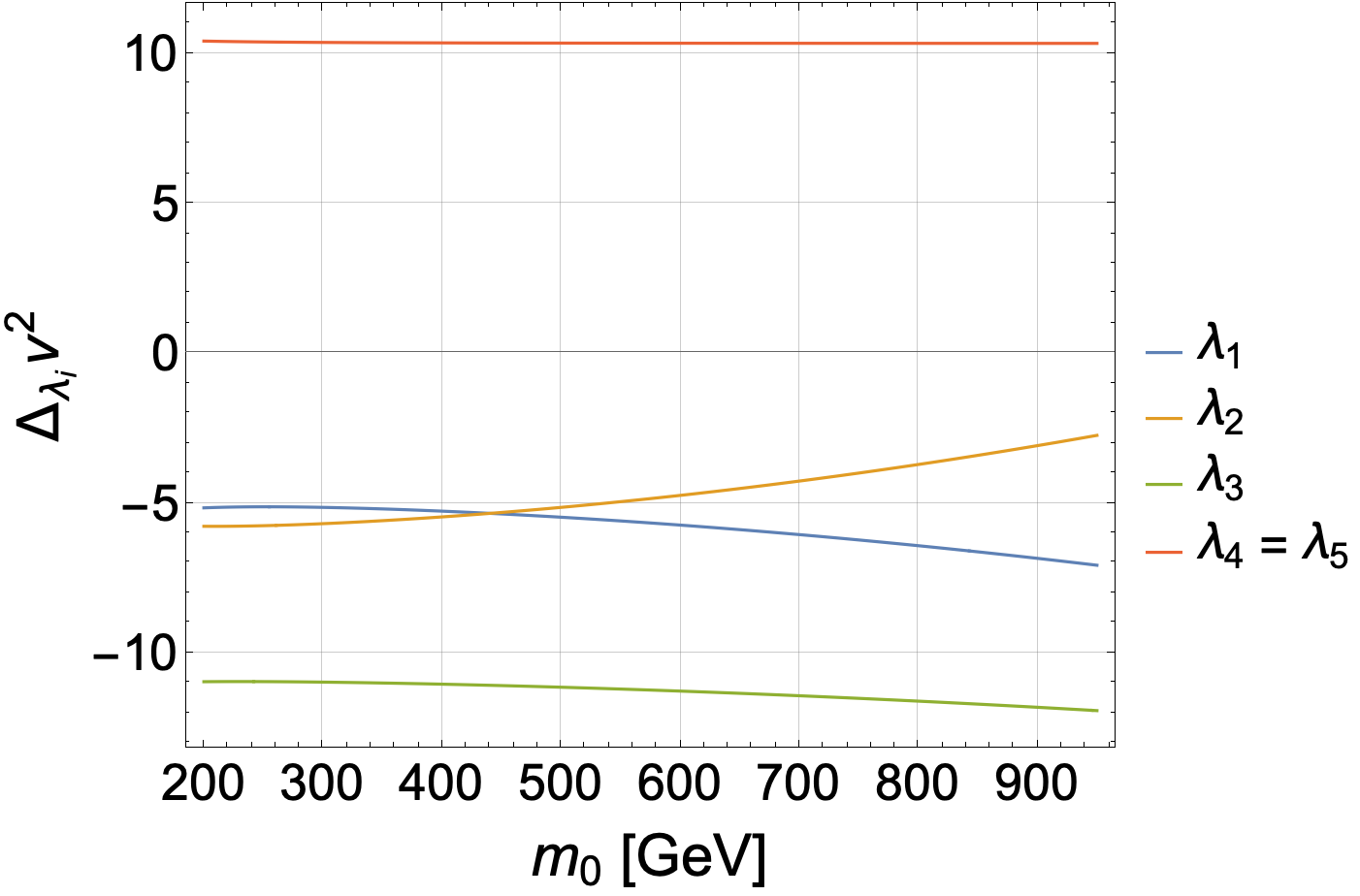}}
    \caption{The same as Fig. \ref{Fig:Van1_FTv} for Scenario 2.}
    \label{Fig:Van2_FTv}
\end{figure}

\begin{figure} [h!]
    \centering
    \subfigure
    {\includegraphics[height=4.5cm]{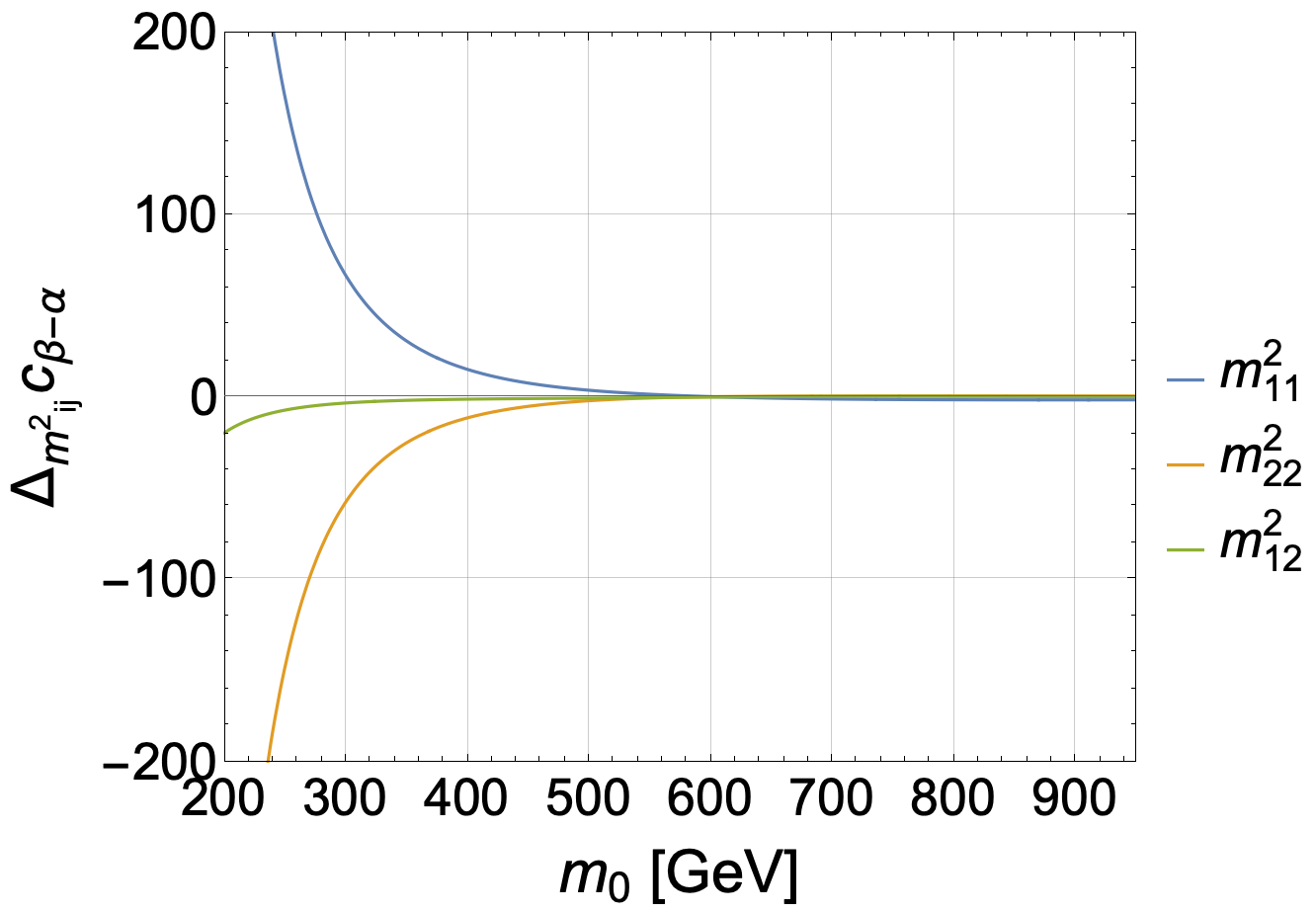}}\hspace{5mm}
    \subfigure
    {\includegraphics[height=4.5cm]{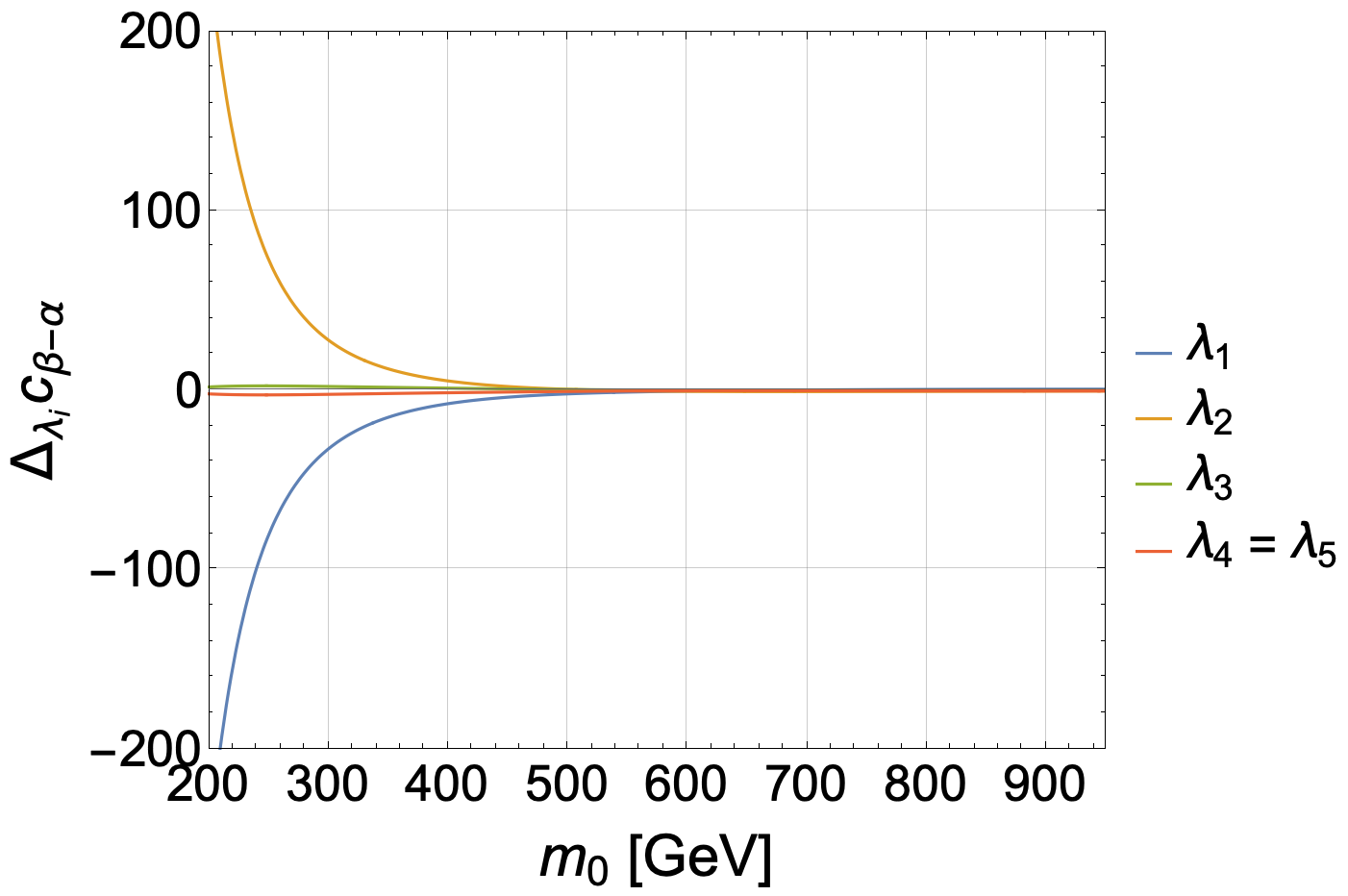}}
    \caption{The same as Fig. \ref{Fig:Van1_FTcos} for Scenario 2. }
    \label{Fig:Van2_FTcos}
\end{figure}

\begin{figure} [h!]
    \centering
    \subfigure{\includegraphics[height=4.5cm]{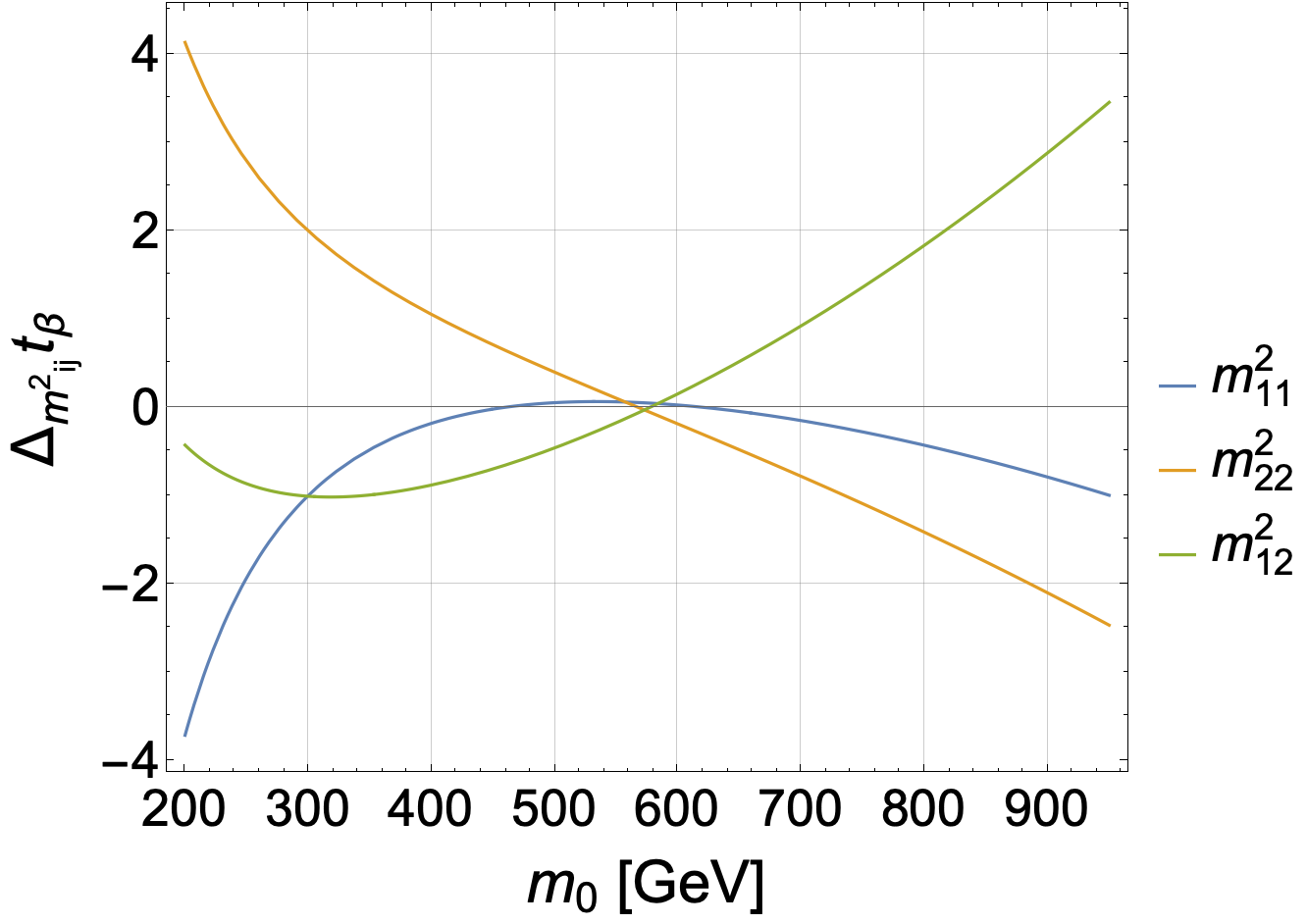}}\hspace{5mm}
    \subfigure{\includegraphics[height=4.5cm]{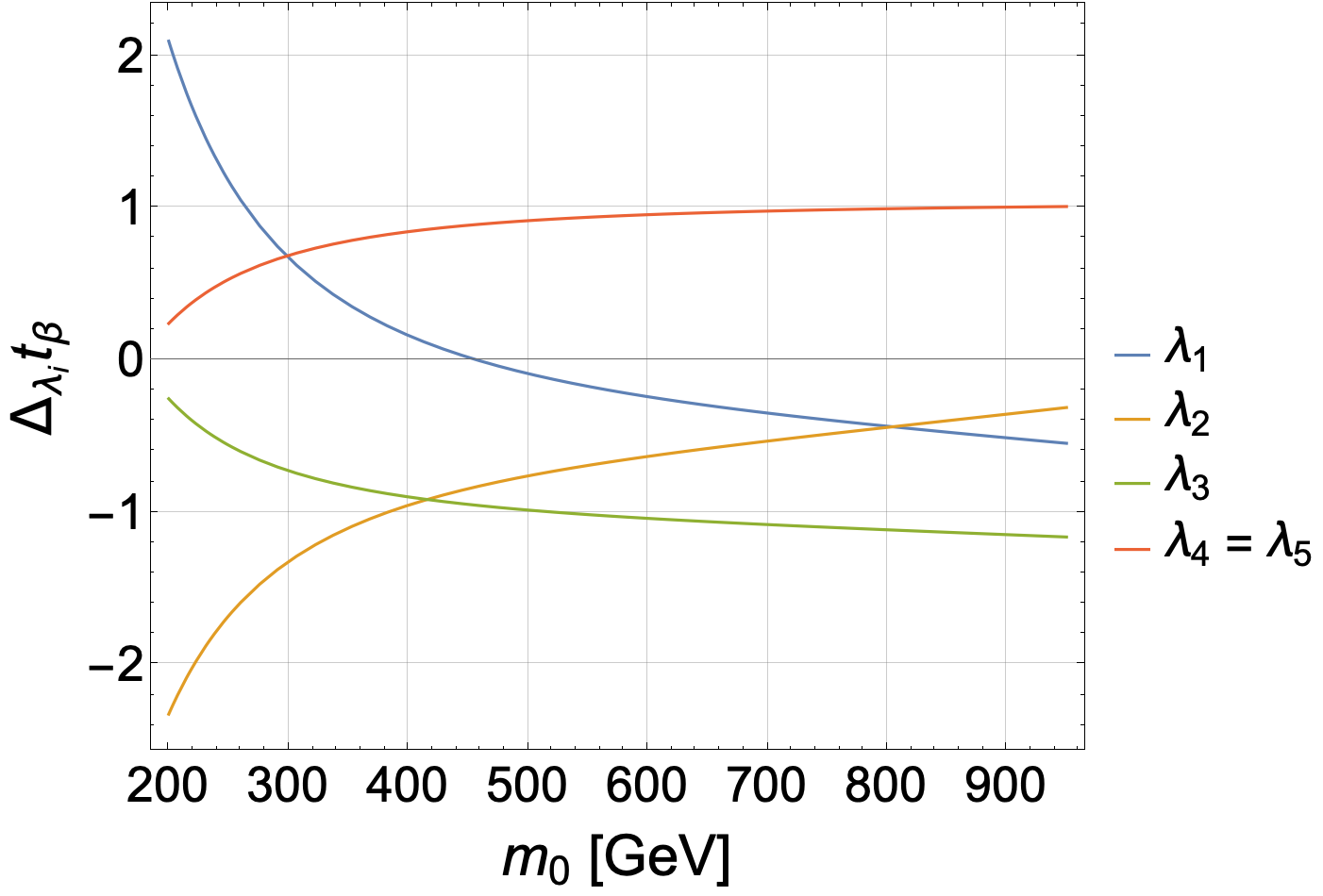}}
    \caption{The same as Fig. \ref{Fig:Van1_FTtan} for Scenario 2.}
    \label{Fig:Van2_FTtan}
\end{figure}

Apart from this limitation, the trends of the fine-tuning are similar to those for Scenario 1, as shown in Figs. \ref{Fig:Van2_FTv}-\ref{Fig:Van2_FTtan}.
The main difference has to do with the alignment fine-tuning, $\Delta c_{\beta-\alpha}$, which is less severe than before (compare Figs.~\ref{Fig:Van1_FTcos} and \ref{Fig:Van2_FTcos}) because the required value of $c_{\beta-\alpha}$ is not that small. Again, there is an intermediate region, $500\  {\rm GeV}\simlt m_0\simlt 700\ {\rm GeV}$, where both the electroweak and the alignment fine-tunings remain at acceptable values, $\simlt {\cal O}(10)$.
same as Fig.~\ref{Fig:Van1_FTv} for Scenario 2.

\subsection*{Scenario 3}

This scenario is chosen to show the dependence of the fine-tuning on $t_\beta$ in the case of sharp alignment,  $\epsilon=c_{\beta-\alpha}=0.01$ and low extra-Higgs masses, $\sim 250$ GeV. 
The value of $\lambda_5$ has been picked to ensure perturbativity and stability in a significant range of $t_\beta$, namely
$t_\be\in [1,15]$.

\begin{figure} [h!]
    \centering
    \subfigure
    {\includegraphics[height=4.5cm]{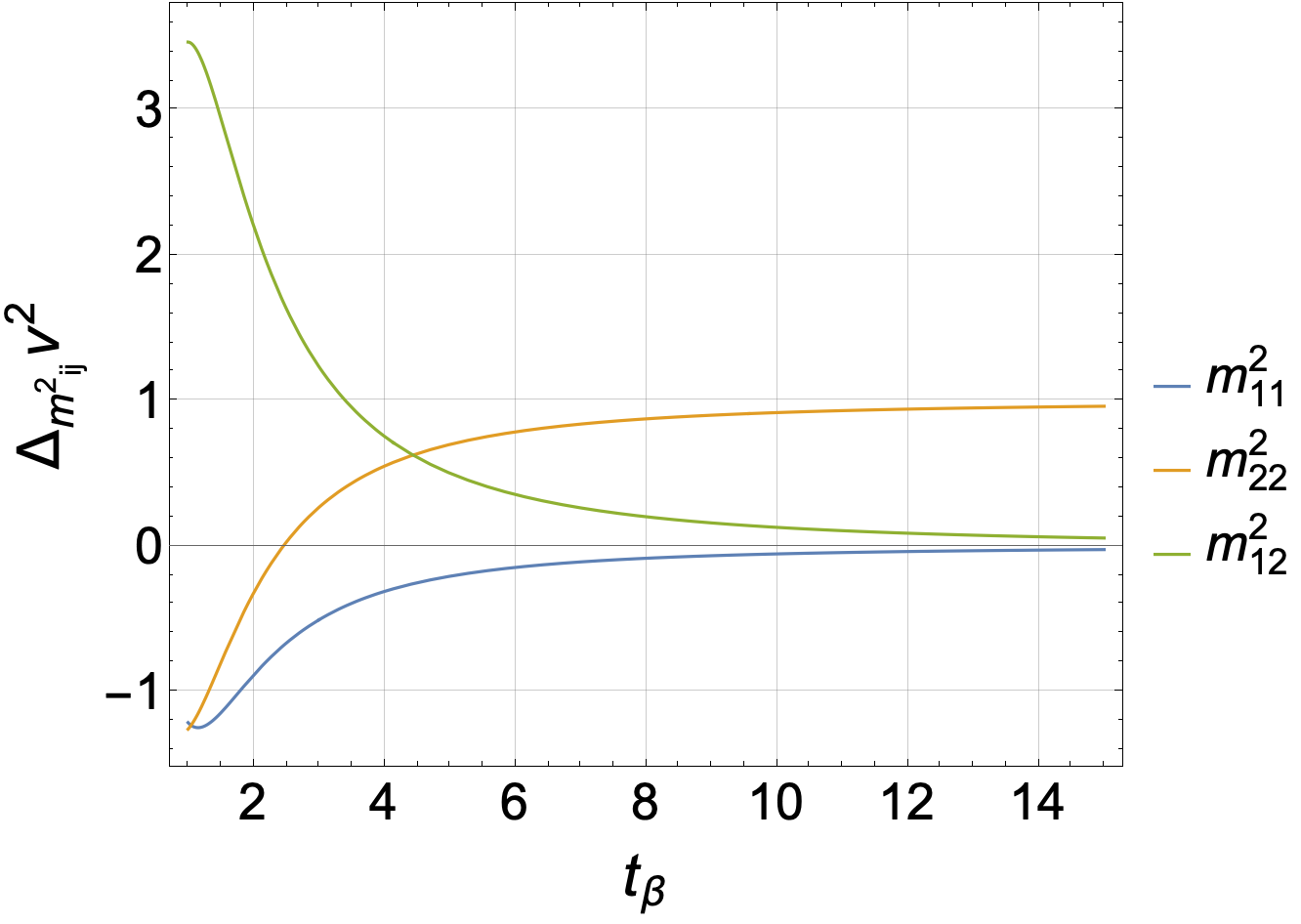}}\hspace{5mm}
    \subfigure
    {\includegraphics[height=4.5cm]{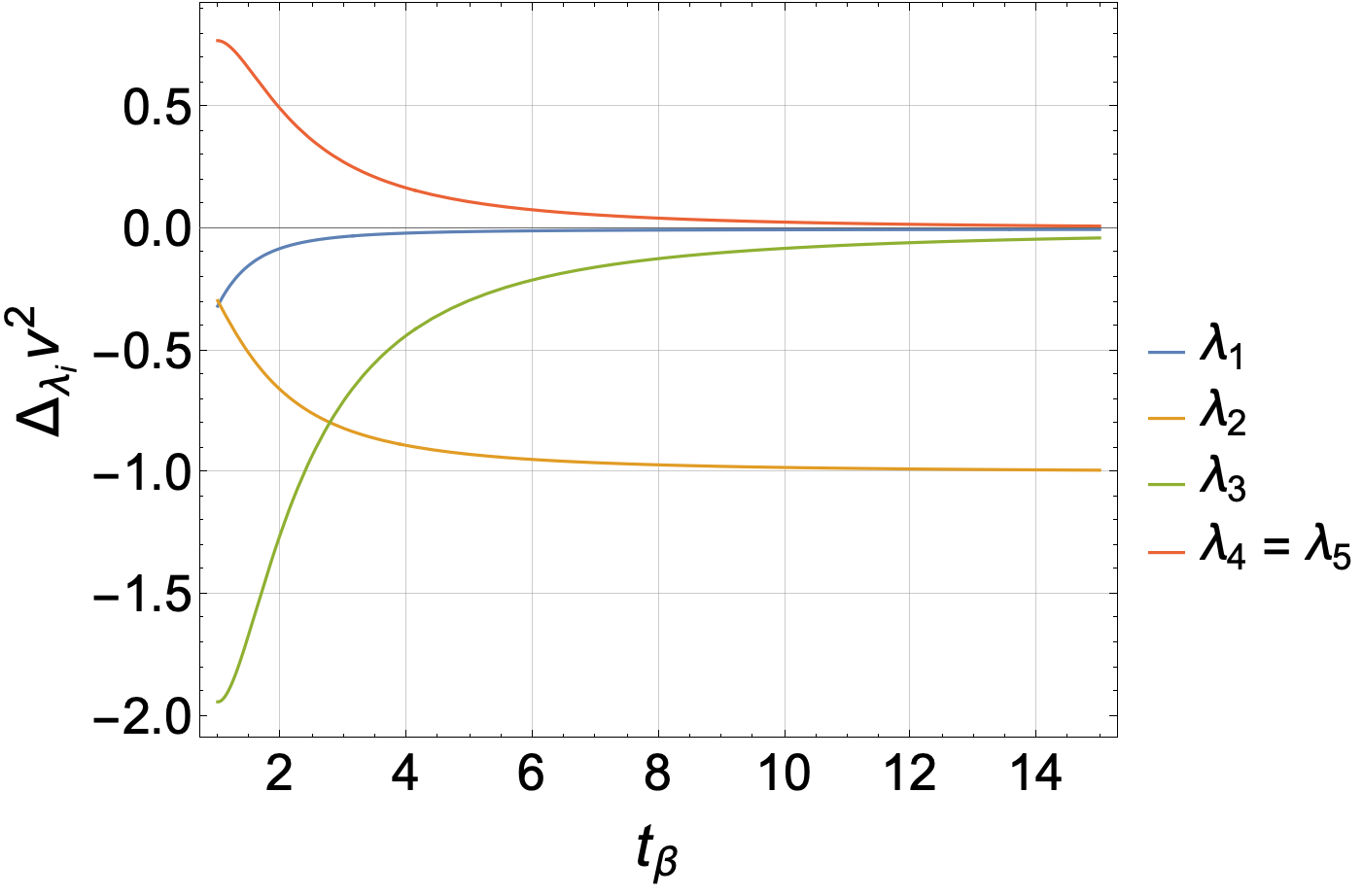}}
    \caption{Dependence of the electroweak fine-tuning,  $\Delta_{\theta_i}v^2$ on $t_\beta$ for Scenario 3.}
    \label{Fig:Van3_FTv}
\end{figure}

The electroweak fine-tuning, $\Delta_{\theta_i}v^2$ vs. $t_\beta$ is shown in Fig.~\ref{Fig:Van3_FTv}. As discussed in the previous section, the fine-tuning decreases with increasing $t_\beta$. Since this scenario has a rather low value of $m_0$, the electroweak fine-tuning is never large or even significant. 

\begin{figure} [h!]
    \centering
    \subfigure
    {\includegraphics[height=4.5cm]{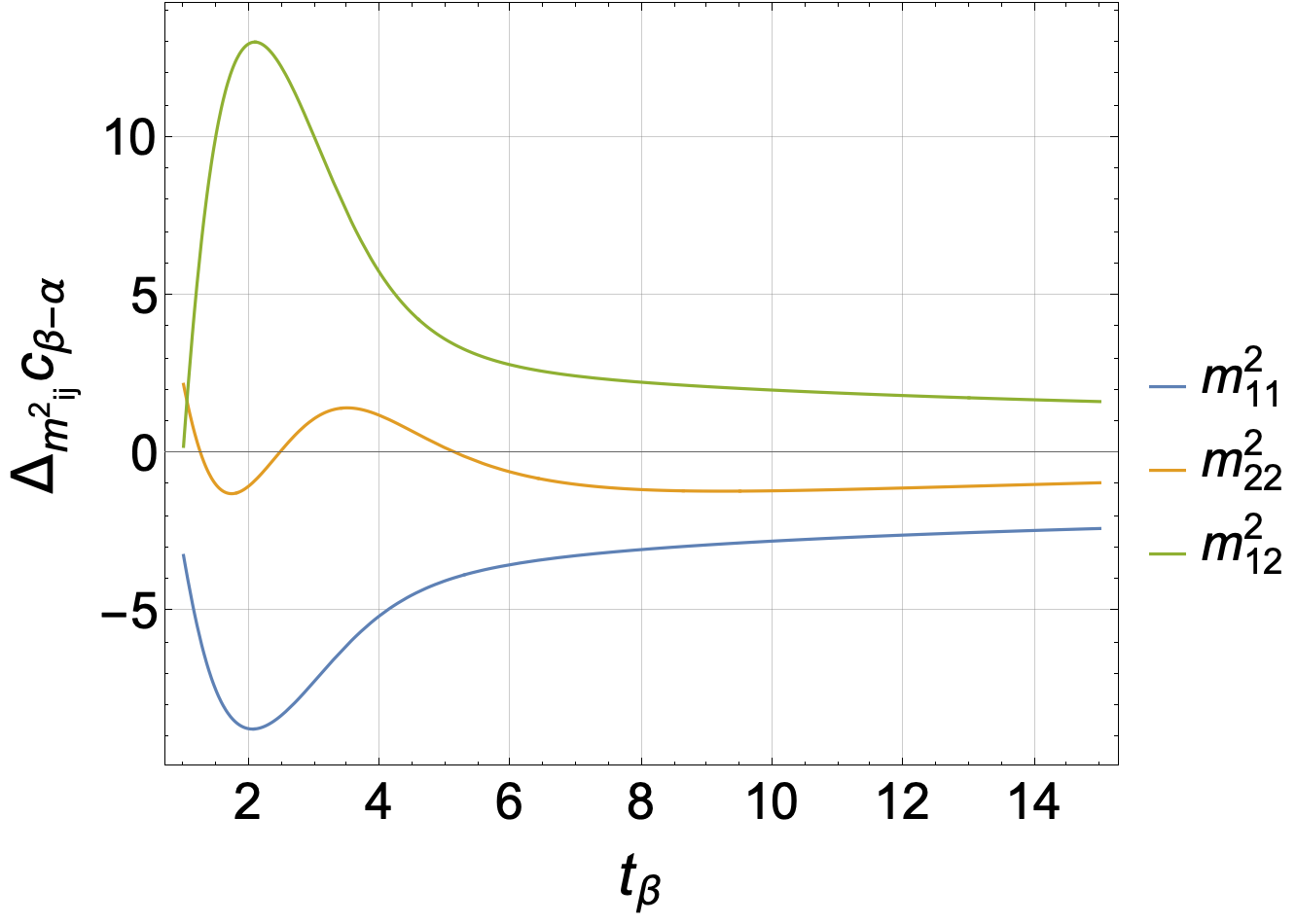}}\hspace{5mm}
    \subfigure
    {\includegraphics[height=4.5cm]{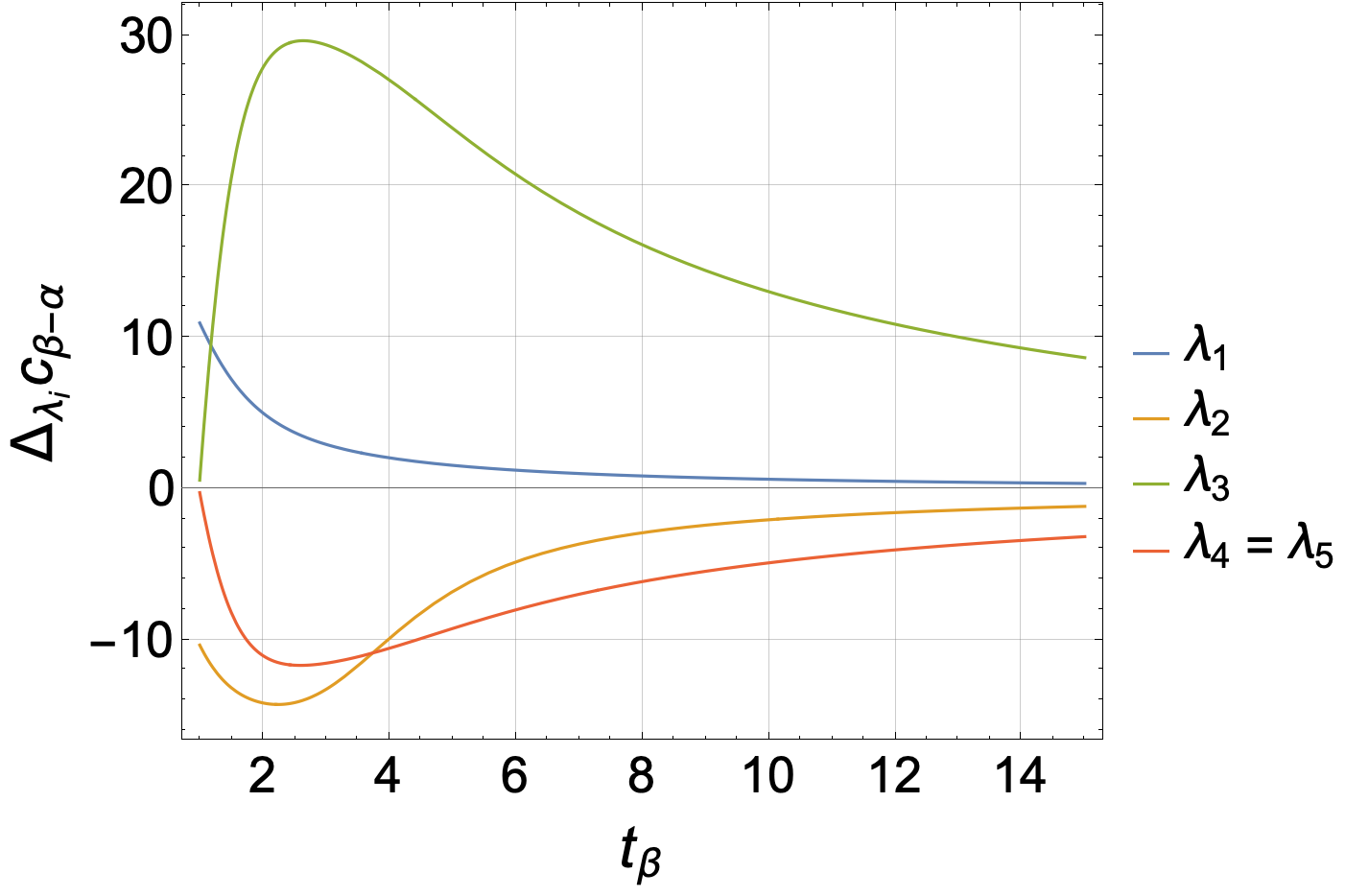}}
    \caption{The same as Fig. \ref{Fig:Van3_FTv} for  the independent $\Delta_{\theta_i} c_{\be-\al}$ fine-tuning (projected onto the once the $v^2={\rm const.}$ hypersurface). }
    \label{Fig:Van3_FTcos}
\end{figure}

The fine-tuning in the alignment,  $\Delta_{\theta_i}c_{\beta-\al}$ vs. $t_\beta$, is shown in Fig.~\ref{Fig:Van3_FTcos}. As discussed above, this fine-tuning also gets less severe as $t_\beta$ increases, and for $t_\beta\simgt 10$ it becomes $\leq {\cal O} (10)$. 
It is worth-noticing that
several fine-tuning parameters, namely
$\Delta_{m^2_{12}}c_{\beta-\al}$, $\Delta_{\lambda_3}c_{\beta-\al}$, $\Delta_{\lambda_4}c_{\beta-\al}$ and $\Delta_{\lambda_5}c_{\beta-\al}$, abruptly fall to zero for $t_\beta\rightarrow 1$.
This is a general fact, as can be verified from the general expressions
(\ref{cba-lowestorder}) for these quantities, which have a $(t_\beta^2-1)$ factor. The origin of this feature is the following. 
It is easy to check from Eqs.(\ref{dV1}, \ref{dV2}, \ref{tbetaalign}) that $t_\beta\simeq 1$ {\em and} a fine alignment require $\lambda_1\simeq \lambda_2$, $m_{11}^2\simeq m_{22}^2$. This corresponds to a situation where the consistency condition for perfect alignment, (\ref{eq:constraint}) is approximately fulfilled in an obvious way. In that case, it is clear  that changes in $m_{12}^2, \lambda_3, \lambda_4, \lambda_5$ do not spoil  condition (\ref{eq:constraint}); in other words, the fine alignment is insensitive to these changes, thus the vanishing of the associated fine-tuning parameters at $t_\beta=1$. This rule does not apply to the other initial parameters, $m_{11}^2, m_{22}^2, \lambda_1, \lambda_2$.
As it was commented after Eq.(\ref{eq:constraint}), in the $\lambda_1= \lambda_2$, $m_{11}^2= m_{22}^2$. limit (which corresponds to $t_\beta=1$ and perfect alignment) there arises a symmetry in the potential, which is however broken by the Yukawa couplings.

\begin{figure} [h!]
    \centering
    \subfigure{\includegraphics[height=4.5cm]{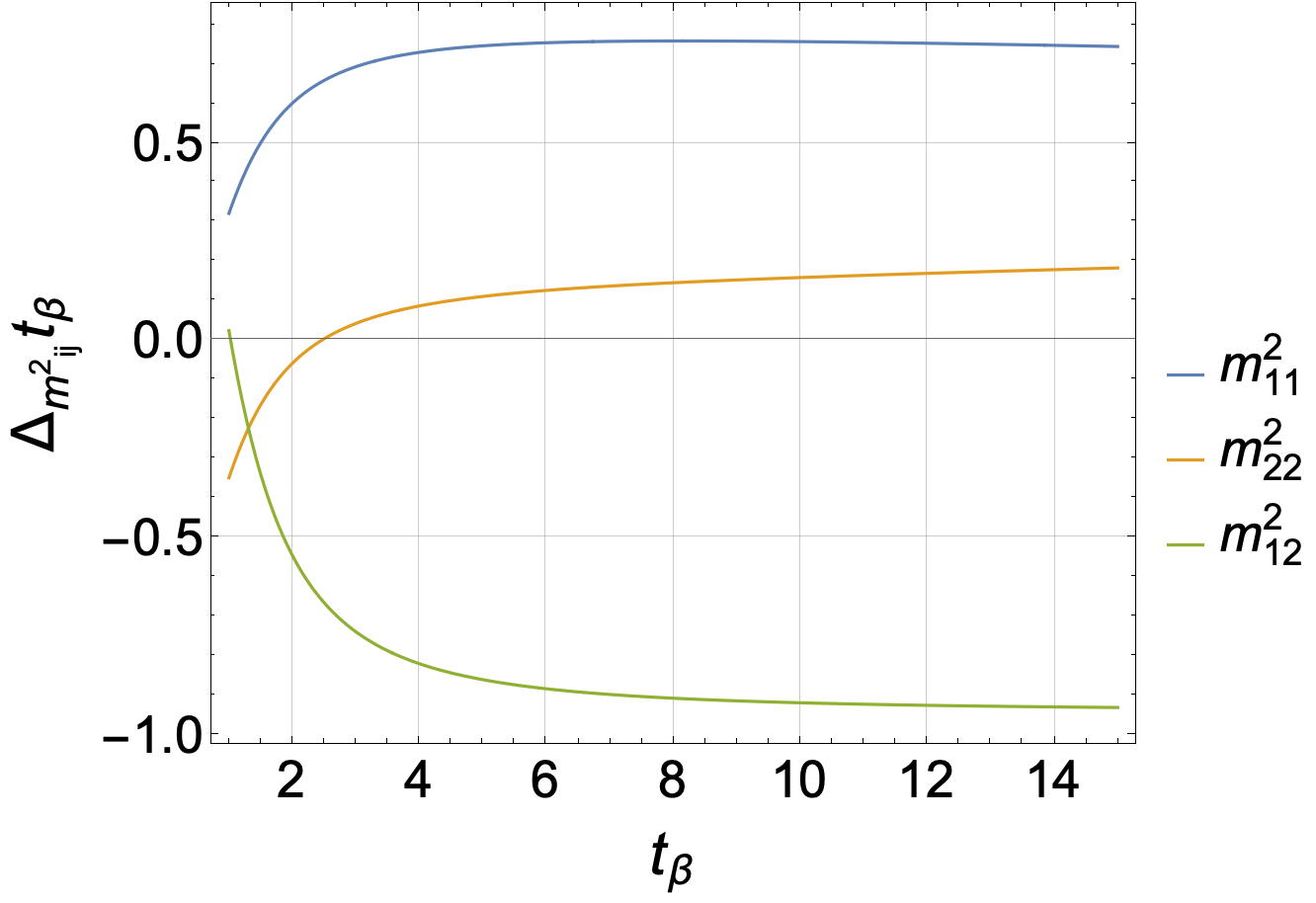}}\hspace{5mm}
    \subfigure{\includegraphics[height=4.5cm]{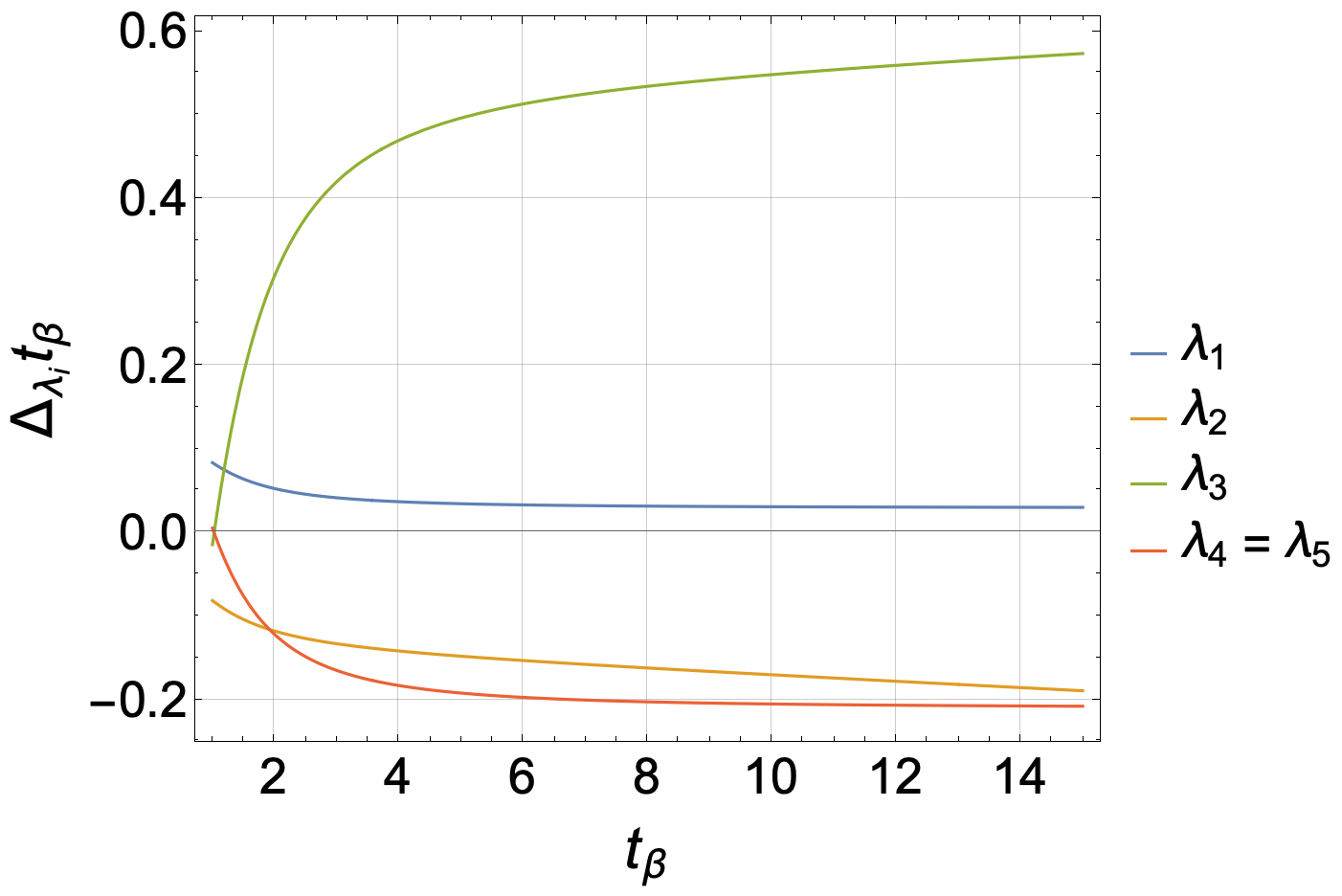}}
    \caption{
    The same as Fig. \ref{Fig:Van3_FTv} for  the $\Delta_{\theta_i} t_\beta$ fine-tuning.}
    \label{Fig:Van3_FTtan}
\end{figure}

Finally, as usual, there is no fine-tuning in $t_\beta$, as shown in Fig.~\ref{Fig:Van3_FTtan}.

\newpage
\subsection*{Scenario 4}

\begin{figure} [h!]
    \centering
    \subfigure
    {\includegraphics[height=4.5cm]{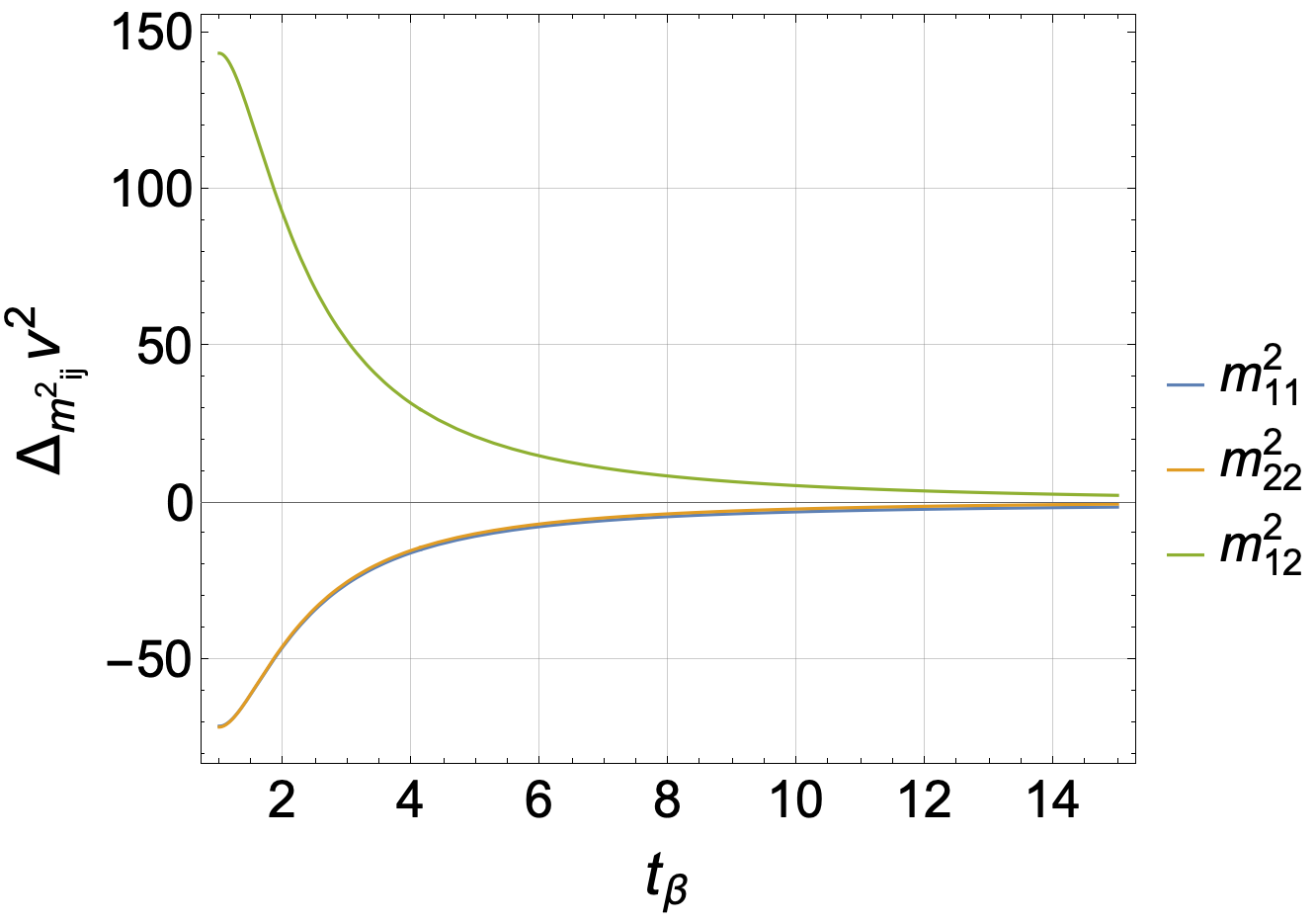}}\hspace{5mm}
    \subfigure
    {\includegraphics[height=4.5cm]{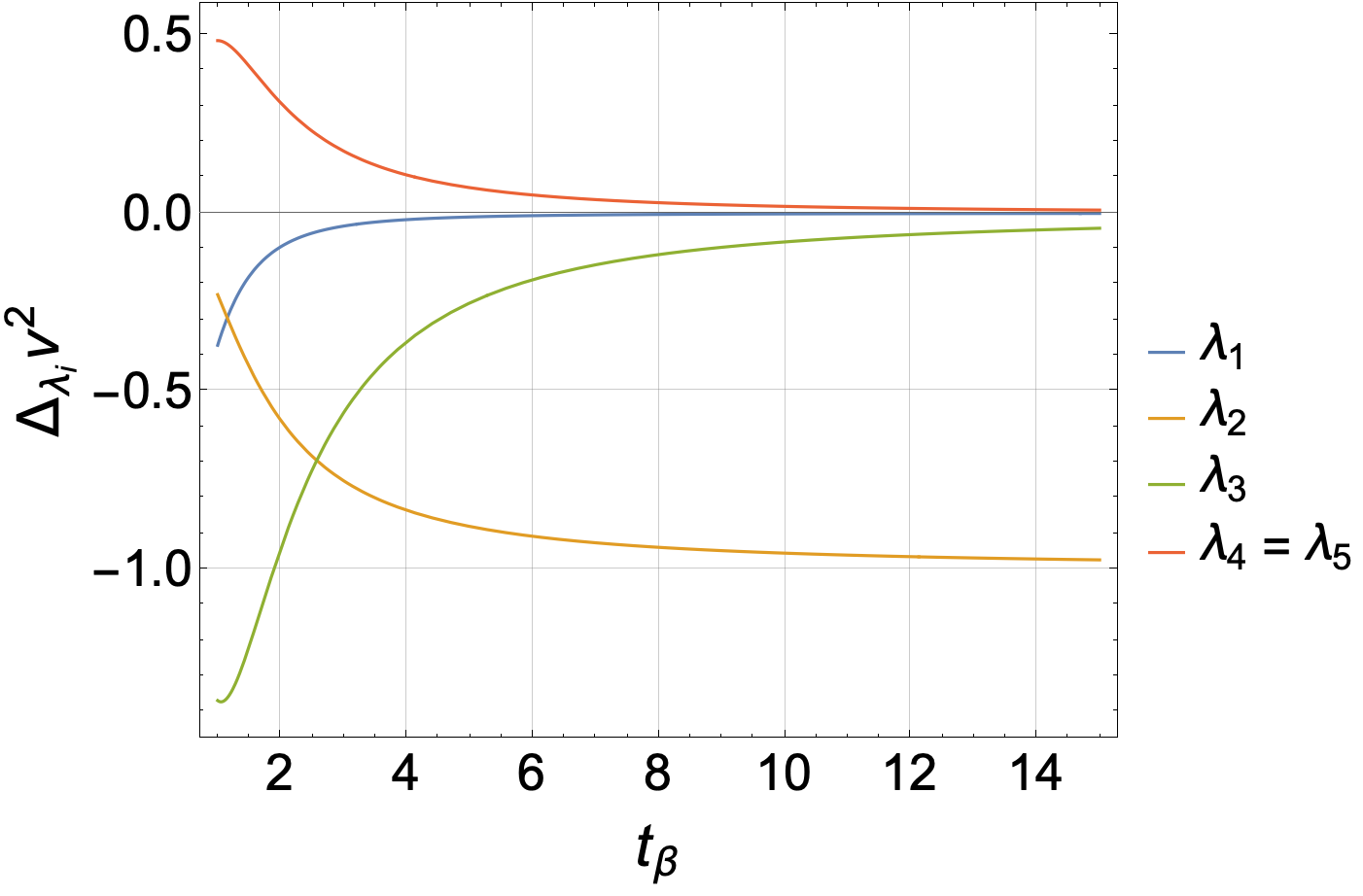}}
    \caption{The same as Fig. \ref{Fig:Van3_FTv} for Scenario 4.}
    \label{Fig:Van4_FTv}
\end{figure}

\begin{figure} [h!]
    \centering
    \subfigure
    {\includegraphics[height=4.5cm]{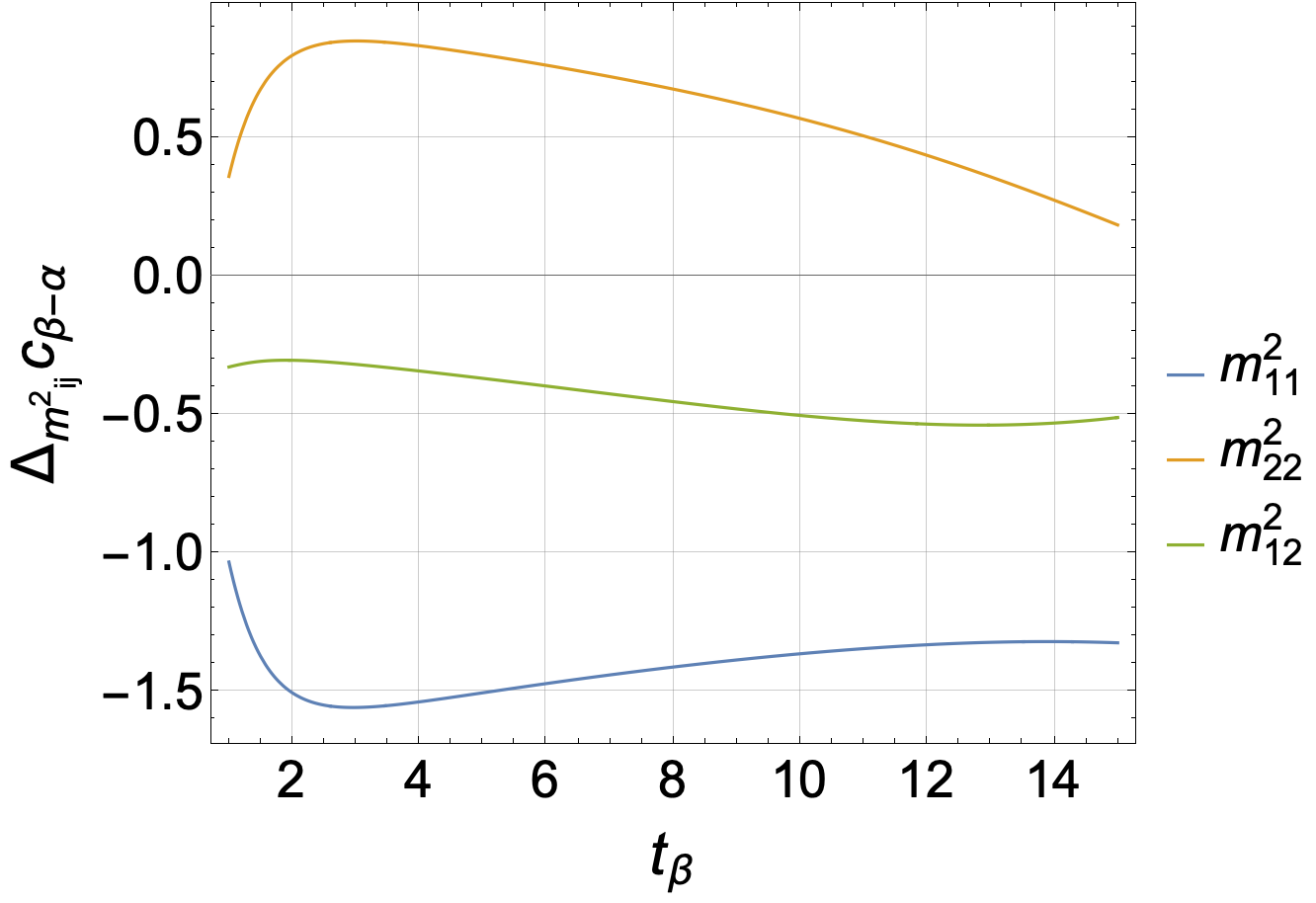}}\hspace{5mm}
    \subfigure
    {\includegraphics[height=4.5cm]{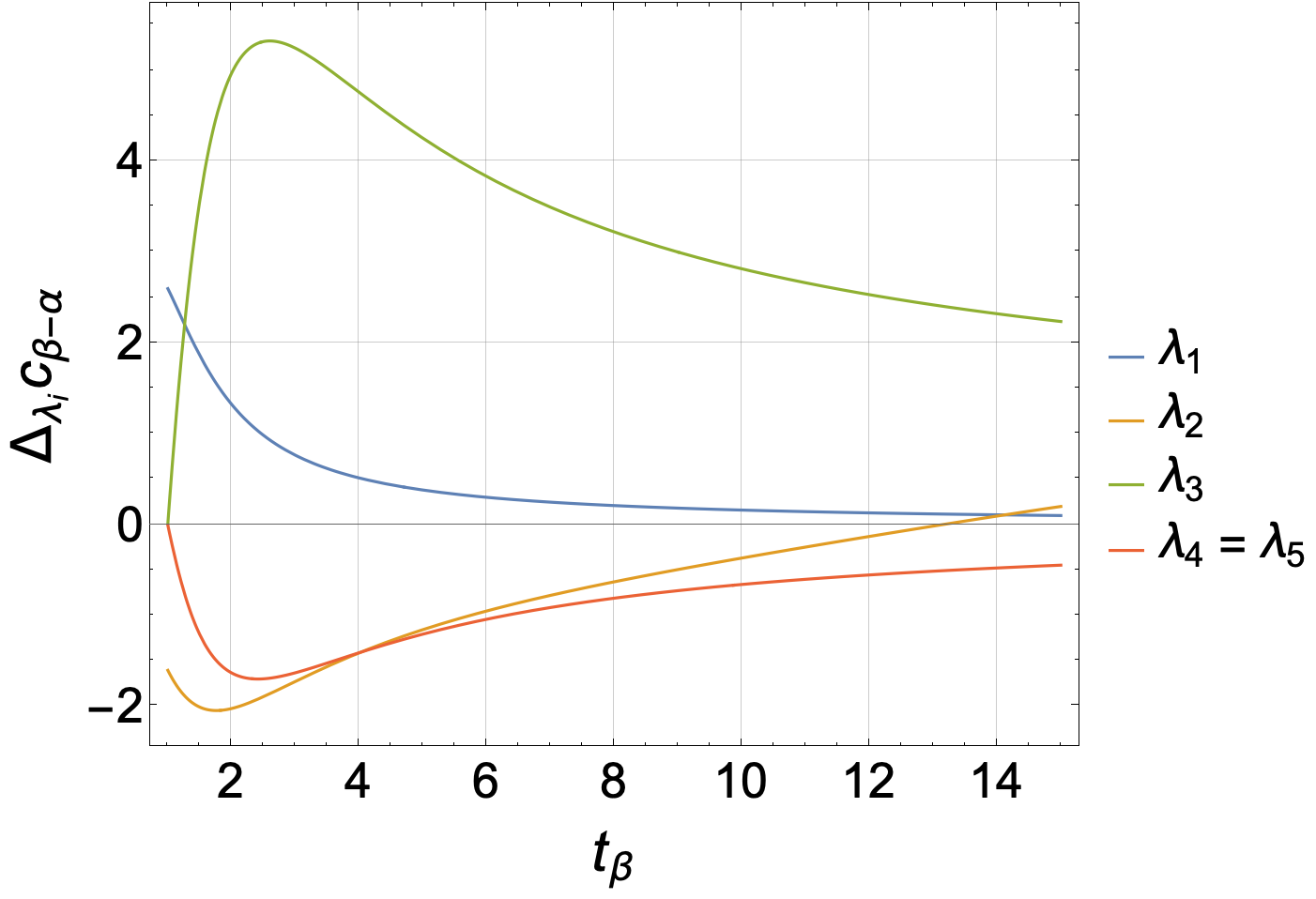}}
    \caption{The same as Fig. \ref{Fig:Van3_FTcos} for Scenario 4.}
    \label{Fig:Van4_FTcos}
\end{figure}

\begin{figure} [h!]
    \centering
    \subfigure{\includegraphics[height=4.5cm]{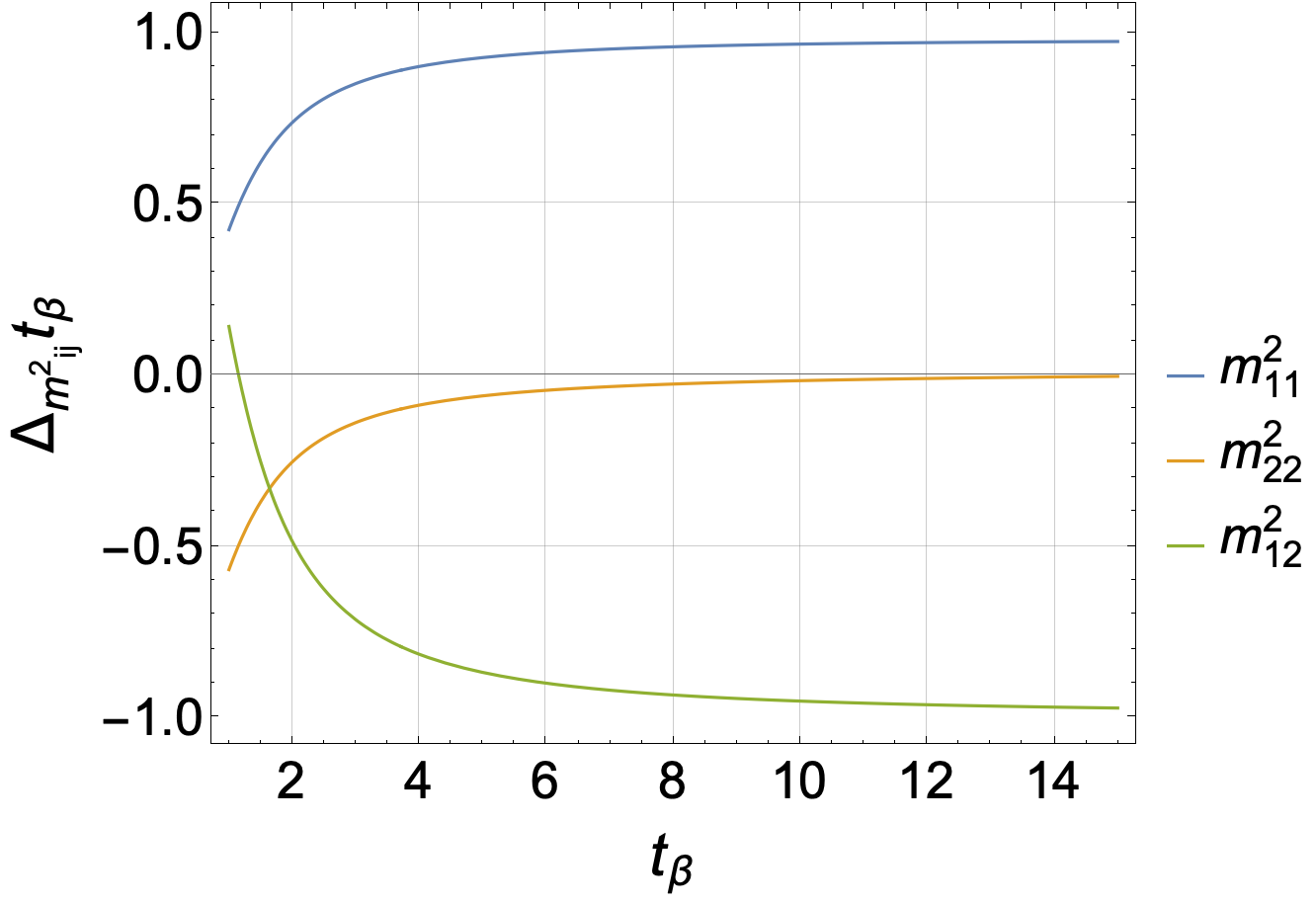}}\hspace{5mm}
    \subfigure{\includegraphics[height=4.5cm]{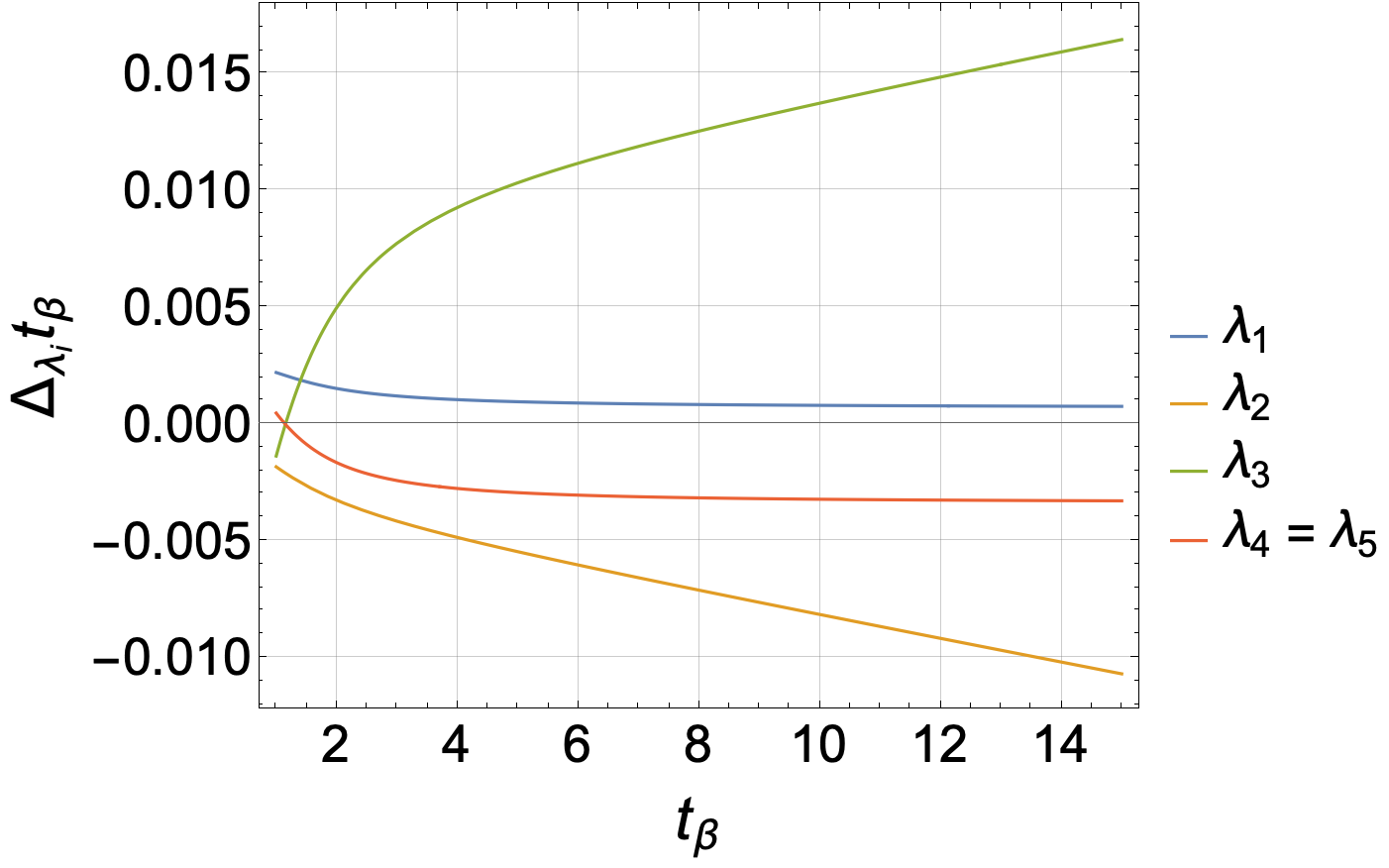}}
    \caption{
    The same as Fig. \ref{Fig:Van3_FTtan} for Scenario 4.}
    \label{Fig:Van4_FTtan}
\end{figure}

This scenario explores the dependence of the various fine-tunings on $t_\beta$, in a scenario with large extra-Higgs masses, $\sim 1500$ GeV, and very sharp alignment, $\epsilon=10^{-3}$.
The results are shown in Figs. \ref{Fig:Van4_FTv}-\ref{Fig:Van4_FTtan}.

Such a strong alignment is actually the natural consequence of two  features that push in that direction: the decoupling due to the large extra masses, and the sizeable $t_\beta$. This is demonstrated by the fact that, even with such alignment, there is no relevant $\Delta c_{\beta-\alpha}$ fine-tuning (Fig.~\ref{Fig:Van4_FTcos}).
For $t_\beta\leq {O}(10)$, there is still a (strong) electroweak fine-tuning, $\Delta v^2$, see Fig.~\ref{Fig:Van4_FTv}, due to the large extra Higgs masses. 
This fine-tuning is compensated by the competing effect of a large $t_\beta$ for $t_\beta\geq {O}(10)$.

Finally, once more, $t_\beta$ is not fine-tuned, as shown in Fig. \ref{Fig:Van4_FTtan}.

\clearpage
\subsection*{Scenario 5}

\begin{figure} [h!]
    \centering
    \subfigure
    {\includegraphics[height=4.5cm]{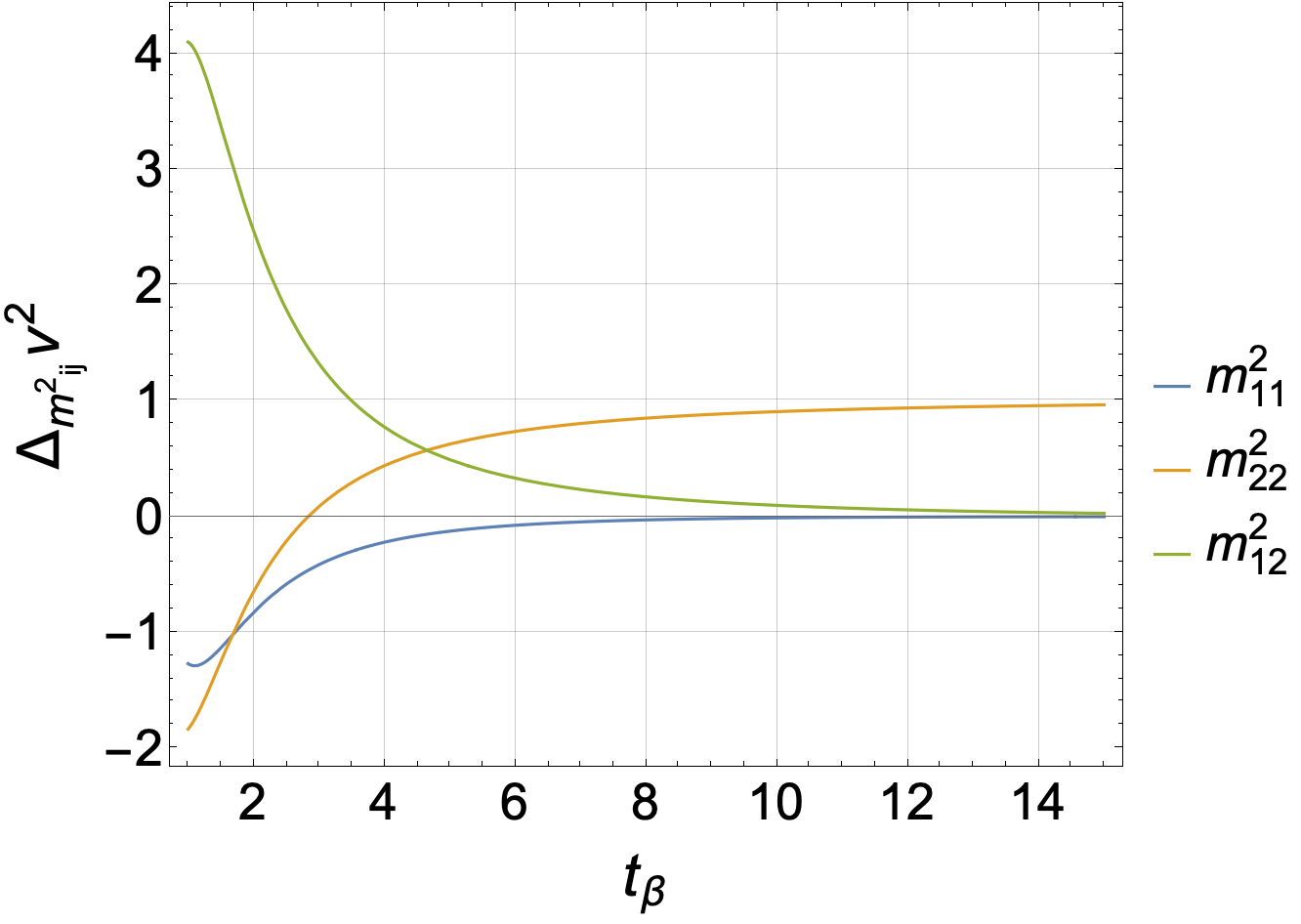}}\hspace{5mm}
    \subfigure
    {\includegraphics[height=4.5cm]{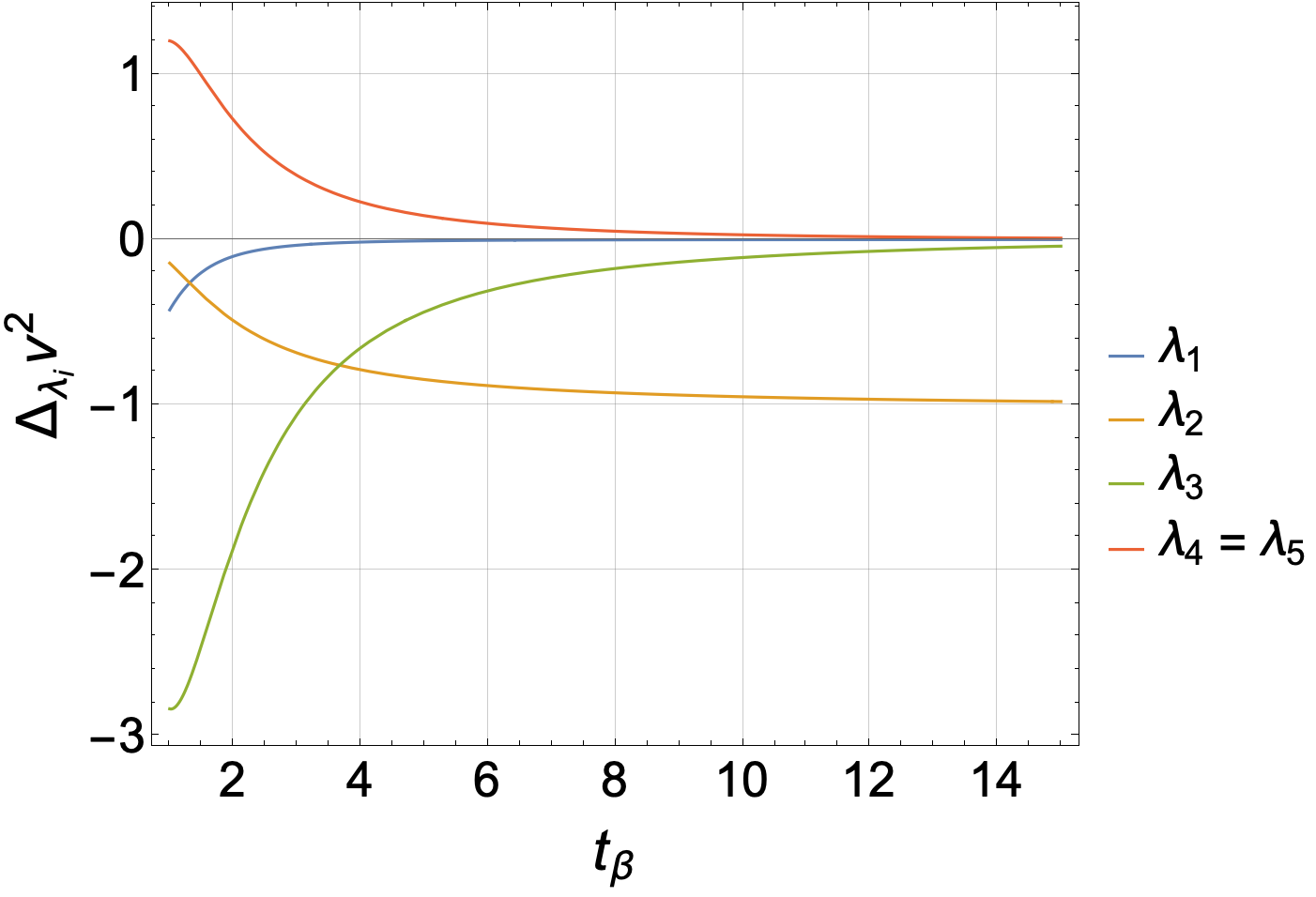}}
    \caption{The same as Fig. \ref{Fig:Van3_FTv} for Scenario 5.}
    \label{Fig:Van5_FTv}
\end{figure}

\begin{figure} [h!]
    \centering
    \subfigure
    {\includegraphics[height=4.5cm]{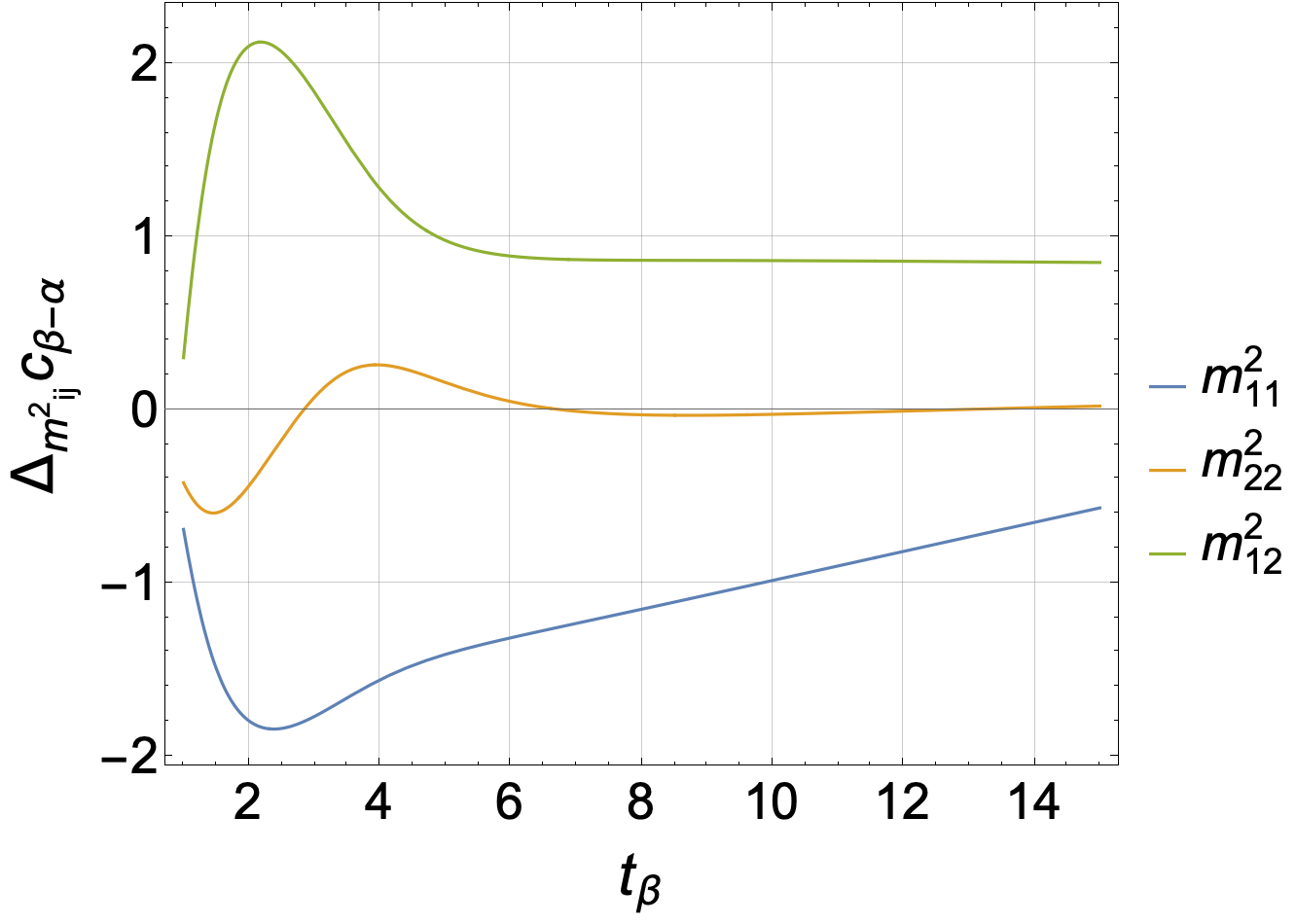}}\hspace{5mm}
    \subfigure
    {\includegraphics[height=4.5cm]{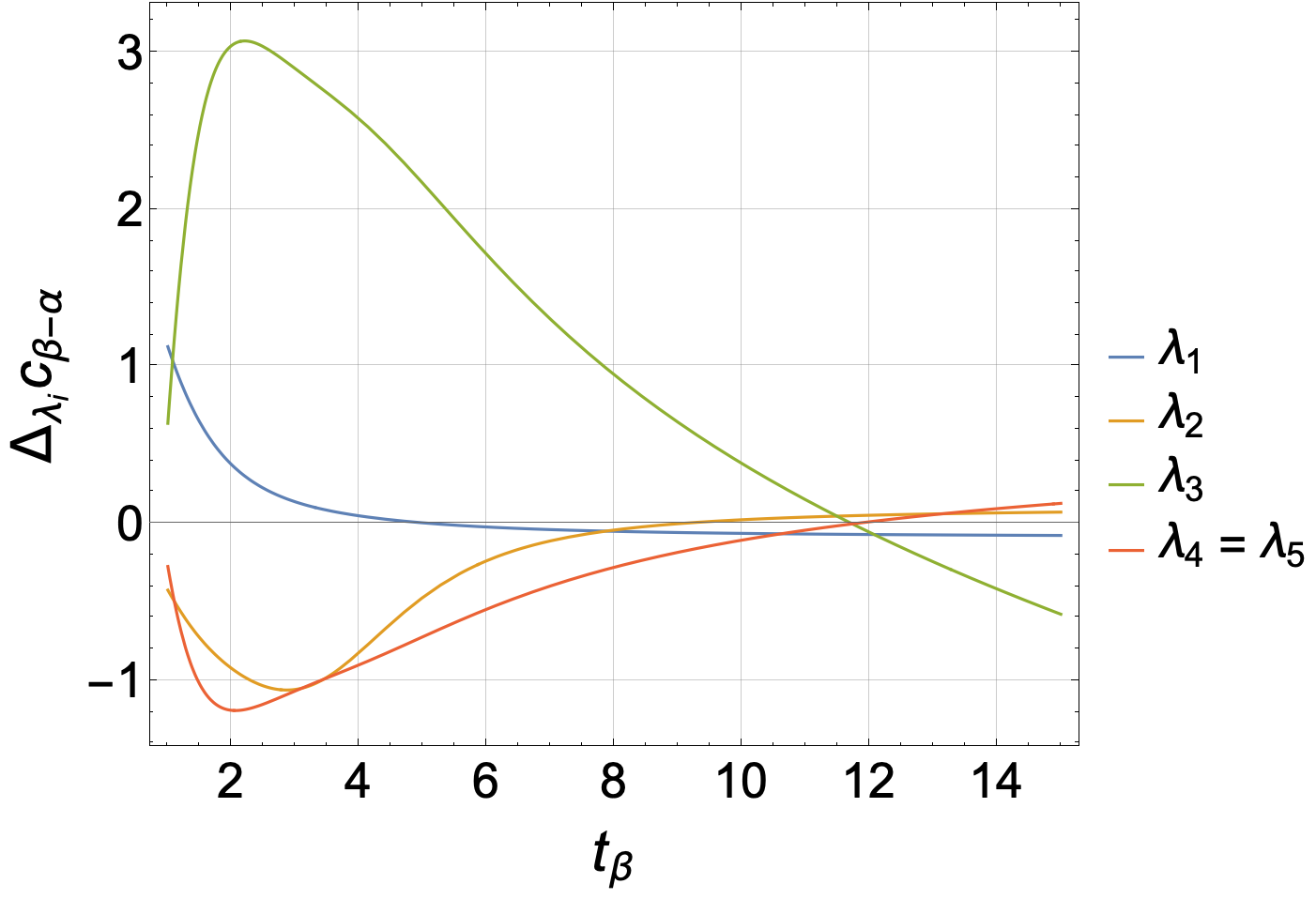}}
    \caption{The same as Fig. \ref{Fig:Van3_FTcos} for Scenario 5.}
    \label{Fig:Van5_FTcos}
\end{figure}

\begin{figure} [h!]
    \centering
    \subfigure{\includegraphics[height=4.5cm]{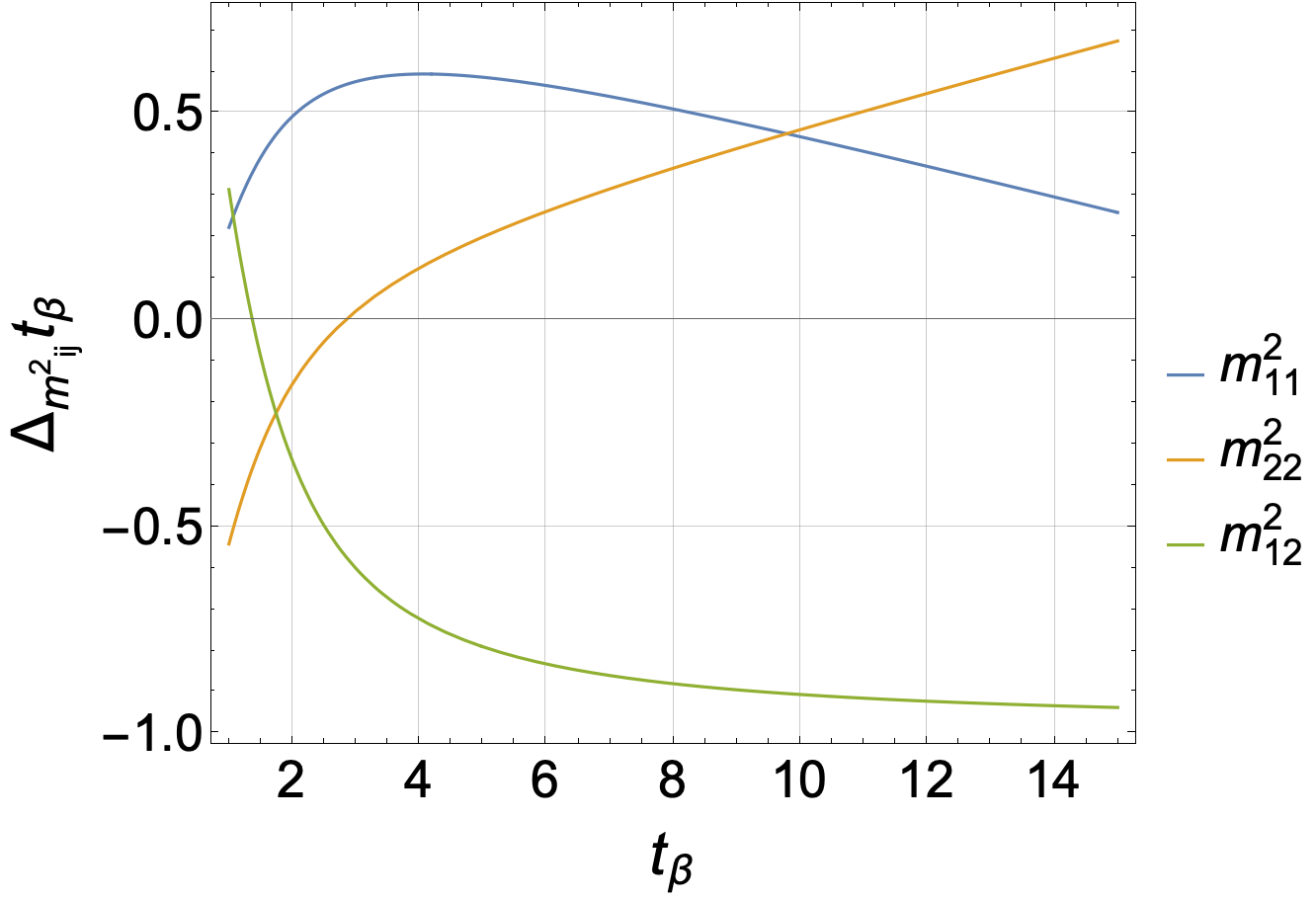}}\hspace{5mm}
    \subfigure{\includegraphics[height=4.5cm]{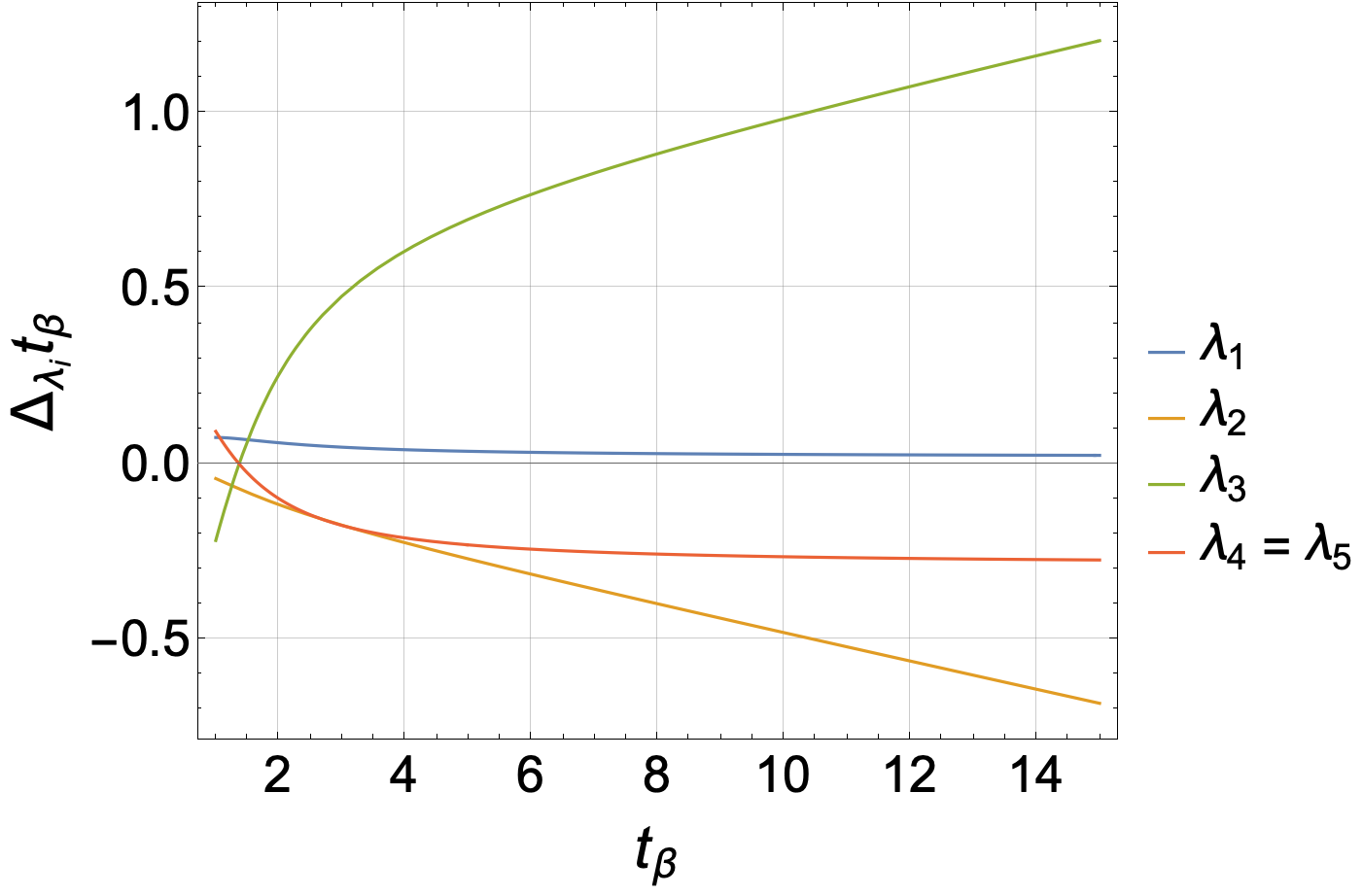}}
    \caption{
    The same as Fig. \ref{Fig:Van3_FTtan} for Scenario 5.}
    \label{Fig:Van5_FTtan}
\end{figure}

 This scenario explores the dependence of the various fine-tunings on $t_\beta$, for small extra-Higgs masses, $\sim 300$ GeV, and mild alignment, $\epsilon=0.1$ (the opposite to scenario 4).
The results are shown in Figs.~\ref{Fig:Van5_FTv}-\ref{Fig:Van5_FTtan}.
In this case, due to the low extra-Higgs masses, there is no appreciable electroweak fine-tuning, Fig.~\ref{Fig:Van5_FTv}. Likewise, due to the (as large as possible) value of $c_{\beta-\alpha}$, there is no fine-tuning for the alignment, Fig.~\ref{Fig:Van5_FTcos}. Of course, $t_\beta$ is not fine-tuned either, Fig.~\ref{Fig:Van5_FTtan}.

\newpage
\section{Conclusions}\label{Conclusions}

The Two-Higgs Doublet Model is one of the most popular and natural extensions of the Higgs sector. It arises in many BSM scenarios, such as grand unification, supersymmetry and axion models. But, besides its theoretical appeal, the 2HDM has two potential fine-tuning problems.

 The first one is related to the electroweak breaking, when the magnitude of the VEV, $v^2=(246\ {\rm GeV})^2$, is much smaller than  the  squared-mass terms entering the theory. This is the familiar electroweak fine-tuning or little-hierarchy problem. The second one is related to the alignment condition. Namely, current experimental data require that the properties of one of the Higgs fields are equal or very similar to those of the SM Higgs boson, i.e. the model must be close to the alignment limit.
 
 In this paper we have explored how these fine-tunings arise in the 2HDM (with the usual $\ZZ_2$ parity implemented to avoid danegerous FCNC), looking for regions of the parameter space where both are mild or irrelevant. In order to quantify the fine-tunings we have used the fairly standard Barbieri-Giudice criterion; e.g.
 the fine-tuning in $v^2$ respect to some initial parameter $\theta$ is quantified by the fine-tuning parameter
 $\Delta_\theta v^2=\partial \log v^2/\partial \log \theta$.
 
 Concerning the electroweak fine-tuning, since there are two vacuum expectation values, $v_1^2+v_2^2=v^2$, one may wonder if there are fine-tunings associated with both $v_1^2, v_2^2$, or equivalently with $v^2$ and $t_\beta=v_2/v_1$. We have shown that, in all cases, $t_\beta$ is not a fine-tuned parameter, so the whole electroweak fine-tuning is captured by $\Delta v^2$. 
 Regarding the alignment, the relevant fine-tuned observable is  $c_{\beta-\alpha}
 =\cos (\beta-\alpha)$, which vanishes for perfect alignment and is currently constrained
by experimental bounds to be  $|c_{\beta-\alpha}|\simlt 0.1$.

We have obtained analytical expressions for $\Delta_{\theta_i}v^2$, $\Delta_{\theta_i}t_\beta$, $\Delta_{\theta_i}c_{\beta-\alpha}$ both in terms of the initial parameters of the theory, $\{m^2_{i,j}, \lambda_i\}$ and on the physical observables, $\{m_H, m_A, m_{H^\pm},t_\beta,\ {\rm etc.} \}$. As already mentioned, there is no fine-tuning for $t_\beta$, i.e. $\Delta_{\theta_i}t_\beta = {\cal O}(1)$ in all cases. On the other hand, the electroweak and alignment fine-tunings are not independent and they should not be multiplied without further ado to get the ``total" fine-tuning.

In particular, in the decoupling limit, once the initial parameters have been adjusted to get the right  $v^2$, there is no need of extra adjustments to get alignment. To remove  the ``double counting" in the evaluation of the $c_{\beta-\alpha}$ fine-tuning one has to discard variations in the $\theta_i$ parameters that change the value of $v^2$. This is equivalent to project the gradient $\Delta_{\theta_i}  c_{\beta-\alpha}=\partial \log c_{\beta-\alpha}/\partial \log\theta_i$ onto the constant-$v^2$ hypersurface. This is the procedure we have followed to present the $\Delta_{\theta_i}v^2$, $\Delta_{\theta_i}c_{\beta-\alpha}$ fine-tunings as independent parameters.

We have also obtained approximate expressions for the various fine-tunings performing an expansion in $c_{\beta-\alpha}$. The general trend for them is:
\beq
\label{aprox}
\De v^2\sim  \frac{4}{t_\beta^2}\frac{m_A^2}{m_h^2},\ \ \ 
\De c_{\beta-\alpha}\sim  \frac{1}{c_{\beta-\alpha}}\frac{1}{t_\beta}\frac{v^2}{m_H^2},\ \ \ \Delta_{\theta_i}t_\beta = {\cal O}(1)
\eeq

We have investigated numerically  the behaviour of the  fine-tunings in the   the 2HDM parameter space, by using representative benchmark scenarios that explore different relevant regions. 

As it is clear from the previous general trends, for moderate values of $t_\beta$, the electroweak and the alignment fine-tunings, $\Delta v^2$ and $\Delta c_{\beta-\alpha}$, become severe in different regions of the parameter space, namely in the regimes of large and small extra-Higgs masses, respectively. The first one corresponds to the decoupling limit. This means that there is an intermediate region, more precisely $500\ {\rm GeV} \simlt \{m_H, m_A, m_{H^\pm}\} \simlt 700\ {\rm GeV}$, where both tunings are acceptably small. 

It should be mentioned here that, if one does not take into account the electroweak fine-tuning it is not correct to say that the alignment becomes natural in the decoupling limit. In that case, one should not project $\Delta c_{\beta-\alpha}$ onto the constant-$v^2$ hypersurface.
Then it contains a contribution of order  $\Delta_{\theta_i}v^2$ that shows that it is in fact fine-tuned. This occurs because the Barbieri-Giudice parameters incorporate all the adjustments required to get the desired size for the observable under consideration.

The analytical expressions and the numerical results show an interesting trend that is not obvious at first glance. Namely, for large $t_\beta$ both the electroweak and the alignment fine-tunings become mitigated. We have discussed in the paper the reason for this feature. In consequence, the 2HDM becomes quite natural for $t_\beta\geq {\cal O} (10)$, even if $m_H, m_A, m_{H^\pm}$ are as large as 1500 GeV.

This situation contrasts with the supersymmetric case, where the $t_\beta\gg 1$ regime implies a notorious fine-tuning.
The reason for that was that the mass parameter $m_{22}^2$ at low energy contained several contributions proportional to the soft supersymmetric masses and the $\mu-$term. Then, the smallness of $|m_{22}^2|$ could only be achieved by a fine cancellation of such  contributions. This does not need to be the case in a generic 2HDM.

In summary, the 2HDM has potentially severe fine-tunings associated with the smallness of the electroweak scale and the sharpness of the alignment, but there are interesting regions in the parameter space where both remain at acceptable levels.

\section*{Acknowledgements}

This work is supported by the grants IFT Centro de Excelencia Severo Ochoa SEV-2016-0597, CEX2020-001007-S and by PID2019-110058GB-C22 funded by MCIN/AEI/10.13039/501100011033 and by ERDF. 
The work of A.B. is supported through the FPI grant PRE2020-095867  funded by MCIN/AEI/10.13039/501100011033.

\newpage

\section{Appendix: Exact Expressions for the Fine-Tuning}\label{Appendix}
\phantom{...}
 \subsection*{Fine-Tuning in $c_{\be-\al}$}
\begin{eqnarray*}
 \Delta_{m^2_{11}} c_{\beta - \alpha}  &=&
 \frac{1}{2 m_h^2 m_H^2 \left(t_\alpha^2+1\right)^3 t_\beta \left(t_\beta^2+1\right)^2 \left(m_h^2-m_H^2\right) (t_\alpha t_\beta+1)}\left[(t_\alpha-t_\beta)\right.\times \\[2mm]
 & & \left. \left(-2 \left(t_\alpha^2+1\right) t_\beta^2 \left(\lambda_5 v^2+m_A^2\right)+m_h^2 t_\alpha \left(t_\beta^2+1\right) (t_\alpha-t_\beta)+m_H^2 \left(t_\beta^2+1\right) (t_\alpha t_\beta+1)\right) \right.\times \\[1mm]
 & &\left.\left(-m_h^2 \left(\lambda_5 v^2 \left(t_\alpha^2 \left(2 t_\beta^2+3\right)+t_\alpha^3 t_\beta-t_\alpha t_\beta \left(3 t_\beta^2+2\right)-t_\beta^2\right) + \right.\right.\right. \\[1mm]
& & \left. \left. \left. m_H^2 \left(t_\alpha^2+1\right) t_\beta \left(t_\alpha^2 t_\beta+t_\alpha \left(t_\beta^2+3\right)-t_\beta\right)\right)+m_H^2 t_\alpha (t_\alpha t_\beta+1)\times \right.\right.\\[1mm]
& & \left.\left.\left(m_H^2 \left(t_\alpha^2 t_\beta-t_\alpha \left(t_\beta^2+1\right)+t_\beta\right)-\lambda_5 v^2 \left(t_\alpha^2 t_\beta-3 t_\alpha \left(t_\beta^2+1\right)+t_\beta\right)\right)+\right.\right.\\[1mm]
& & \left.\left. m_A^2 \left(m_h^2 \left(-t_\alpha^2 \left(2 t_\beta^2+3\right)+t_\alpha^3 (-t_\beta)+t_\alpha t_\beta \left(3 t_\beta^2+2\right)+t_\beta^2\right)+\right.\right.\right.\\[1mm]
& & \left.\left.\left. m_H^2 t_\alpha \left(t_\alpha^3 \left(-t_\beta^2\right)+t_\alpha^2 t_\beta \left(3 t_\beta^2+2\right)+t_\alpha \left(2 t_\beta^2+3\right)-t_\beta\right)\right)+m_h^4 \left(t_\alpha^3 t_\beta+t_\alpha^2-t_\alpha t_\beta^3-t_\beta^2\right)\right)\right]\,,
\end{eqnarray*}
\begin{eqnarray*}
 \Delta_{m^2_{22}} c_{\beta - \alpha}  &=&
 -\frac{1}{2 m_h^2 m_H^2 \left(t_\alpha^2+1\right)^3 t_\beta \left(t_\beta^2+1\right)^2 \left(m_h^2-m_H^2\right) (t_\alpha t_\beta+1)}\left[(t_\alpha-t_\beta) \right.\times \\[2mm] 
 & &\left.\left(2 \left(t_\alpha^2+1\right) t_\beta \left(\lambda_5 v^2+m_A^2\right)+m_h^2 \left(t_\beta^2+1\right) (t_\alpha-t_\beta)-m_H^2 t_\alpha \left(t_\beta^2+1\right) (t_\alpha t_\beta+1)\right)\right.\times\\[1mm]
 & &\left. \left(m_h^2 \left(\lambda_5 t_\alpha v^2 \left(-t_\alpha^2 \left(2 t_\beta^2+3\right)+t_\alpha^3 (-t_\beta)+t_\alpha t_\beta \left(3 t_\beta^2+2\right)+t_\beta^2\right)+\right.\right.\right.\\[1mm]
 & &\left.\left.\left. m_H^2 \left(t_\alpha^3 \left(3 t_\beta^2+1\right)+t_\alpha^4 (-t_\beta)+3 t_\alpha t_\beta^2+t_\alpha+t_\beta\right)\right)-m_H^2 (t_\alpha t_\beta+1)\right.\right. \times\\[1mm]
 & & \left.\left. \left(m_H^2 \left(t_\alpha^2 t_\beta-t_\alpha \left(t_\beta^2+1\right)+t_\beta\right)-\lambda_5 v^2 \left(t_\alpha^2 t_\beta-3 t_\alpha \left(t_\beta^2+1\right)+t_\beta\right)\right)+\right.\right.\\[1mm]
& & \left.\left. m_A^2 \left(m_h^2 t_\alpha \left(-t_\alpha^2 \left(2 t_\beta^2+3\right)+t_\alpha^3 (-t_\beta)+t_\alpha t_\beta \left(3 t_\beta^2+2\right)+t_\beta^2\right)+\right.\right.\right. \\[1mm]
& & \left.\left.\left. m_H^2 \left(t_\alpha^3 t_\beta^2-t_\alpha^2 t_\beta \left(3 t_\beta^2+2\right)-t_\alpha \left(2 t_\beta^2+3\right)+t_\beta\right)\right)+m_h^4 t_\alpha \left(t_\alpha^3 t_\beta+t_\alpha^2-t_\alpha t_\beta^3-t_\beta^2\right)\right)\right]\,,
\end{eqnarray*}
\begin{eqnarray*}
 \Delta_{m^2_{12}} c_{\beta - \alpha}  &=&
 -\frac{1}{m_h^2 m_H^2 \left(t_\alpha^2+1\right)^2 \left(t_\beta^2+1\right)^2 \left(m_h^2-m_H^2\right) (t_\alpha t_\beta+1)}\left[(t_\alpha-t_\beta) \left(\lambda_5 v^2+m_A^2\right)\right.\times\\[2mm]
 & &\left. \left(m_h^2 \left(\lambda_5 v^2 \left(3 t_\alpha^3 \left(t_\beta^2+1\right)+t_\alpha^2 \left(t_\beta-t_\beta^3\right)+t_\alpha^4 t_\beta-3 t_\alpha \left(t_\beta^4+t_\beta^2\right)-t_\beta^3\right)+2 m_H^2 t_\alpha \left(t_\alpha^2+1\right)\right.\right.\right.\times \\[1mm]
 & &\left.\left.\left. \left(t_\beta^4-1\right)\right)-m_H^2 \left(t_\alpha^2 t_\beta^2-1\right) \left(m_H^2 \left(t_\alpha^2 t_\beta-t_\alpha \left(t_\beta^2+1\right)+t_\beta\right)-\right.\right.\right. \\[1mm]
 & & \left.\left.\left.\lambda_5 v^2 \left(t_\alpha^2 t_\beta-3 t_\alpha \left(t_\beta^2+1\right)+t_\beta\right)\right)+m_A^2 \left(m_h^2 \left(3 t_\alpha^3 \left(t_\beta^2+1\right)+t_\alpha^2 \left(t_\beta-t_\beta^3\right)+t_\alpha^4 t_\beta-\right.\right.\right.\right.\\[1mm]
& & \left.\left.\left.\left. 3 t_\alpha \left(t_\beta^4+t_\beta^2\right)-t_\beta^3\right)+m_H^2 \left(t_\alpha^4 t_\beta^3-3 t_\alpha^3 \left(t_\beta^4+t_\beta^2\right)+t_\alpha^2 t_\beta \left(t_\beta^2-1\right)+3 t_\alpha \left(t_\beta^2+1\right)-t_\beta\right)\right)+\right.\right.\\[1mm]
& & \left.\left. m_h^4 \left(-(t_\alpha+t_\beta)^2\right) \left(t_\alpha^2 t_\beta-t_\alpha t_\beta^2+t_\alpha-t_\beta\right)\right)\right]\,,
\end{eqnarray*}
\begin{eqnarray*}
 \Delta_{\lambda_1} c_{\beta - \alpha}  & = & 
\frac{1}{2 m_h^2 m_H^2 \left(t_\alpha^2+1\right)^3 t_\beta \left(t_\beta^2+1\right)^2 \left(m_h^2-m_H^2\right) (t_\alpha t_\beta+1)}\left[(t_\alpha-t_\beta)\right.\times\\[2mm]
& & \left. \left(-\left(t_\alpha^2+1\right) t_\beta^2 \left(\lambda_5 v^2+m_A^2\right)+m_h^2 t_\alpha^2 \left(t_\beta^2+1\right)+m_H^2 \left(t_\beta^2+1\right)\right) \right.\times\\[1mm]
& & \left. \left(m_h^2 (t_\alpha-t_\beta) \left(\lambda_5 v^2 \left(t_\alpha^2 t_\beta+3 t_\alpha \left(t_\beta^2+1\right)+t_\beta\right)+m_H^2 \left(t_\alpha^2+1\right) t_\beta (t_\alpha t_\beta+1)\right)-\right.\right.\\[1mm]
& &\left.\left. m_H^2 t_\alpha (t_\alpha t_\beta+1) \left(m_H^2 \left(t_\alpha^2 t_\beta-t_\alpha \left(t_\beta^2+1\right)+t_\beta\right)-\lambda_5 v^2 \left(t_\alpha^2 t_\beta-3 t_\alpha \left(t_\beta^2+1\right)+t_\beta\right)\right)+\right.\right.\\[1mm]
& &\left.\left. m_A^2 \left(m_h^2 \left(t_\alpha^2 \left(2 t_\beta^2+3\right)+t_\alpha^3 t_\beta-t_\alpha t_\beta \left(3 t_\beta^2+2\right)-t_\beta^2\right)+m_H^2 t_\alpha \left(t_\alpha^3 t_\beta^2-t_\alpha^2 t_\beta \left(3 t_\beta^2+2\right)-\right.\right.\right.\right. \\[1mm]
& & \left.\left.\left.\left. t_\alpha \left(2 t_\beta^2+3\right)+t_\beta\right)\right)+m_h^4 \left(t_\alpha^3 (-t_\beta)-t_\alpha^2+t_\alpha t_\beta^3+t_\beta^2\right)\right)\right]\,,
\end{eqnarray*}
\begin{eqnarray*} 
\Delta_{\lambda_2} c_{\beta - \alpha}   & = & 
\frac{1}{2 m_h^2 m_H^2 \left(t_\alpha^2+1\right)^3 \left(t_\beta^2+1\right)^2 \left(m_h^2-m_H^2\right) (t_\alpha t_\beta+1)}\left[(t_\alpha-t_\beta)\right.\times\\[2mm]
& &  \left. \left(-\lambda_5 t_\alpha^2 v^2-\lambda_5 v^2+m_A^2 \left(-\left(t_\alpha^2+1\right)\right)+m_h^2 \left(t_\beta^2+1\right)+m_H^2 t_\alpha^2 t_\beta^2+m_H^2 t_\alpha^2\right)\right.\times\\[1mm]
& & \left. \left(m_h^2 (t_\alpha-t_\beta) \left(\lambda_5 t_\alpha v^2 \left(t_\alpha^2 t_\beta+3 t_\alpha \left(t_\beta^2+1\right)+t_\beta\right)+m_H^2 \left(t_\alpha^2+1\right) (t_\alpha t_\beta+1)\right)+\right.\right.\\[1mm]
& &\left.\left. m_H^2 (t_\alpha t_\beta+1) \left(m_H^2 \left(t_\alpha^2 t_\beta-t_\alpha \left(t_\beta^2+1\right)+t_\beta\right)-\lambda_5 v^2 \left(t_\alpha^2 t_\beta-3 t_\alpha \left(t_\beta^2+1\right)+t_\beta\right)\right)+\right.\right.\\[1mm]
& & \left.\left. m_A^2 \left(m_h^2 t_\alpha \left(t_\alpha^2 \left(2 t_\beta^2+3\right)+t_\alpha^3 t_\beta-t_\alpha t_\beta \left(3 t_\beta^2+2\right)-t_\beta^2\right)+m_H^2 \left(t_\alpha^3 \left(-t_\beta^2\right)+t_\alpha^2 t_\beta \left(3 t_\beta^2+2\right)+\right.\right.\right.\right.\\[1mm]
& & \left.\left.\left.\left. t_\alpha \left(2 t_\beta^2+3\right)-t_\beta\right)\right)+m_h^4 t_\alpha \left(t_\alpha^3 (-t_\beta)-t_\alpha^2+t_\alpha t_\beta^3+t_\beta^2\right)\right)\right]\,,
\end{eqnarray*}
\begin{eqnarray*}
\Delta_{\lambda_3} c_{\beta - \alpha}   & = & 
-\frac{1}{2 m_h^2 m_H^2 \left(t_\alpha^2+1\right)^3 t_\beta \left(t_\beta^2+1\right)^2 \left(m_h^2-m_H^2\right) (t_\alpha t_\beta+1)}\left[(t_\alpha-t_\beta)\right.\times\\[2mm]
& & \left. \left(-m_h^2 (t_\alpha-t_\beta) \left(m_H^2 \left(t_\alpha^2+1\right) \left(t_\beta^2+1\right) (t_\alpha t_\beta+1)-\lambda_5 v^2 \left(t_\alpha^2 \left(4 t_\beta^2+3\right)+t_\alpha^3 t_\beta+\right.\right.\right.\right.\\[1mm]
& & \left.\left.\left.\left. t_\alpha t_\beta \left(3 t_\beta^2+4\right)+t_\beta^2\right)\right)-m_H^2 \left(t_\alpha^2 t_\beta^2-1\right) \left(m_H^2 \left(t_\alpha^2 t_\beta-t_\alpha \left(t_\beta^2+1\right)+t_\beta\right)-\right.\right.\right.\\[1mm]
& & \left.\left.\left. \lambda_5 v^2 \left(t_\alpha^2 t_\beta-3 t_\alpha \left(t_\beta^2+1\right)+t_\beta\right)\right)+m_A^2 \left(m_h^2 \left(3 t_\alpha^3 \left(t_\beta^2+1\right)+t_\alpha^2 \left(t_\beta-t_\beta^3\right)+t_\alpha^4 t_\beta-\right.\right.\right.\right.\\[1mm]
& &\left.\left.\left.\left. 3 t_\alpha \left(t_\beta^4+t_\beta^2\right)-t_\beta^3\right)+m_H^2 \left(t_\alpha^4 t_\beta^3-3 t_\alpha^3 \left(t_\beta^4+t_\beta^2\right)+t_\alpha^2 t_\beta \left(t_\beta^2-1\right)+3 t_\alpha \left(t_\beta^2+1\right)-t_\beta\right)\right)+\right.\right.\\[1mm]
& & \left.\left. m_h^4 \left(-(t_\alpha+t_\beta)^2\right) \left(t_\alpha^2 t_\beta-t_\alpha t_\beta^2+t_\alpha-t_\beta\right)\right) \left(\left(t_\alpha^2+1\right) t_\beta \left(\lambda_5 v^2+m_A^2-2 m_{H^\pm}^2\right)+\right.\right.\\[1mm]
& &\left.\left. m_h^2 t_\alpha \left(t_\beta^2+1\right)-m_H^2 t_\alpha \left(t_\beta^2+1\right)\right)\right]\,,
\end{eqnarray*}
\begin{eqnarray*}
 \Delta_{\lambda_4} c_{\beta - \alpha}  & = & 
\frac{1}{2 m_h^2 m_H^2 \left(t_\alpha^2+1\right)^2 \left(t_\beta^2+1\right)^2 \left(m_h^2-m_H^2\right) (t_\alpha t_\beta+1)}\left[(t_\alpha-t_\beta)\left(\lambda_5 v^2+2 m_A^2-2 m_{H^\pm}^2\right)\right.\times\\[2mm]
& &\left.  \left(-m_h^2 (t_\alpha-t_\beta) \left(m_H^2 \left(t_\alpha^2+1\right) \left(t_\beta^2+1\right) (t_\alpha t_\beta+1)-\lambda_5 v^2 \left(t_\alpha^2 \left(4 t_\beta^2+3\right)+t_\alpha^3 t_\beta+\right.\right.\right.\right.\\[1mm]
& & \left.\left.\left.\left. t_\alpha t_\beta \left(3 t_\beta^2+4\right)+t_\beta^2\right)\right)-m_H^2 \left(t_\alpha^2 t_\beta^2-1\right) \left(m_H^2 \left(t_\alpha^2 t_\beta-t_\alpha \left(t_\beta^2+1\right)+t_\beta\right)-\right.\right.\right.\\[1mm]
& & \left.\left.\left.\lambda_5 v^2 \left(t_\alpha^2 t_\beta-3 t_\alpha \left(t_\beta^2+1\right)+t_\beta\right)\right)+m_A^2 \left(m_h^2 \left(3 t_\alpha^3 \left(t_\beta^2+1\right)+t_\alpha^2 \left(t_\beta-t_\beta^3\right)+t_\alpha^4 t_\beta-\right.\right.\right.\right.\\[1mm]
& &\left.\left.\left.\left. 3 t_\alpha \left(t_\beta^4+t_\beta^2\right)-t_\beta^3\right)+m_H^2 \left(t_\alpha^4 t_\beta^3-3 t_\alpha^3 \left(t_\beta^4+t_\beta^2\right)+t_\alpha^2 t_\beta \left(t_\beta^2-1\right)+3 t_\alpha \left(t_\beta^2+1\right)-t_\beta\right)\right)+\right.\right. \\[1mm]
& &\left.\left. m_h^4 \left(-(t_\alpha+t_\beta)^2\right) \left(t_\alpha^2 t_\beta-t_\alpha t_\beta^2+t_\alpha-t_\beta\right)\right)\right]\,,
\end{eqnarray*}
\begin{eqnarray*}
  \Delta_{\lambda_5} c_{\beta - \alpha}   & = & 
\frac{1}{2 m_h^2 m_H^2 \left(t_\alpha^2+1\right)^2 \left(t_\beta^2+1\right)^2 \left(m_h^2-m_H^2\right) (t_\alpha t_\beta+1)}\left[\lambda_5 v^2 (t_\alpha-t_\beta) \right.\times\\[2mm]
& & \left. \left(-m_h^2 (t_\alpha-t_\beta) \left(m_H^2 \left(t_\alpha^2+1\right) \left(t_\beta^2+1\right) (t_\alpha t_\beta+1)-\lambda_5 v^2 \left(t_\alpha^2 \left(4 t_\beta^2+3\right)+t_\alpha^3 t_\beta+\right.\right.\right.\right.\\[1mm]
& &\left.\left.\left.\left. t_\alpha t_\beta \left(3 t_\beta^2+4\right)+t_\beta^2\right)\right)-m_H^2 \left(t_\alpha^2 t_\beta^2-1\right) \left(m_H^2 \left(t_\alpha^2 t_\beta-t_\alpha \left(t_\beta^2+1\right)+t_\beta\right)-\right.\right.\right.\\[1mm]
& &\left.\left.\left. \lambda_5 v^2 \left(t_\alpha^2 t_\beta-3 t_\alpha \left(t_\beta^2+1\right)+t_\beta\right)\right)+m_A^2 \left(m_h^2 \left(3 t_\alpha^3 \left(t_\beta^2+1\right)+t_\alpha^2 \left(t_\beta-t_\beta^3\right)+t_\alpha^4 t_\beta-\right.\right.\right.\right.\\[1mm]
& & \left.\left.\left.\left. 3 t_\alpha \left(t_\beta^4+t_\beta^2\right)-t_\beta^3\right)+m_H^2 \left(t_\alpha^4 t_\beta^3-3 t_\alpha^3 \left(t_\beta^4+t_\beta^2\right)+t_\alpha^2 t_\beta \left(t_\beta^2-1\right)+3 t_\alpha \left(t_\beta^2+1\right)-t_\beta\right)\right)+\right.\right.\\[1mm]
& & \left.\left. m_h^4 \left(-(t_\alpha+t_\beta)^2\right) \left(t_\alpha^2 t_\beta-t_\alpha t_\beta^2+t_\alpha-t_\beta\right)\right)\right]\,.
\end{eqnarray*}
\newpage
\subsection*{Fine-tuning in $v^2$}
\begin{eqnarray*}
\nonumber
 \Delta_{m^2_{11}}  v^2   & = &
\frac{1}{m_h^2 m_H^2 \left(t_\alpha^2+1\right)^2 \left(t_\beta^2+1\right)^2}\left[\left(m_h^2 (t_\alpha t_\beta+1)+m_H^2 t_\alpha (t_\alpha-t_\beta)\right)\times\right.\\[2mm]\nonumber 
& &\left.\left(-2 \left(t_\alpha^2+1\right) t_\beta^2 \left(\lambda_5 v^2+m_A^2\right)+m_h^2 t_\alpha \left(t_\beta^2+1\right) (t_\alpha-t_\beta)+m_H^2 \left(t_\beta^2+1\right) (t_\alpha t_\beta+1)\right)\right]\,,\\[5mm] \nonumber
 \Delta_{m^2_{22}}   v^2  & = &
-\frac{1}{m_h^2 m_H^2 \left(t_\alpha^2+1\right)^2 \left(t_\beta^2+1\right)^2}\left[\left(m_h^2 t_\alpha (t_\alpha t_\beta+1)+m_H^2 (t_\beta-t_\alpha)\right)\right.\times\\[2mm]\nonumber
& & \left. \left(2 \left(t_\alpha^2+1\right) t_\beta \left(\lambda_5 v^2+m_A^2\right)+m_h^2 \left(t_\beta^2+1\right) (t_\alpha-t_\beta)-m_H^2 t_\alpha \left(t_\beta^2+1\right) (t_\alpha t_\beta+1)\right)\right]\,,\\[5mm] \nonumber
    \Delta_{m^2_{12}}  v^2  & = &
 \frac{2 t_\beta \left(\lambda_5 v^2+m_A^2\right) \left(m_h^2 \left(t_\alpha^2 t_\beta+t_\alpha t_\beta^2+t_\alpha+t_\beta\right)+m_H^2 \left(t_\alpha^2 t_\beta-t_\alpha \left(t_\beta^2+1\right)+t_\beta\right)\right)}{m_h^2 m_H^2 \left(t_\alpha^2+1\right) \left(t_\beta^2+1\right)^2}\,,\\[5mm]\nonumber
  \Delta_{\lambda_1}   v^2   & = &
-\frac{1}{m_h^2 m_H^2 \left(t_\alpha^2+1\right)^2 \left(t_\beta^2+1\right)^2}\left[\left(m_h^2 (t_\alpha t_\beta+1)+m_H^2 t_\alpha (t_\alpha-t_\beta)\right)\right.\times\\[2mm]\nonumber
& &\left. \left(-\left(t_\alpha^2+1\right) t_\beta^2 \left(\lambda_5 v^2+m_A^2\right)+m_h^2 t_\alpha^2 \left(t_\beta^2+1\right)+m_H^2 \left(t_\beta^2+1\right)\right)\right]\,,\\[5mm] \nonumber
  \Delta_{\lambda_2}  v^2   & = &
-\frac{1}{m_h^2 m_H^2 \left(t_\alpha^2+1\right)^2 \left(t_\beta^2+1\right)^2}\left[t_\beta \left(m_h^2 t_\alpha (t_\alpha t_\beta+1)+m_H^2 (t_\beta-t_\alpha)\right)\right.\times\\[2mm]\nonumber
& & \left. \left(-\lambda_5 t_\alpha^2 v^2-\lambda_5 v^2+m_A^2 \left(-\left(t_\alpha^2+1\right)\right)+m_h^2 \left(t_\beta^2+1\right)+m_H^2 t_\alpha^2 t_\beta^2+m_H^2 t_\alpha^2\right)\right]\,,\\[5mm] \nonumber
  \Delta_{\lambda_3}   v^2   & = &
\frac{1}{m_h^2 m_H^2 \left(t_\alpha^2+1\right)^2 \left(t_\beta^2+1\right)^2}\left[\left(m_h^2 \left(t_\alpha^2 t_\beta+t_\alpha t_\beta^2+t_\alpha+t_\beta\right)+m_H^2 \left(t_\alpha^2 t_\beta-t_\alpha \left(t_\beta^2+1\right)+t_\beta\right)\right)\right.\times\\[2mm]\nonumber
& &\left. \left(\left(t_\alpha^2+1\right) t_\beta \left(\lambda_5 v^2+m_A^2-2 m_{H^\pm}^2\right)+m_h^2 t_\alpha \left(t_\beta^2+1\right)-m_H^2 t_\alpha \left(t_\beta^2+1\right)\right)\right]\,,\\[5mm] \nonumber
  \Delta_{\lambda_4}   v^2   & = &
-\frac{t_\beta \left(\lambda_5 v^2+2 m_A^2-2 m_{H^\pm}^2\right) \left(m_h^2 \left(t_\alpha^2 t_\beta+t_\alpha t_\beta^2+t_\alpha+t_\beta\right)+m_H^2 \left(t_\alpha^2 t_\beta-t_\alpha \left(t_\beta^2+1\right)+t_\beta\right)\right)}{m_h^2 m_H^2 \left(t_\alpha^2+1\right) \left(t_\beta^2+1\right)^2}\,,\\[5mm]\nonumber
  \Delta_{\lambda_5}   v^2   & = &
-\frac{\lambda_5 t_\beta v^2 \left(m_h^2 \left(t_\alpha^2 t_\beta+t_\alpha t_\beta^2+t_\alpha+t_\beta\right)+m_H^2 \left(t_\alpha^2 t_\beta-t_\alpha \left(t_\beta^2+1\right)+t_\beta\right)\right)}{m_h^2 m_H^2 \left(t_\alpha^2+1\right) \left(t_\beta^2+1\right)^2}\,.\nonumber
 \end{eqnarray*} 

 \subsection*{Fine-tuning in $t_\be$}
 
 \begin{eqnarray*} 
 \nonumber
 \Delta_{m^2_{11}} t_\be   & = &
\frac{1}{2 m_h^2 m_H^2 \left(t_\alpha^2+1\right)^2 \left(t_\beta^3+t_\beta\right)}\left[\left(m_h^2 (t_\alpha-t_\beta)-m_H^2 t_\alpha (t_\alpha t_\beta+1)\right)\right.\times\\[2mm]\nonumber
& &\left.\left(-2 \left(t_\alpha^2+1\right) t_\beta^2 \left(\lambda_5 v^2+m_A^2\right)+m_h^2 t_\alpha \left(t_\beta^2+1\right) (t_\alpha-t_\beta)+m_H^2 \left(t_\beta^2+1\right) (t_\alpha t_\beta+1)\right)\right]\,,\\[5mm] \nonumber
 \Delta_{m^2_{22}} t_\be   & = &
\frac{1}{2 m_h^2 m_H^2 \left(t_\alpha^2+1\right)^2 \left(t_\beta^3+t_\beta\right)}\left[\left(m_h^2 t_\alpha (t_\alpha-t_\beta)+m_H^2 (t_\alpha t_\beta+1)\right)\right.\times\\[2mm]\nonumber
& &\left. \left(-2 \left(t_\alpha^2+1\right) t_\beta \left(\lambda_5 v^2+m_A^2\right)+m_h^2 \left(t_\beta^2+1\right) (-(t_\alpha-t_\beta))+m_H^2 t_\alpha \left(t_\beta^2+1\right) (t_\alpha t_\beta+1)\right)\right]\,,\\[5mm] \nonumber
 \Delta_{m^2_{12}} t_\be   & = &
\frac{\left(\lambda_5 v^2+m_A^2\right) \left(m_h^2 \left(t_\alpha^2-t_\beta^2\right)+m_H^2 \left(1-t_\alpha^2 t_\beta^2\right)\right)}{m_h^2 m_H^2 \left(t_\alpha^2+1\right) \left(t_\beta^2+1\right)}\,,\\[5mm] 
    \Delta_{\lambda_1} t_\be   & = & 
-\frac{1}{2 m_h^2 m_H^2 \left(t_\alpha^2+1\right)^2 \left(t_\beta^3+t_\beta\right)}\left[\left(m_h^2 (t_\alpha-t_\beta)-m_H^2 t_\alpha (t_\alpha t_\beta+1)\right)\right.\times\\[2mm]\nonumber
& & \left. \left(-\left(t_\alpha^2+1\right) t_\beta^2 \left(\lambda_5 v^2+m_A^2\right)+m_h^2 t_\alpha^2 \left(t_\beta^2+1\right)+m_H^2 \left(t_\beta^2+1\right)\right)\right]\,,\\[5mm] \nonumber
     \Delta_{\lambda_2} t_\be   & = & 
-\frac{1}{2 m_h^2 m_H^2 \left(t_\alpha^2+1\right)^2 \left(t_\beta^2+1\right)}\left[\left(m_h^2 t_\alpha (t_\alpha-t_\beta)+m_H^2 (t_\alpha t_\beta+1)\right)\right.\times\\[2mm]\nonumber
& & \left. \left(-\lambda_5 t_\alpha^2 v^2-\lambda_5 v^2+m_A^2 \left(-\left(t_\alpha^2+1\right)\right)+m_h^2 \left(t_\beta^2+1\right)+m_H^2 t_\alpha^2 t_\beta^2+m_H^2 t_\alpha^2\right)\right]\,,\\[5mm] \nonumber
\Delta_{\lambda_3} t_\be  & = & 
\frac{1}{2 m_h^2 m_H^2 \left(t_\alpha^2+1\right)^2 \left(t_\beta^3+t_\beta\right)}\left[\left(m_h^2 \left(t_\alpha^2-t_\beta^2\right)+m_H^2 \left(1-t_\alpha^2 t_\beta^2\right)\right)\right.\times\\[2mm]\nonumber
& & \left. \left(\left(t_\alpha^2+1\right) t_\beta \left(\lambda_5 v^2+m_A^2-2 m_{H^\pm}^2\right)+m_h^2 t_\alpha \left(t_\beta^2+1\right)-m_H^2 t_\alpha \left(t_\beta^2+1\right)\right)\right]\,,\\[5mm] \nonumber
 \Delta_{\lambda_4} t_\be  & = &  
\frac{\left(\lambda_5 v^2+2 m_A^2-2 m_{H^\pm}^2\right) \left(m_h^2 \left(t_\beta^2-t_\alpha^2\right)+m_H^2 \left(t_\alpha^2 t_\beta^2-1\right)\right)}{2 m_h^2 m_H^2 \left(t_\alpha^2+1\right) \left(t_\beta^2+1\right)}\,,\\[5mm] \nonumber
  \Delta_{\lambda_5} t_\be  & = &  
\frac{\lambda_5 v^2 \left(m_h^2 \left(t_\beta^2-t_\alpha^2\right)+m_H^2 \left(t_\alpha^2 t_\beta^2-1\right)\right)}{2 m_h^2 m_H^2 \left(t_\alpha^2+1\right) \left(t_\beta^2+1\right)}\,.\nonumber
 \end{eqnarray*}

\newpage

  \bibliographystyle{JHEP.bst} 
  \bibliography{refs.bib} 

\end{document}